\documentclass[aps,a4paper,superscriptaddress,showpacs,preprintnumbers,amsmath,amssymb]{revtex4}

\usepackage{ulem}

\usepackage{psfrag} \usepackage{graphicx} \usepackage{dcolumn}
\usepackage{color} \usepackage{latexsym,amsfonts} \usepackage{bm}
\usepackage{amssymb}
\begin{document}

\title{Theoretical description \\ of the neutron beta decay in the
  standard model at the level of $10^{-5}$}

\author{A. N. Ivanov}\email{ivanov@kph.tuwien.ac.at}
\affiliation{Atominstitut, Technische Universit\"at Wien, Stadionallee
  2, A-1020 Wien, Austria}
\author{R. H\"ollwieser}\email{roman.hoellwieser@gmail.com}
\affiliation{Atominstitut, Technische Universit\"at Wien, Stadionallee
  2, A-1020 Wien, Austria}\affiliation{Department of Physics,
  Bergische Universit\"at Wuppertal, Gaussstr. 20, D-42119 Wuppertal,
  Germany} \author{N. I. Troitskaya}\email{natroitskaya@yandex.ru}
\affiliation{Atominstitut, Technische Universit\"at Wien, Stadionallee
  2, A-1020 Wien, Austria}
\author{M. Wellenzohn}\email{max.wellenzohn@gmail.com}
\affiliation{Atominstitut, Technische Universit\"at Wien, Stadionallee
  2, A-1020 Wien, Austria} \affiliation{FH Campus Wien, University of
  Applied Sciences, Favoritenstra\ss e 226, 1100 Wien, Austria}
\author{Ya. A. Berdnikov}\email{berdnikov@spbstu.ru}\affiliation{Peter
  the Great St. Petersburg Polytechnic University, Polytechnicheskaya
  29, 195251, Russian Federation}

\date{\today}

\begin{abstract}
  In the framework of the Standard Model (SM) a theoretical
  description of the neutron beta decay is given at the level of
  $10^{-5}$. The neutron lifetime and correlation coefficients of the
  neutron beta decay for a polarized neutron, a polarized electron and
  an unpolarized proton are calculated at the account for i) the
  radiative corrections (RC) of order $O(\alpha E_e/m_N) \sim
  10^{-5}$, i.e. $O(\alpha E_e/m_N)$ RC, to Sirlin's {\it outer} and
  {\it inner} $O(\alpha/\pi)$ RC, where $\alpha$ and $E_e$ are the
  fine-structure constant and the electron energy, respectively, ii)
  the outer $O(\alpha E_e/m_N)$ RC, caused by Sirlin's outer
  $O(\alpha/\pi)$ RC and the phase-volume of the neutron beta decay,
  calculated to next-to-leading order in the large nucleon mass $m_N$
  expansion, iii) the corrections of order $O(E^2_e/m^2_N) \sim
  10^{-5}$, caused by weak magnetism and proton recoil and iv)
  Wilkinson's corrections of order $10^{-5}$ (Wilkinson, Nucl. Phys. A
  {\bf 377}, 474 (1982)). These corrections define the SM background
  of the theoretical description of the neutron beta decay at the
  level of $10^{-5}$, which is required by experimental searches of
  interactions beyond the SM with experimental uncertainties of a few
  parts of $10^{-5}$.
\end{abstract}
\pacs{12.15.Ff, 13.15.+g, 23.40.Bw, 26.65.+t}

\maketitle

\section{Introduction}
\label{sec:introduction}

A contemporary level of sensitivity of about $10^{-4}$ or even better
for experimental investigations of the neutron beta decay
\cite{Abele2008, Nico2009, Paul2009} with a polarized neutron and
unpolarized electron and proton \cite{Abele2016, Abele2018,
  Sirlin2018, Dubbers2021} and with a polarized neutron, a polarized
electron and an unpolarized proton \cite{Bodek2019} demand the
theoretical description of the neutron beta decay within the Standard
Model (SM) \cite{DGH2014, PDG2020} at the level of $10^{-5}$. As has
been shown in \cite{Ivanov2013, Ivanov2017, Ivanov2018, Ivanov2019a}
Wilkinson's corrections \cite{Wilkinson1982} provide the SM
contributions to the neutron lifetime and correlation coefficients of
the neutron beta decay of order $10^{-5}$. Of course, they do not
exhaust a complete set of the SM corrections of order $10^{-5}$.

In Refs. \cite{Ivanov2019b} and \cite{Ivanov2020a} we have calculated
radiative corrections (RC) of order $O(\alpha E_e/m_N) \sim 10^{-5}$,
i.e.  $O(\alpha E_e/m_N)$ RC, where $\alpha$, $E_e$ and $m_N$ are the
fine--structure constant \cite{PDG2020}, the electron energy and the
nucleon mass, respectively, to Sirlin's {\it outer} and {\it inner}
(see \cite{Wilkinson1970}) $O(\alpha/\pi)$ RC \cite{Sirlin1967,
  Sirlin1978} (see also \cite{Shann1971, Ando2004, Gudkov2006}), which
are independent of the hadronic structure of the neutron (see
\cite{Ivanov2019b} ) and induced by the hadronic structure of the
neutron (see \cite{Ivanov2020a}), respectively. In turn, in
\cite{Ivanov2020b} we have calculated a complete set of corrections of
order $O(E^2_e/m^2_N) \sim 10^{-5}$, caused by weak magnetism and
proton recoil. Together with Wilkinson's corrections
\cite{Wilkinson1982} (see also \cite{Ivanov2013, Ivanov2017,
  Ivanov2018, Ivanov2019a}) the corrections, calculated in
\cite{Ivanov2019b, Ivanov2020a, Ivanov2020b}, define the SM background
of the theoretical description of the neutron beta decay at the level
of $10^{-5}$.In this work we supplement this SM theoretical background
of the neutron beta decay by the outer $O(\alpha E_e/m_N) \sim
10^{-5}$ RC, caused by Sirlin's outer $O(\alpha/\pi)$ RC and the
phase-volume of the neutron beta decay, calculated to next-to-leading
order (NLO) in the large nucleon mass $m_N$ expansion (see Appendix
D).

For the electron-energy and angular distribution of the neutron beta
decay for a polarized neutron, a polarized electron and an unpolarized
proton we use the most general form, proposed by Jackson {\it et al.}
\cite{Jackson1957a, Jackson1957b, Jackson1958} and Ebel and Feldman
\cite{Ebel1957}:
\begin{eqnarray}\label{eq:1}
\hspace{-0.15in}&&\frac{d^5 \lambda_n(E_e, \vec{k}_e,
  \vec{k}_{\bar{\nu}}, \vec{\xi}_n, \vec{\xi}_e)}{dE_e d\Omega_e
  d\Omega_{\bar{\nu}}} = (1 + 3 g^2_A)\,\frac{|G_V|^2}{16 \pi^5}\,(E_0
- E_e)^2 \,\sqrt{E^2_e - m^2_e}\, E_e\,F(E_e, Z =
1)\,\zeta(E_e)\,\Big\{1 + b(E_e)\,\frac{m_e}{E_e}\nonumber\\
\hspace{-0.15in}&& + a(E_e)\,\frac{\vec{k}_e\cdot
  \vec{k}_{\bar{\nu}}}{E_e E_{\bar{\nu}}} +
A(E_e)\,\frac{\vec{\xi}_n\cdot \vec{k}_e}{E_e} + B(E_e)\,
\frac{\vec{\xi}_n\cdot \vec{k}_{\bar{\nu}}}{E_{\bar{\nu}}} +
K_n(E_e)\,\frac{(\vec{\xi}_n\cdot \vec{k}_e)(\vec{k}_e\cdot
  \vec{k}_{\bar{\nu}})}{E^2_e E_{\bar{\nu}}}+
Q_n(E_e)\,\frac{(\vec{\xi}_n\cdot \vec{k}_{\bar{\nu}})(\vec{k}_e\cdot
  \vec{k}_{\bar{\nu}})}{ E_e E^2_{\bar{\nu}}}\nonumber\\
\hspace{-0.15in}&& + D(E_e)\,\frac{\vec{\xi}_n\cdot (\vec{k}_e\times
  \vec{k}_{\bar{\nu}})}{E_e E_{\bar{\nu}}} + G(E_e)\,\frac{\vec{\xi}_e
  \cdot \vec{k}_e}{E_e} + H(E_e)\,\frac{\vec{\xi}_e \cdot
  \vec{k}_{\bar{\nu}}}{E_{\bar{\nu}}} + N(E_e)\,\vec{\xi}_n\cdot
\vec{\xi}_e + Q_e(E_e)\,\frac{(\vec{\xi}_n\cdot \vec{k}_e)(
  \vec{k}_e\cdot \vec{\xi}_e)}{(E_e + m_e) E_e}\nonumber\\
\hspace{-0.15in}&& + K_e(E_e)\,\frac{(\vec{\xi}_e\cdot \vec{k}_e)(
  \vec{k}_e\cdot \vec{k}_{\bar{\nu}})}{(E_e + m_e)E_e E_{\bar{\nu}}} +
R(E_e)\,\frac{\vec{\xi}_n\cdot(\vec{k}_e \times \vec{\xi}_e)}{E_e} +
L(E_e)\,\frac{\vec{\xi}_e\cdot(\vec{k}_e \times
  \vec{k}_{\bar{\nu}})}{E_eE_{\bar{\nu}}} +
S(E_e)\,\frac{(\vec{\xi}_n\cdot \vec{\xi}_e)(\vec{k}_e \cdot
  \vec{k}_{\bar{\nu}})}{E_e E_{\bar{\nu}}} \nonumber\\
\hspace{-0.15in}&& + T(E_e)\,\frac{(\vec{\xi}_n \cdot
  \vec{k}_{\bar{\nu}})(\vec{\xi}_e \cdot \vec{k}_e)}{E_e
  E_{\bar{\nu}}} + U(E_e)\, \frac{(\vec{\xi}_n\cdot
  \vec{k}_e)(\vec{\xi}_e \cdot \vec{k}_{\bar{\nu}})}{E_e
  E_{\bar{\nu}}} + V(E_e)\, \frac{\vec{\xi}_n\cdot (\vec{\xi}_e \times
  \vec{k}_{\bar{\nu}})}{E_{\bar{\nu}}} + W(E_e)\,
\frac{\vec{\xi}_n\cdot (\vec{k}_e \times
  \vec{k}_{\bar{\nu}})(\vec{\xi}_e \cdot \vec{k}_e)}{(E_e + m_e) E_e
  E_{\bar{\nu}}} \Big\},
\end{eqnarray}
where we have used the notations in Refs.  \cite{Ivanov2013} --
\cite{Ivanov2020b, Ivanov2021a, Ivanov2021b}. Then, $g_A$ and $G_V$
are the axial and vector coupling constants, respectively,
\cite{PDG2020, Abele2008, Nico2009, Abele2018, Sirlin2018}, the
Cabibbo--Kobayashi--Maskawa (CKM) matrix element $V_{ud}$ is included
in the defintion of the vector coupling constant $G_V$, $\vec{\xi}_n$
and $\vec{\xi}_e$ are unit spin--polarization vectors of the neutron
and electron \cite{Ivanov2013, Ivanov2017, Ivanov2019a} (see also
\cite{Itzykson1980}), respectively, $d\Omega_e$ and
$d\Omega_{\bar{\nu}}$ are infinitesimal solid angles in the directions
of electron $\vec{k}_e$ and antineutrino $\vec{k}_{\bar{\nu}}$
3--momenta, respectively, $E_0 = (m^2_n - m^2_p + m^2_e)/2m_n =
1.2926\,{\rm MeV}$ is the end--point energy of the electron--energy
spectrum \cite{Abele2008, Nico2009}, $F(E_e, Z = 1)$ is the
relativistic Fermi function, describing the electron--proton
final--state Coulomb interaction, is equal to \cite{Blatt1952}(see
also \cite{Wilkinson1982} and a discussion in \cite{Ivanov2017})
\begin{eqnarray}\label{eq:2}
\hspace{-0.3in}F(E_e, Z = 1 ) = \Big(1 +
\frac{1}{2}\gamma\Big)\,\frac{4(2 r_pm_e\beta)^{2\gamma}}{\Gamma^2(3 +
  2\gamma)}\,\frac{\displaystyle e^{\,\pi \alpha/\beta}}{(1 -
  \beta^2)^{\gamma}}\,\Big|\Gamma\Big(1 + \gamma +
i\,\frac{\alpha}{\beta}\Big)\Big|^2,
\end{eqnarray}
where $\beta = k_e/E_e = \sqrt{E^2_e - m^2_e}/E_e$ is the electron
velocity, $\gamma = \sqrt{1 - \alpha^2} - 1$, $r_p = 0.841\,{\rm fm}$
is the electric radius of the proton \cite{Antognini2013}. The
correlation coefficient $b(E_e)$ is the Fierz interference term
\cite{Fierz1937}. The structure and the value of the Fierz
interference term may depend on interactions beyond the SM
\cite{Fierz1937}. An information of a contemporary theoretical and
experimental status of the Fierz interference term can be found in
\cite{ Hardy2020, Severijns2019, Abele2019, Young2019, Sun2020} (see
also \cite{Ivanov2019y,Ivanov2019x}).

We would like to notice that recently \cite{Ivanov2021a, Ivanov2021b}
the correlation coefficients $T(E_e)$, $S(E_e)$ and $U(E_e)$ have been
investigated theoretically within the SM at the level of $10^{-3}$ by
taking into account i) the outer model-independent $O(\alpha/\pi)$ RC,
calculated to leading order (LO) in the large nucleon mass $m_N$
expansion, ii) the $O(E_e/m_N)$ corrections, caused by weak magnetism
and proton recoil, and iii) the corrections, caused by interactions
beyond the SM \cite{Jackson1957a}, including the contributions of the
{\it second class} currents or the $G$-odd correlations (as for
$G$--parity invariance of strong interactions, we refer to the paper
by Lee and Yang \cite{Lee1956a}) by Weinberg \cite{Weinberg1958} (see
also \cite{Gardner2001, Gardner2013} and \cite{Ivanov2017, Ivanov2018,
  Ivanov2019a}).

The paper is organized as follows. In section \ref{sec:coefficient} we
give the analytical expressions for the correlation function
$\zeta(E_e)$, which is responsible for the correct electron-energy
spectrum of the neutron beta decay and correct value of the neutron
lifetime, and the correlation coefficients $X(E_e)$ for $X = a, A, B,
\ldots, T$ and $U$, including i) the $O(\alpha/\pi)$ and $O(\alpha
E_e/m_N) \sim 10^{-5}$ RC, ii) the $O(E_e/m_N)$ and $O(E^2_e/m^2_N)
\sim 10^{-5}$ corrections, caused by weak magnetism and proton recoil,
and iii) Wilkinson's corrections of order of a few parts of $10^{-5}$,
which we have calculated in Appendices A, B and C. The results,
represented in section \ref{sec:coefficient}, illustrate the SM
theoretical description of the neutron beta decay at level of
$10^{-5}$ with a theoretical accuracy of a few parts of $10^{-6}$. In
section \ref{sec:Abschluss} we discuss the obtained results and some
problems of the analysis of the contributions of the neutron radiative
beta decay. In Appendices A, B, C and D we give detailed calculations
  of the correlation function $\zeta(E_e)$ and correlation
  coefficients $X(E_e)$ for $X = a, A, B, \ldots,T$ and $U$ adduced in
  section \ref{sec:coefficient}. In Appendix E we give the analytical
  expressions of the correlation function $\zeta(E_e)$ and correlation
  coefficients $X(E_e)$ for $X = a, A, B, \ldots, U$ as functions of
  the electron energy $E_e$ and the axial coupling constant $g_A$. For
  the practical applications and numerical analysis the correlation
  function $\zeta(E_e)$ and correlation coefficients $X(E_e)$ for $X =
  a, A, B, \ldots, U$ are programmed in \cite{MathW}. In Appendix F
    we give the contributions to the electron-energy and angular
    distribution of the neutron beta decay with correlations
    structures, which go beyond the standard correlation structures in
    Eq.(\ref{eq:1}) by Jackson {\it et al.}  \cite{Jackson1957a,
      Jackson1957b, Jackson1958} and Ebel and Feldman \cite{Ebel1957}.

\section{Correlation function and coefficients of the electron-energy
  and angular distribution Eq.(\ref{eq:1})}
\label{sec:coefficient}

In Appendices A, B, C and D at the level of $10^{-5}$ with a
theoretical accuracy of a few parts of $10^{-6}$ we give a detailed SM
calculation of the correlation function $\zeta(E_e)$ and correlation
coefficients in Eq.(\ref{eq:1}), the correlation structures of which
are invariant under time-reversal transformation,
i.e. T-even. According to our analysis carried out in Appendices A, B,
C and D, the correlation function $\zeta(E_e)$ and correlation
coefficients can be represented in the following form
\begin{eqnarray}\label{eq:3}
\hspace{-0.3in}\zeta(E_e) &=& \zeta(E_e)_{\rm RC} +
  \zeta(E_e)_{\rm RC-PhV} + \zeta(E_e)_{\rm WP} + \zeta(E_e)_{\rm
  WC},\nonumber\\
\hspace{-0.3in}X(E_e) &=& X(E_e)_{\rm RC} +
  X(E_e)_{\rm RC-PhV} + X(E_e)_{\rm WP} + X(E_e)_{\rm WC},
\end{eqnarray}
where $X = a, A, B, K_n, Q_n, G, H, N, K_e, Q_e, S, T$ and $U$. Then,
$Y(E_e)_{\rm RC}$, $Y(E_e)_{\rm RC-PhV}$, $Y(E_e)_{\rm WP}$ and
$Y(E_e)_{\rm W}$ for $Y = \zeta, X$ are i) the sum of the outer
$O(\alpha/\pi)$ RC, calculated to LO in the large nucleon mass $m_N$
expansion, and $O(\alpha E_e/m_N) \sim 10^{-5}$ RC, which are treated
as NLO corrections in the large nucleon mass $m_N$ expansion to the
outer and inner $O(\alpha/\pi)$ RC and denoted as $Y_{\rm RC-NLO}$
(see Appendix A), ii) the outer $O(\alpha E_e/m_N)$ RC, caused by the
outer $O(\alpha/\pi)$ RC and the phase-volume of the neutron beta
decay taken to NLO in the large nucleon mass expansion, iii) the sum
of the $O(E_e/m_N)$ and $O(E^2_e/m^2_N) \sim 10^{-5}$ corrections,
caused by weak magnetism and proton recoil, and iv) Wilkinson's
corrections of order $10^{-5}$.  For the practical applications and
numerical analysis the analytical expressions of the correlation
function $\zeta(E_e)$ and correlation coefficients $X(E_e)$ for $X =
a, A, B, K_n, Q_n, G, H, N, K_e, Q_e, S, T$ and $U$ are programmed in
\cite{MathW}.

  In order to illustrate the SM description of the neutron beta decay
  at the level $10^{-5}$ we represent the correlation function
  $\zeta(E_e)$ and correlation coefficients $a(E_e), A(E_e), B(E_e),
  \ldots, U(E_e)$ with the contributions of the corrections, caused by
  weak magnetism and proton recoil of order $O(E_e/m_N)$ and
  $O(E^2_e/m^2_N)$ in Appendix B, and Wilkinson's corrections in
  Appendix C, as functions of the variable $E_e/E_0$. According to our
  calculation in \cite{MathW}, we get
\begin{eqnarray*}
\hspace{-0.3in}\zeta(E_e) &=& \zeta(E_e)_{\rm RC} +
  \zeta(E_e)_{\rm RC-PhV} - 5.57 \times 10^{-4}\,\frac{E_0}{E_e} -
3.20 \times 10^{-3} + 9.81\times 10^{-3}\,\frac{E_e}{E_0} \nonumber\\
\hspace{-0.3in}&+& 7.13 \times 10^{-5}\,\frac{E^2_e}{E^2_0} - 3.16
\times 10^{-5}\, \frac{E_e}{\beta E_0},\nonumber\\
\hspace{-0.3in}a(E_e) &=& a(E_e)_{\rm RC} +
  a(E_e)_{\rm RC-PhV}- 5.81 \times 10^{-5}\,\frac{E_0}{E_e} + 3.24
\times 10^{-3} - 9.16 \times 10^{-3} \,\frac{E_e}{E_0} \nonumber\\
\hspace{-0.3in}&-& 3.16 \times 10^{-5}\, \frac{E_0 - E_e}{\beta^3
  E_0},\nonumber\\
\hspace{-0.3in}A(E_e) &=& A(E_e)_{\rm RC} - 6.71 \times
10^{-5}\,\frac{E_0}{E_e} - 1.75 \times 10^{-3}
\,\frac{E_e}{E_0},\nonumber\\
\hspace{-0.3in}B(E_e) &=& B(E_e)_{\rm RC} + 5.26 \times
10^{-5}\,\frac{E_0}{E_e} + 3.27 \times 10^{-4} - 3.42 \times 10^{-4}
\,\frac{E_e}{E_0} - 2.87 \times
10^{-5}\,\frac{E^2_e}{E^2_0},\nonumber\\
\hspace{-0.3in}K_n(E_e) &=& K_n(E_e)_{\rm RC} + 7.15 \times
10^{-4}\,\frac{E_e}{E_0} + 1.51\times 10^{-5}
\,\frac{E^2_e}{E^2_0},\nonumber\\
\hspace{-0.3in}Q_n(E_e) &=& Q_n(E_e)_{\rm RC-PhV} +
3.19 \times 10^{-3} - 7.29 \times 10^{-3} \,\frac{E_e}{E_0} - 1.97
\times 10^{-5}\,\frac{E^2_e}{E^2_0} - 3.12 \times 10^{-5}\, \frac{E_0
  - E_e}{\beta^3 E_0},\nonumber\\
\hspace{-0.3in}G(E_e) &=& G(E_e)_{\rm RC} - 5.59 \times
10^{-4}\,\frac{E_0}{E_e} + 3.78 \times 10^{-4} - 5.26 \times 10^{-5}
\,\frac{E_e}{E_0} - 1.26 \times
10^{-4}\,\frac{E^2_e}{E^2_0},\nonumber\\
\hspace{-0.3in}H(E_e) &=& H(E_e)_{\rm RC} + 2.28 \times
10^{-5}\,\frac{E^2_0}{E^2_e} - 1.28 \times 10^{-3}\,\frac{E_0}{E_e} +
1.13 \times 10^{-3} - 1.46 \times 10^{-5} \,\frac{E_e}{E_0},\nonumber\\
\hspace{-0.3in}N(E_e) &=& N(E_e)_{\rm RC} + 2.33 \times
10^{-5}\,\frac{E^2_0}{E^2_e} - 3.21 \times 10^{-4}\,\frac{E_0}{E_e} +
5.73 \times 10^{-5},\nonumber\\
\hspace{-0.3in}Q_e(E_e) &=&  Q_e(E_e)_{\rm RC} +  Q_e(E_e)_{\rm RC-PhV} +
2.24 \times 10^{-5}\,\frac{E_0}{E_e} + 7.00 \times 10^{-4} + 3.57
\times 10^{-3} \,\frac{E_e}{E_0} - 1.57 \times
10^{-5}\,\frac{E^2_e}{E^2_0} \nonumber\\
\hspace{-0.3in} &+& 1.04 \times 10^{-4}\,\big(1 + \sqrt{1 -
  \beta^2}\,\big)\, \frac{E_0 - E_e}{\beta^3 E_0},\nonumber\\
\hspace{-0.3in}K_e(E_e) &=& K_e(E_e)_{\rm RC} +
  K_e(E_e)_{\rm RC-PhV} + 1.93 \times 10^{-5}\,\frac{E_0}{E_e} - 6.93
\times 10^{-4} + 9.17 \times 10^{-3} \,\frac{E_e}{E_0} + 2.95 \times
10^{-5}\,\frac{E^2_e}{E^2_0} \nonumber\\
\hspace{-0.3in} &+& 3.16 \times 10^{-5}\,\big(1 + \sqrt{1 -
  \beta^2}\,\big)\, \frac{E_0 - E_e}{\beta^3 E_0},\nonumber\\
\hspace{-0.3in}S(E_e) &=& S(E_e)_{\rm RC} - 2.82 \times 10^{-4},\nonumber\\
\end{eqnarray*}
\begin{eqnarray}\label{eq:4}
\hspace{-0.3in}T(E_e) &=& T(E_e)_{\rm RC} - 1.81 \times
10^{-4}\,\frac{E_0}{E_e} - 1.25 \times 10^{-5} + 2.77 \times 10^{-4}
\,\frac{E_e}{E_0}+ 4.69 \times
10^{-5}\,\frac{E^2_e}{E^2_0},\nonumber\\
\hspace{-0.3in}U(E_e) &=& U(E_e)_{\rm RC} + 1.19 \times 10^{-5},
\end{eqnarray}
where the numerical coefficients are calculated at the axial coupling
constant $g_A = 1.2764$ \cite{Abele2018}. The
  contributions of the corrections of the $O(\alpha E_e/m_N)$ RC are
  plotted in \cite{MathW}.

Then, the analytical expressions of $\zeta(E_e)_{\rm RC}$ and
$X(E_e)_{\rm RC}$ for $X = a,A, \ldots,T$ and $U$ are given in
Appendix A (see Eq.(\ref{eq:A.20})) and in \cite{MathW}. At $\alpha =
0$ the correlation function $\zeta(E_e)_{\rm RC}$ and the correlation
coefficients $X(E_e)_{\rm RC}$ for $X = a,A, \ldots,U$ reduce to their
values, calculated to LO in the large nucleon mass $m_N$ expansion
(see Appendix A). The outer RC $\zeta(E_e)_{\rm RC-PhV}$ and
$X(E_e)_{\rm RC-PhV}$ are calculated in Appendix D (see
Eq.(\ref{eq:D.10})).

In addition to the correlation function $\zeta(E_e)$ and the
correlation coefficients in Eq.(\ref{eq:4}) we give the correlation
coefficient $A^{(\beta)}(E_e) = A(E_e) + \frac{1}{3}\,Q_n(E_e)$
\cite{Wilkinson1982} that measures the electron (beta) asymmetry of
the neutron beta decay \cite{Abele2018}:
\begin{eqnarray}\label{eq:5}
\hspace{-0.3in}A^{(\beta)}(E_e) &=& A^{(\beta)}(E_e)_{\rm RC} +
\frac{1}{3}\,Q_n(E_e)_{\rm RC-PhV} - 6.67 \times 10^{-5}\,
\frac{E_0}{E_e} + 9.78 \times 10^{-4} - 4.18 \times 10^{-3}\,
\frac{E_e}{E_0} \nonumber\\
\hspace{-0.3in} &-& 1.04 \times 10^{-5}\, \frac{E_0 - E_e}{\beta^3
  E_0},
\end{eqnarray}
calculated at $g_A = 1.2764$ \cite{Abele2018}, where
$A^{(\beta)}(E_e)_{\rm RC} = A(E_e)_{\rm RC}$, since $Q_n(E_e)_{\rm
  RC} = 0$ (see Eq.(\ref{eq:A.20})).  The correlation function
$\zeta(E_e)$ and the correlation coefficients $X(E_e)$ for $X = a,A,B,
\ldots,U$ in Eq.(\ref{eq:4}) and $A^{(\beta)}(E_e)$ in Eq.(\ref{eq:5})
describe the neutron beta decay for a polarized neutron, a polarized
electron and an unpolarized proton at the level of $10^{-5}$ in the
framework of the SM with a theoretical accuracy of a few parts of
$10^{-6}$.

\section{Discussion}
\label{sec:Abschluss}

We have given a SM theoretical description of the neutron beta decay
for a polarized neutron, a polarized electron and an unpolarized
proton at the level of $10^{-5}$ with a theoretical accuracy of a few
parts of $10^{-6}$. To the well-known $O(\alpha/\pi)$ RC
\cite{Sirlin1967, Sirlin1978, Shann1971, Ando2004, Gudkov2006} (see
also \cite{Ivanov2013, Ivanov2017, Ivanov2019a, Ivanov2021a,
  Ivanov2021b}) and $O(E_e/m_N)$ corrections \cite{Bilenky1959} and
\cite{Wilkinson1982, Ando2004, Gudkov2006} (see also \cite{Ivanov2013,
  Ivanov2017, Ivanov2019a, Ivanov2021a, Ivanov2021b}) we have added i)
the inner $O(\alpha E_e/m_N) \sim 10^{-5}$ RC \cite{Ivanov2019b,
  Ivanov2020a}, which are treated as NLO corrections in the large
nucleon mass $m_N$ expansion to Sirlin's outer and inner
$O(\alpha/\pi)$ RC, calculated to LO in the large nucleon mass $m_N$,
expansion, ii) the outer $O(\alpha E_e/m_N) \sim
  10^{-5}$ RC, induced by Sirlin's outer $O(\alpha/\pi)$ RC and the
  phase-volume of the neutron beta decay, calculated to NLO in the
  large nucleon mass $m_N$ expansion, iii) the $O(E^2_e/m^2_N) \sim
10^{-5}$ corrections \cite{Ivanov2020b}, caused by weak magnetism and
proton recoil, and iv) Wilkinson's corrections \cite{Wilkinson1982}
(see also \cite{Ivanov2013, Ivanov2017, Ivanov2019a}) of order
$10^{-5}$. As has been shown in Eq.(\ref{eq:4}), all of these
corrections define the SM background of the theoretical description of
the neutron beta decay at the level of $10^{-5}$ with a theoretical
accuracy of about a few parts of $10^{-6}$ \cite{MathW}.

Having accepted the value of the axial coupling constant $g_A =
1.2764$ \cite{Abele2018, Sirlin2018}, the correlation function
$\zeta(E_e)$ and correlation coefficients, given in Eq.(\ref{eq:4})
and Eq.(\ref{eq:5}), can be used as the SM theoretical background of
the neutron beta decay for experimental searches of contributions of
interactions beyond the SM with experimental uncertainties of a few
parts of $10^{-5}$ \cite{Abele2016, Dubbers2021, Bodek2019} (see also
\cite{Ivanov2017, Ivanov2020b}). Because of Wilkinson's corrections,
induced by the proton recoil in the electron-proton final-state
Coulomb interaction (see Eq.(\ref{eq:C.8}) in Appendix C), and the
outer $O(\alpha E_e/m_N)$ RC (see Eq.(\ref{eq:D.10}) in Appendix D)
the correlation function $\zeta(E_e)$ and correlation coefficients in
Eq.(\ref{eq:4}) and Eq.(\ref{eq:5}) are well defined in the
experimental electron-energy region $0.811\,{\rm MeV} \le E_e \le
1.211\,{\rm MeV}$ \cite{Abele2018}.

In Appendix E we give the analytical expressions for the correlation
function $\zeta(E_e)$ and correlation coefficients $ X(E_e)$ for $X =
a, A, B, \ldots, U$ as functions of the electron energy $E_e$ and the
axial coupling constant $g_A$. These expressions can be used as a SM
theoretical background for processing experimental data on the neutron
lifetime, the electron-antineutrino angular correlations, and electron
and antineutrino asymmetries with experimental uncertainties of about
a few parts of $10^{-5}$. Such a SM theoretical background and
experimental data, obtained with experimental uncertainties of about a
few parts of $10^{-5}$, should allow to improve the currently
available experimental value of the axial coupling constant $g_A$
\cite{Abele2018, Sirlin2018}. They can be also used for searches of
contributions of interactions beyond the SM in experiments with
polarized neutrons and electrons \cite{Bodek2019}.

We have also to emphasize that for the correct description of the
neutron lifetime one has to add the inner radiative corrections
$\Delta^V_R$ and $\Delta^A_R$ of order $O(\alpha/\pi)$, defined by the
Feynman $\gamma W$-box diagrams, to the rates of the neutron beta
decay and superallowed nuclear beta decays, which have been calculated
to LO in the large nucleon mass $m_N$ expansion in \cite{Sirlin1986,
  Sirlin2004, Sirlin2006, Seng2018, Seng2018a, Sirlin2019, Hayen2020,
  Hayen2021, Gorchtein2021}. These corrections are very important for
the correct extraction of the value of the Cabibbo-Kobayashi-Maskawa
(CKM) matrix element $V_{ud}$.

Finally we would like to discuss the problem of the removal of
infrared divergences for the calculation of the outer RC in the
neutron beta decay. Since the virtual photon exchange leads to the
dependence of the amplitude of the neutron beta decay on the infrared
cut-off $\mu$, which is an infinitesimal photon mass $\mu$ in the
covariant regularization \cite{Sirlin1967, Berman1958, Kinoshita1959,
  Berman1962, Kaellen1967, Abers1968}, one has to take into account
the contribution of the neutron radiative beta decay \cite{Sirlin1967,
  Berman1958, Kinoshita1959, Berman1962, Kaellen1967, Abers1968}. For
this aim the energy and angular distribution of the rate of the
neutron radiative decay should be summed with the energy and angular
distribution of the rate of the neutron beta decay \cite{Sirlin1967,
  Berman1958, Kinoshita1959, Berman1962, Kaellen1967, Abers1968}. In
case of the investigation of the electron-energy and angular
distribution of the neutron beta decay (see, for example,
Eq.(\ref{eq:1})), the standard procedure for the calculation of
distributions for both the neutron beta decay and neutron radiative
beta decay is to integrate, first, over the proton 3-momenta and then
over the energy of the antineutrino \cite{Sirlin1967, Shann1971,
  Ando2004, Gudkov2006, Berman1958, Kinoshita1959, Berman1962,
  Kaellen1967, Abers1968, Garcia1978, Gaponov1996, Bernard2004,
  Gluck1993, Gluck1995, Gluck1996, Gluck1997, Gluck1998} (see also
\cite{Ivanov2013, Ivanov2017, Ivanov2018, Ivanov2019a, Ivanov2021a,
  Ivanov2021b, Ivanov2019y}). In the rest frame of the neutron and
after the integration of the proton 3-momentum, the latter appears in
the distributions in the form $\vec{k}_p = - \vec{k}_e -
\vec{k}_{\bar{\nu}_e}$ imposed by momentum conservation, where
$\vec{k}_e$ and $\vec{k}_{\bar{\nu}_e}$ are 3-momenta of the electron
and antineutrino, respectively. In this case for the calculation of
the electron-energy and angular distribution of the neutron radiative
beta decay the integration over directions of the photon 3-momentum
takes into account correlations of the photon 3-momentum with
3-momenta of the electron and antineutrino and implicitly with the
proton 3-momentum (for the details of these calculations we refer to
\cite{Ivanov2013, Ivanov2017, Ivanov2019a, Ivanov2021a,
  Ivanov2021b}). The electron-energy and angular distribution
Eq.(\ref{eq:1}) is usually used for the measurements of the electron
(beta) asymmetry, which is characterized by the correlation
coefficient $A^{(\beta)}(E_e)$ \cite{Abele2018}. In these measurements
the electron asymmetry defines the asymmetry of the emission of decay
electrons relative to the neutron spin polarization into solid angles
related by the polar angle $\theta \to \pi - \theta$ \cite{Abele2018,
  Mund2013, Mendenhall2013, Brown2018, Serebrov2019} (for the details
of the calculation we refer to \cite{Ivanov2013}). The electron-energy
and angular distribution Eq.(\ref{eq:1}) can be also applied to the
measurement of the antineutrino asymmetry, which is practically
defined by the correlation coefficient $B(E_e)$. Formally, the
antineutrino asymmetry $B_{\rm exp}(E_e)$ defines the asymmetry of the
emission of the antineutrino relative to the neutron spin polarization
into solid angles related by the polar angle $\theta \to \pi -
\theta$. However, in experiments \cite{Dubbers2021, Serebrov1998,
  Abele2005, Schumann2007} because of the electroneutrality of the
antineutrino such an asymmetry is equivalent to the asymmetry of the
emission of the electron-proton pairs into the solid angles related by
the polar angle $\theta \to \pi - \theta$. For the first time, the
asymmetry $B_{\rm exp}(E_e)$ has been calculated by Gl\"uck {\it et
  al.}  \cite{Gluck1995, Gluck1998} in terms of the correlation
coefficients $a(E_e)$, $A(E_e)$ and $B(E_e)$ (for the details of the
calculation we refer to \cite{Ivanov2013}).

In turn, for the measurements of the electron-antineutrino angular
correlations \cite{Nico2017, Beck2020, Nico2021} and the proton recoil
asymmetry, defined by the correlation coefficient $C$
\cite{Schumann2008}, one has to use the electron-proton-energy and
angular distribution (or the proton-energy and angular distribution)
\cite{ Gluck1993, Gluck1995, Gluck1996, Gluck1997, Gluck1998}. For the
calculation of the electron-proton-energy and angular distribution one
has to integrate over the antineutrino 3-momentum. Then, having
integrated over the electron energy one obtains the proton-energy and
angular distribution (for the details of the calculation we refer to
\cite{Ivanov2013}). The same procedure should be used for the neutron
radiative beta decay \cite{Ivanov2013}. In case of the neutron
radiative beta decay for the calculation of the
electron-proton-photon-energy and angular distribution one deals with
direct photon-proton correlations \cite{Gluck1993, Gluck1995,
  Gluck1996, Gluck1997, Gluck1998}. However, as has been shown in
\cite{Ivanov2013a}, the contributions of these correlations do not
destroy the radiative corrections, defined by the functions
$(\alpha/\pi)\,\bar{g}_n(E_e)$ and $(\alpha/\pi)\,f_n(E_e)$. As has
been found in \cite{Ivanov2013a}, the contributions of the
photon-proton correlations in the neutron radiative beta decay to the
proton recoil asymmetry $C$ are of order of $10^{-4}$ . They make the
contributions of the radiative corrections to the proton recoil
asymmetry $C$ symmetric with respect to a change $A_0 \leftrightarrow
B_0$, where $A_0$ and $B_0$ are the correlation coefficients $A(E_e)$
and $B(E_e)$ calculated to LO in the large nucleon mass $m_N$
expansion \cite{Abele2008, Nico2009}. They depend on the axial
coupling constant only (see also Eq.(\ref{eq:A.16}) in Appendix A).
We are planning to carry out the analysis of the
electron-proton-energy and angular distributions of the neutron beta
decay at the SM theoretical level of about $10^{-5}$ in our
forthcoming publication.

We would like to note that for practical applications and numerical
analysis of the correlation function $\zeta(E_e)$ and correlation
coefficients $a(E_e), A(E_e), B(E_e), \ldots, U(E_e)$ we have
programmed their analytical expressions in \cite{MathW}. We have
carried out numerical calculations for the axial coupling constant
$g_A = 1.2764$ \cite{Abele2018} and plotted the $O(\alpha E_e/m_N)$
corrections.

\section{Acknowledgements}

We thank Hartmut Abele for discussions stimulating this work. The work
of A. N. Ivanov was supported by the Austrian ``Fonds zur F\"orderung
der Wissenschaftlichen Forschung'' (FWF) under contracts P31702-N27
and P26636-N20, and ``Deutsche F\"orderungsgemeinschaft'' (DFG) AB
128/5-2. The work of R. H\"ollwieser was supported by the Deutsche
Forschungsgemeinschaft in the SFB/TR 55. The work of M. Wellenzohn was
supported by the MA 23.

\newpage

\section*{Appendix A: The electron-energy and angular distribution
  of the neutron beta decay with the account for the radiative
  corrections of order $O(\alpha E_e/m_N)$}
\renewcommand{\theequation}{A-\arabic{equation}}
\setcounter{equation}{0}

According to \cite{Ivanov2013, Ivanov2021a, Ivanov2021b}, the
electron--energy and angular distribution of the neutron beta decay
for a polarized neutron, a polarized electron and an unpolarized
proton in Eq.(\ref{eq:1}) is determined by
 \begin{eqnarray}\label{eq:A.1}
\hspace{-0.3in}\frac{d^5\lambda_n(E_e, \vec{k}_e,
  \vec{k}_{\bar{\nu}},\vec{\xi}_n, \vec{\xi}_e)}{ d E_e
  d\Omega_ed\Omega_{\bar{\nu}}} &=& (1 + 3 g^2_A) \, \frac{|G_V|^2}{16
  \pi^5}\,(E_0 - E_e)^2\,\sqrt{E^2_e - m^2_e}\,E_e\,F(E_e, Z = 1)\nonumber\\
\hspace{-0.3in}&&\times \, \Phi_n(\vec{k}_e, \vec{k}_{\bar{\nu}})
\sum_{\rm pol.}\frac{|M(n \to p e^- \bar{\nu}_e)|^2}{(1 + 3 g^2_A)
  |G_V|^2 64 m^2_n E_e E_{\bar{\nu}}},
\end{eqnarray}
where we sum over polarizations of the massive fermions.  The
function $\Phi_n(\vec{k}_e, \vec{k}_{\bar{\nu}})$ defines the
contribution of the phase-volume of the neutron beta decay
\cite{Ivanov2013, Ivanov2020b}. It is equal to \cite{Ivanov2013,
  Ivanov2020b}
\begin{eqnarray}\label{eq:A.2}
\hspace{-0.3in}\Phi_n(\vec{k}_e, \vec{k}_{\bar{\nu}}) = 1 + 3
\frac{E_e}{m_N}\Big(1 - \frac{\vec{k}_e\cdot
  \vec{k}_{\bar{\nu}}}{E_e E_{\bar{\nu}}}\Big),
\end{eqnarray}
taken to NLO in the large nucleon mass $m_N$ expansion. The amplitude
of the neutron beta decay, taking into account the radiative
corrections of order $O(\alpha/\pi)$ and $O(\alpha E_e/m_N)$, is
defined by \cite{Ivanov2013, Ivanov2019b, Ivanov2020a}
\begin{eqnarray}\label{eq:A.3}
\hspace{-0.3in}&&M(n\to p\,e^-\bar{\nu}_e) = - 2 m_n G_V\Big\{\big(1 +
U_1\big) \,[\varphi^{\dagger}_p\varphi_n][\bar{u}_e\gamma^0(1 -
  \gamma^5)v_{\bar{\nu}}] + g_A \big(1 + U_2\big) \,
[\varphi^{\dagger}_p\vec{\sigma}\,\varphi_n]\cdot
[\bar{u}_e\vec{\gamma}\, (1 - \gamma^5)v_{\bar{\nu}}]\nonumber\\
\hspace{-0.3in}&&+ U_3  \,[\varphi^{\dagger}_p\varphi_n][\bar{u}_e (1 -
  \gamma^5)v_{\bar{\nu}}] + g_A U_4\,
       [\varphi^{\dagger}_p\vec{\sigma}\,\varphi_n] \cdot
       [\bar{u}_e\gamma^0 \vec{\gamma} \,(1 - \gamma^5)v_{\bar{\nu}}] +
         U_5 \,[\varphi^{\dagger}_p(\vec{k}_e\cdot \vec{\sigma}\,)
           \varphi_n][\bar{u}_e(1 - \gamma^5)v_{\bar{\nu}}] \nonumber\\
\hspace{-0.3in}&& + U_6
\,[\varphi^{\dagger}_p(\vec{k}_{\bar{\nu}}\cdot \vec{\sigma}\,)
  \varphi_n ] [\bar{u}_e(1 - \gamma^5)v_{\bar{\nu}}] + U_7
\,[\varphi^{\dagger}_p (\vec{k}_e\cdot \vec{\sigma}\,)
  \varphi_n][\bar{u}_e\gamma^0(1 - \gamma^5)v_{\bar{\nu}}] + U_8 \,
  [\varphi^{\dagger}_p(\vec{k}_e\cdot
    \vec{\sigma}\,)\vec{\sigma}\,\varphi_n]\cdot [\bar{u}_e
    \vec{\gamma}\,(1 - \gamma^5)v_{\bar{\nu}}]. \nonumber\\
\hspace{-0.3in}&&
\end{eqnarray}
The functions $U_j$ for $j = 1,2,\ldots,8$ are given by
\begin{eqnarray}\label{eq:A.4}
\hspace{-0.3in}U_1 &=& \frac{\alpha}{2\pi}\Big(f_{\beta^-_c}(E_e,\mu)
+ \frac{E_e}{m_N}\,f_V(E_e) + \bar{g}_{\rm st}(E_e)\Big), \nonumber\\
\hspace{-0.3in}U_2 &=& \frac{\alpha}{2\pi}\Big(
  f_{\beta^-_c}(E_e,\mu) + \frac{E_e}{m_N} f_A(E_e) + \frac{5}{2}
  \frac{m^2_N}{M^2_W} {\ell n}\frac{M^2_W}{m^2_N} + \frac{1}{g_A}\,
  \bar{f}_{\rm st}(E_e)\Big), \nonumber\\
\hspace{-0.3in}U_3 &=& \frac{\alpha}{2\pi}\, \Big(- \frac{\sqrt{1 -
      \beta^2}}{2\beta}{\ell n}\Big(\frac{1 + \beta}{1 - \beta}\Big) +
  \frac{E_e}{m_N} f_S(E_e)\Big), \nonumber\\
\hspace{-0.3in}U_4 &=&\frac{\alpha}{2\pi}\, \Big(- \frac{\sqrt{1 -
    \beta^2}}{2\beta}\,{\ell n}\Big(\frac{1 + \beta}{1 - \beta}\Big) +
\frac{E_e}{m_N}\,f_T(E_e)\Big),\nonumber\\
\hspace{-0.3in}U_5 &=&\frac{\alpha}{2\pi}\,\frac{1}{E_e}\,
\frac{E_e}{m_N}\,\,g_S(E_e)\;,\; U_6 = \frac{\alpha}{2\pi}\,\frac{1}{E_e}\,
\frac{E_e}{m_N}\,h_S(E_e), \nonumber\\
\hspace{-0.3in}U_7 &=&\frac{\alpha}{2\pi}\,\frac{1}{E_e}\,
\frac{E_e}{m_N}\,g_V(E_e) \;,\; U_8 =
\frac{\alpha}{2\pi}\,\frac{1}{E_e}\,\frac{E_e}{m_N}\,h_A(E_e),
\end{eqnarray}
where $\beta = k_e/E_e = \sqrt{1 - m^2_e/E^2_e}$ is the electron
velocity, the function $f_{\beta^-_c}(E_e, \mu)$, where $\mu$ is a
covariant infrared cut-off having a meaning of a photon mass
\cite{Berman1958, Kinoshita1959, Sirlin1967}, was calculated by Sirlin
\cite{Sirlin1967} to LO in the large nucleon $m_N$ expansion (for the
details of the calculation of the function $f_{\beta^-_c}(E_e, \mu)$
we refer to \cite{Ivanov2013}). It defines so-called outer
model-independent radiative corrections \cite{Wilkinson1970}. Then,
the functions $f_V(E_e)$, $f_A(E_e)$, $f_S(E_e)$, $f_T(E_e)$,
$g_V(E_e)$, $g_S(E_e)$, $h_S(E_e)$ and $h_A(E_e)$ determine the inner
radiative corrections dependent on the axial coupling constant $g_A$
to Sirlin's outer radiative corrections $O(\alpha/\pi)$, calculated to
NLO in the large nucleon mass $m_N$ expansion in
\cite{Ivanov2019b}. In turn, the functions $\bar{g}_{\rm st}(E_e)$ and
$\bar{f}_{\rm st}(E_e)$ describe the inner radiative corrections,
caused by the hadronic structure of the neutron and calculated to NLO
in the large nucleon $m_N$ mass expansion in \cite{Ivanov2020a} as NLO
corrections to Sirlin's inner radiative corrections $O(\alpha/\pi)$,
caused by the hadronic structure of the neutron and calculated to LO
in the large nucleon mass expansion \cite{Sirlin1967, Sirlin1978}. The
analytical expressions of these functions are equal to
\cite{Ivanov2013, Ivanov2019b, Ivanov2020a}
\begin{eqnarray*}
\hspace{-0.15in}f_{\beta^-_c}(E_e, \mu) &=& \frac{3}{4}\,{\ell
  n}\frac{m^2_N}{m^2_e} - \frac{11}{8} + {\ell
  n}\Big(\frac{\mu}{m_e}\Big) \Big[\frac{1}{\beta} {\ell
    n}\Big(\frac{1 + \beta}{1 - \beta}\Big) - 2 \Big] +
\frac{1}{2\beta}\, {\ell n}\Big(\frac{1 + \beta}{1 - \beta}\Big) -
\frac{1}{4\beta}\, {\ell n}^2\Big(\frac{1 + \beta}{1 -
  \beta}\Big) - \frac{1}{\beta}\,{\rm Li}_2\Big(\frac{2\beta}{1 +
  \beta}\Big),\nonumber\\
\hspace{-0.15in}f_V(E_e) &=& 1 + \frac{1}{2}\,{\ell
  n}\frac{m^2_N}{m^2_e} + \frac{2 - 3\beta^2}{2 \beta}\,{\ell
  n}\Big(\frac{1 + \beta}{1 - \beta}\Big) + (g_A - 1)\,\Big[-
  \frac{1}{4} - \frac{5}{8}\,{\ell n}\frac{m^2_N}{m^2_e} - \frac{2 -
    5\beta^2}{8 \beta}\,{\ell n}\Big(\frac{1 + \beta}{1 -
    \beta}\Big)\Big],\nonumber\\
\end{eqnarray*}
\begin{eqnarray}\label{eq:A.5}
\hspace{-0.15in}f_A(E_e) &=&1 + \frac{1}{2}\,{\ell
  n}\frac{m^2_N}{m^2_e} + \frac{2 - 3\beta^2}{2 \beta}\,{\ell
  n}\Big(\frac{1 + \beta}{1 - \beta}\Big) + \frac{g_A - 1}{g_A}\,\Big[
  \frac{3}{4}\,{\ell n}\frac{m^2_N}{m^2_e} - \frac{3}{4}\,\beta\,
       {\ell n}\Big(\frac{1 + \beta}{1 - \beta}\Big)\Big],\nonumber\\
\hspace{-0.15in}f_S(E_e) &=&\sqrt{1 - \beta^2}\Big\{-
\frac{1}{2}\,{\ell n}\frac{m^2_N}{m^2_e} + \frac{2E_0 -
  E_e}{E_e}\,\frac{1}{2 \beta}\,{\ell n}\Big(\frac{1 + \beta}{1 -
  \beta}\Big) + (g_A - 1)\,\Big[ \frac{1}{4} - \frac{1}{8}\,{\ell
    n}\frac{m^2_N}{m^2_e} - \frac{1}{4 \beta}\,{\ell n}\Big(\frac{1 +
    \beta}{1 - \beta}\Big)\Big]\Big\},\nonumber\\
\hspace{-0.15in}f_T(E_e) &=&\sqrt{1 - \beta^2}\Big[-
  \frac{1}{2}\,{\ell n}\frac{m^2_N}{m^2_e} + \frac{1}{2 \beta}\,{\ell
    n}\Big(\frac{1 + \beta}{1 - \beta}\Big)\Big],\nonumber\\
\hspace{-0.15in}g_S(E_e) &=&(g_A - 1)\,\frac{\sqrt{1 - \beta^2}}{8
  \beta}\,{\ell n}\Big(\frac{1 + \beta}{1 - \beta}\Big),\nonumber\\
\hspace{-0.15in}h_S(E_e) &=& \frac{\sqrt{1 - \beta^2}}{\beta}\,{\ell
  n}\Big(\frac{1 + \beta}{1 - \beta}\Big),\nonumber\\
\hspace{-0.15in}g_V(E_e) &=&- \frac{1}{2}\,{\ell n}\frac{m^2_N}{m^2_e}
+ \frac{1}{2 \beta}\,{\ell n}\Big(\frac{1 + \beta}{1 - \beta}\Big) +
(g_A - 1)\,\Big[- \frac{1}{4} - \frac{3}{8}\,{\ell
    n}\frac{m^2_N}{m^2_e} + \frac{5}{8 \beta}\,{\ell n}\Big(\frac{1 +
    \beta}{1 - \beta}\Big)\Big],\nonumber\\
\hspace{-0.15in}h_A(E_e) &=& - \frac{1}{2}\,{\ell
  n}\frac{m^2_N}{m^2_e} + \frac{1}{2 \beta}\,{\ell n}\Big(\frac{1 +
  \beta}{1 - \beta}\Big),
\end{eqnarray}
where ${\rm Li}_2(z)$ is the PolyLogarithmic function
\cite{Mitchell1949}, and
\begin{eqnarray}\label{eq:A.6}
 \hspace{-0.30in}\bar{g}_{\rm st}(E_e) &=& - G^{(V)}_{\rm
   st}\frac{E_0}{m_N} + \big( G^{(W)}_{\rm st} + F^{(W)}_{\rm
   st}\big)\, \frac{m^2_N}{M^2_W} + H^{(V)}_{\rm st}\frac{E_e}{m_N} =
 0.098 \,\Big(1 + 0.95\,\frac{E_e}{E_0}\Big), \nonumber\\
 \hspace{-0.30in}\bar{f}_{\rm st}(E_e) &=& + G^{(A)}_{\rm
   st}\frac{E_0}{m_N} - H^{(W)}_{\rm st}\frac{m^2_N}{M^2_W} -
 H^{(A)}_{\rm st}\frac{E_e}{m_N} = 0.057\, \Big(1 +
 \frac{E_e}{E_0}\Big),
\end{eqnarray}
where $E_0 = (m^2_n - m^2_p + m^2_e)/2 m_n = 1.2926\, {\rm MeV}$ is
the end--point energy of the electron--energy spectrum
\cite{Abele2008, Nico2009}, calculated for $m_n = 939.9564\,{\rm
  MeV}$, $m_p = 938.2721\,{\rm MeV}$ and $m_e = 0.5110\,{\rm MeV}$
\cite{PDG2020}. Then, $m_N = (m_n + m_p)/2 = 938.9188\,{\rm MeV}$ and
$M_W = 80.379\,{\rm GeV}$ are nucleon  and electroweak $W^-$-boson
masses \cite{PDG2020}, respectively. The structure constants
$G^{(V)}_{\rm st}$ and so on, calculated in \cite{Ivanov2020a}, are
equal to $G^{(V)}_{\rm st} = - 70.71$, $H^{(V)}_{\rm st} = 67.75$,
$G^{(W)}_{st} = 8.94$, $G^{(A)}_{\rm st} = 41.95$, $H^{(A)}_{\rm st} =
- 40.78$, $H^{(W)}_{\rm st} = 2.10$ and $F^{(W)}_{\rm st} =
-1.64$. For the subsequent analysis of radiative corrections we follow
\cite{Ivanov2021a} (see also \cite{Ivanov2013}) and represent the
function $f_{\beta^-_c}(E_e, \mu)$ as follows
\begin{eqnarray}\label{eq:A.7}
  \hspace{-0.3in} f_{\beta^-_c}(E_e, \mu) = \bar{g}_n(E_e) + \frac{1 -
    \beta^2}{2\beta}\,{\ell n}\frac{1 + \beta}{1 - \beta} -
  g^{(1)}_{\beta \gamma}(E_e, \mu),
\end{eqnarray}
where $2 \bar{g}_n(E_e)$ is Sirlin's function, defining the {\it
  outer} radiative corrections of order $O(\alpha/\pi)$ to the neutron
lifetime \cite{Sirlin1967}. The function $g^{(1)}_{\beta\gamma}(E_e,
\mu)$ can be removed by the contribution of the neutron radiative beta
decay $n \to p + e^- + \bar{\nu}_e + \gamma$ with a real photon
$\gamma$, which should be added, according to Berman \cite{Berman1958}
and Kinoshita and Sirlin \cite{Kinoshita1959} (see also Sirlin
\cite{Sirlin1967}), for the removal of the dependence of the neutron
lifetime on the infrared cut--off. For the detailed calculation of the
function $g^{(1)}_{\beta\gamma}(E_e, \mu)$ and as well as the function
$g^{(1)}_{\beta\gamma}(E_e, \omega_{\rm min})$, describing the
contributions of the neutron radiative beta decay $n \to p + e^- +
\bar{\nu}_e + \gamma$ to the neutron lifetime, where $\omega_{\rm
  min}$ is a non-covariant infrared cut-off having a meaning of the
photon--energy threshold of the detector, we refer to
\cite{Ivanov2013}. We adduce here the analytical expressions of these
functions for completeness (see \cite{Ivanov2013})
\begin{eqnarray}\label{eq:A.8}
\hspace{-0.39in}g^{(1)}_{\beta\gamma}(E_e,\mu) &=& \Big[{\ell
    n}\Big(\frac{2(E_0 - E_e)}{\mu}\Big) - \frac{3}{2} +
  \frac{1}{3}\,\frac{E_0 - E_e}{E_e}\, \Big(1 + \frac{1}{8} \frac{E_0
    - E_e}{E_e} \Big)\Big]\Big[\frac{1}{\beta}\,{\ell n}\Big(\frac{1 +
    \beta}{1 - \beta}\Big) - 2\Big]\nonumber\\
\hspace{-0.3in}&+& 1 + \frac{1}{2\beta}\,{\ell n}\Big(\frac{1 +
  \beta}{1 - \beta}\Big) - \frac{1}{4\beta}\,{\ell n}^2\Big(\frac{1 +
  \beta}{1 - \beta}\Big) - \frac{1}{\beta}\,{\rm Li}_2\Big(\frac{2
  \beta}{1 + \beta} \Big) + \frac{1}{12} \frac{(E_0 -
  E_e)^2}{E^2_e},\nonumber\\
\hspace{-0.39in}g^{(1)}_{\beta\gamma}(E_e,\omega_{\rm min}) &=&
\Big[{\ell n}\Big(\frac{E_0 - E_e}{\omega_{\rm min}}\Big) -
  \frac{3}{2} + \frac{1}{3}\,\frac{E_0 - E_e}{E_e}\Big(1 +
  \frac{1}{8}\,\frac{E_0 - E_e}{E_e}
  \Big)\Big]\Big[\frac{1}{\beta}\,{\ell n}\Big(\frac{1 + \beta}{1 -
    \beta}\Big) - 2\Big] + \frac{1}{12}\,\frac{(E_0 - E_e)^2}{E^2_e}.
\end{eqnarray}
The hermitian conjugate amplitude of the neutron beta decay
Eq.(\ref{eq:A.3}) is equal to
\begin{eqnarray}\label{eq:A.9}
\hspace{-0.3in}&&M^{\dagger}(n\to p\,e^-\bar{\nu}_e) = - 2 m_n
G_V\Big\{\big(1 + U_1\big) \,
[\varphi^{\dagger}_n\varphi_p][\bar{v}_{\bar{\nu}}\gamma^0(1 -
  \gamma^5) u_e] + g_A \big(1 + U_2\big) \,
[\varphi^{\dagger}_n\vec{\sigma}\,\varphi_p]\cdot [\bar{v}_{\bar{\nu}}
  \vec{\gamma}\, (1 - \gamma^5) u_e]\nonumber\\
\hspace{-0.3in}&&+ U_3 \,
       [\varphi^{\dagger}_n\varphi_p][\bar{v}_{\bar{\nu}}(1 +
         \gamma^5) u_e] - g_A U_4 \,
       [\varphi^{\dagger}_p\vec{\sigma}\,\varphi_n] \cdot
       [\bar{v}_{\bar{\nu}}\gamma^0 \vec{\gamma} \,(1 + \gamma^5) u_e]
       + U_5  \,[\varphi^{\dagger}_n (\vec{k}_e\cdot \vec{\sigma}\,)
         \varphi_p][\bar{v}_{\bar{\nu}}(1 + \gamma^5) u_e] \nonumber\\
\hspace{-0.3in}&& + U_6 \,[\varphi^{\dagger}_n
  (\vec{k}_{\bar{\nu}}\cdot \vec{\sigma}\,)\varphi_p ]
       [\bar{v}_{\bar{\nu}}(1 + \gamma^5) u_e] + U_7
       \,[\varphi^{\dagger}_n (\vec{k}_e\cdot \vec{\sigma}\,)
         \varphi_p][\bar{v}_{\bar{\nu}} \gamma^0(1 - \gamma^5) u_e] +
       U_8 \, [\varphi^{\dagger}_n(\vec{k}_e\cdot
         \vec{\sigma}\,)\vec{\sigma}\, \varphi_p]\cdot [
         \bar{v}_{\bar{\nu}} \vec{\gamma}\,(1 - \gamma^5)
         u_e]\Big\}.\nonumber\\
\hspace{-0.3in}&&
\end{eqnarray}
We use this amplitude for the calculation of the square of absolute
value of the amplitude Eq.(\ref{eq:A.3}), summed over polarizations of
massive particles. It is equal to
\begin{eqnarray}\label{eq:A.10}
\hspace{-0.3in}&&\sum_{\rm pol.}\frac{|M(n \to p e^-
  \bar{\nu}_e)|^2}{(1 + 3 g^2_A)|G_V|^2 64 m^2_n E_e E_{\bar{\nu}}} =
\frac{1}{(1 + 3 g^2_A) 8 E_e E_{\bar{\nu}}}\,\bigg\{(1 + 2U_1)\, {\rm
  tr}\{(1 + \vec{\xi}_n\cdot \vec{\sigma}\,)\}{\rm tr}\{(\hat{k}_e + m_e
  \gamma^5 \hat{\zeta}_e)\gamma^0 \hat{k}_{\bar{\nu}}\gamma^0 (1 -
  \gamma^5)\} \nonumber\\
\hspace{-0.3in}&&+ g_A(1 + U_1 + U_2){\rm tr}\{(1 + \vec{\xi}_n\cdot
\vec{\sigma}\,)\vec{\sigma}\,\} \cdot {\rm tr}\{(\hat{k}_e + m_e
\gamma^5 \hat{\zeta}_e)\vec{\gamma}\,\hat{k}_{\bar{\nu}}\gamma^0 (1 -
\gamma^5)\}+ g_A(1 + U_1 + U_2){\rm tr}\{(1 + \vec{\xi}_n\cdot
\vec{\sigma}\,)\vec{\sigma}\,\}\nonumber\\
\hspace{-0.3in}&& \cdot {\rm tr}\{(\hat{k}_e + m_e \gamma^5
\hat{\zeta}_e)\gamma^0 \hat{k}_{\bar{\nu}}\vec{\gamma}\, (1 -
\gamma^5)\} + g^2_A(1 + 2 U_2){\rm tr}\{(1 + \vec{\xi}_n\cdot
\vec{\sigma}\,)\sigma^a \sigma^b\,\} {\rm tr}\{(\hat{k}_e + m_e
\gamma^5 \hat{\zeta}_e)\gamma^b \hat{k}_{\bar{\nu}}\gamma^a (1 -
\gamma^5)\}\nonumber\\
\hspace{-0.3in}&& + U_3{\rm tr}\{(1 + \vec{\xi}_n\cdot
\vec{\sigma}\,)\}{\rm tr}\{(m_e + \hat{k}_e \gamma^5 \hat{\zeta}_e)
\hat{k}_{\bar{\nu}}\gamma^0 (1 - \gamma^5)\} + U_3{\rm tr}\{(1 +
\vec{\xi}_n\cdot \vec{\sigma}\,)\}{\rm tr}\{(m_e + \hat{k}_e \gamma^5
\hat{\zeta}_e) \gamma^0 \hat{k}_{\bar{\nu}} (1 + \gamma^5)\}\nonumber\\
\hspace{-0.3in}&& + g_A U_4 {\rm tr}\{(1 + \vec{\xi}_n\cdot
\vec{\sigma}\,)\vec{\sigma}\,\} \cdot {\rm tr}\{(m_e + \hat{k}_e
\gamma^5 \hat{\zeta}_e)\gamma^0 \vec{\gamma}\, \hat{k}_{\bar{\nu}}
\gamma^0\, (1 - \gamma^5)\} - g_A U_4 {\rm tr}\{(1 + \vec{\xi}_n\cdot
\vec{\sigma}\,)\vec{\sigma}\,\} \cdot {\rm tr}\{(m_e + \hat{k}_e
\gamma^5 \hat{\zeta}_e)\nonumber\\
\hspace{-0.3in}&& \times \, \gamma^0 \, \hat{k}_{\bar{\nu}}\gamma^0
\vec{\gamma}\, (1 + \gamma^5)\} + U_5 {\rm tr}\{(1 + \vec{\xi}_n\cdot
\vec{\sigma}\,)(\vec{k}_e \cdot \vec{\sigma}\,)\} {\rm tr}\{( m_e +
\hat{k}_e \gamma^5 \hat{\zeta}_e)\,\hat{k}_{\bar{\nu}}\gamma^0 (1 -
\gamma^5)\} + U_5 {\rm tr}\{(1 + \vec{\xi}_n\cdot
\vec{\sigma}\,)(\vec{k}_e \cdot \vec{\sigma}\,)\}\nonumber\\
\hspace{-0.3in}&& \times \, {\rm tr}\{( m_e + \hat{k}_e \gamma^5
\hat{\zeta}_e) \gamma^0 \hat{k}_{\bar{\nu}}(1 + \gamma^5)\} + U_6 {\rm
  tr}\{(1 + \vec{\xi}_n\cdot \vec{\sigma}\,)(\vec{k}_{\bar{\nu}} \cdot
\vec{\sigma}\,)\} {\rm tr}\{( m_e + \hat{k}_e \gamma^5
\hat{\zeta}_e)\,\hat{k}_{\bar{\nu}}\gamma^0 (1 - \gamma^5)\} + U_6
    {\rm tr}\{(1 + \vec{\xi}_n\cdot \vec{\sigma}\,)\nonumber\\
\hspace{-0.3in}&& \times \,(\vec{k}_e \cdot \vec{\sigma}\,)\}\, {\rm
  tr}\{( m_e + \hat{k}_e\gamma^5 \hat{\zeta}_e) \gamma^0
\hat{k}_{\bar{\nu}}(1 + \gamma^5)\} + U_7 {\rm tr}\{(1 +
\vec{\xi}_n\cdot \vec{\sigma}\,)(\vec{k}_e \cdot \vec{\sigma}\,)\}
    {\rm tr}\{(\hat{k}_e + m_e \gamma^5 \hat{\zeta}_e)\,\gamma^0
    \hat{k}_{\bar{\nu}} \gamma^0 (1 - \gamma^5)\} \nonumber\\
\hspace{-0.3in}&& + U_7 {\rm tr}\{(1 + \vec{\xi}_n\cdot
\vec{\sigma}\,)(\vec{k}_e \cdot \vec{\sigma}\,)\, {\rm tr}\{(\hat{k}_e
+ m_e\gamma^5 \hat{\zeta}_e) \gamma^0 \hat{k}_{\bar{\nu}} \gamma^0 (1
- \gamma^5)\} + U_8{\rm tr}\{(1 + \vec{\xi}_n\cdot
\vec{\sigma}\,)(\vec{k}_e \cdot \vec{\sigma}\,) \vec{\sigma}\,\} \cdot
    {\rm tr}\{(\hat{k}_e + m_e \gamma^5
    \hat{\zeta}_e)\nonumber\\
\hspace{-0.3in}&& \times\, \vec{\gamma}\,\hat{k}_{\bar{\nu}}\gamma^0
(1 - \gamma^5)\}+ U_8{\rm tr}\{(1 + \vec{\xi}_n\cdot \vec{\sigma}\,)
\vec{\sigma}\,(\vec{k}_e \cdot \vec{\sigma}\,)\} \cdot {\rm
  tr}\{(\hat{k}_e + m_e \gamma^5 \hat{\zeta}_e) \gamma^0
\hat{k}_{\bar{\nu}} \vec{\gamma}\, (1 - \gamma^5)\} + g_A U_3 {\rm
  tr}\{(1 + \vec{\xi}_n\cdot \vec{\sigma}\,)\vec{\sigma}\,\} \nonumber\\
\hspace{-0.3in}&& \cdot {\rm tr}\{(m_e + \hat{k}_e \gamma^5
\hat{\zeta}_e) \hat{k}_{\bar{\nu}}\vec{\gamma}\, (1 - \gamma^5)\} +
g_A U_3 {\rm tr}\{(1 + \vec{\xi}_n\cdot
\vec{\sigma}\,)\vec{\sigma}\,\} \cdot {\rm tr}\{(m_e + \hat{k}_e
\gamma^5 \hat{\zeta}_e) \vec{\gamma}\,\hat{k}_{\bar{\nu}} (1 +
\gamma^5)\} + g^2_A U_4{\rm tr}\{(1 + \vec{\xi}_n\cdot
\vec{\sigma}\,) \nonumber\\
\hspace{-0.3in}&& \times\, \sigma^a \sigma^b\,\} {\rm tr}\{(m_e +
\hat{k}_e \gamma^5 \hat{\zeta}_e) \gamma^0 \gamma^b
\hat{k}_{\bar{\nu}}\gamma^a (1 - \gamma^5)\}- g^2_A U_4{\rm tr}\{(1 +
\vec{\xi}_n\cdot \vec{\sigma}\,)\sigma^a \sigma^b\,\} {\rm tr}\{(m_e +
\hat{k}_e \gamma^5 \hat{\zeta}_e) \gamma^b \hat{k}_{\bar{\nu}}
\gamma^0 \gamma^a (1 - \gamma^5)\} \nonumber\\
\hspace{-0.3in}&& + g_A U_5 {\rm tr}\{(1 + \vec{\xi}_n\cdot
\vec{\sigma}\,) \vec{\sigma}\,(\vec{k}_e \cdot \vec{\sigma}\,)\} \cdot
    {\rm tr}\{( m_e + \hat{k}_e\gamma^5 \hat{\zeta}_e)
    \hat{k}_{\bar{\nu}} \vec{\gamma}\, (1 - \gamma^5)\} + g_A U_5 {\rm
      tr}\{(1 + \vec{\xi}_n\cdot \vec{\sigma}\,) (\vec{k}_e \cdot
    \vec{\sigma}\,) \vec{\sigma}\,\} \cdot {\rm tr}\{( m_e +
    \hat{k}_e\gamma^5 \hat{\zeta}_e) \nonumber\\
\hspace{-0.3in}&& \times\, \vec{\gamma}\, \hat{k}_{\bar{\nu}} (1 +
\gamma^5)\} + g_A U_6 {\rm tr}\{(1 + \vec{\xi}_n\cdot \vec{\sigma}\,)
\vec{\sigma}\,(\vec{k}_{\bar{\nu}} \cdot \vec{\sigma}\,)\} \cdot {\rm
  tr}\{( m_e + \hat{k}_e\gamma^5 \hat{\zeta}_e) \hat{k}_{\bar{\nu}}
\vec{\gamma}\, (1 - \gamma^5)\} + g_A U_6 {\rm tr}\{(1 +
\vec{\xi}_n\cdot \vec{\sigma}\,) (\vec{k}_{\bar{\nu}} \cdot
\vec{\sigma}\,) \vec{\sigma}\,\} \nonumber\\
\hspace{-0.3in}&& \cdot {\rm tr}\{( m_e + \hat{k}_e\gamma^5
\hat{\zeta}_e) \vec{\gamma}\, \hat{k}_{\bar{\nu}} (1 + \gamma^5)\} +
g_A U_7 {\rm tr}\{(1 + \vec{\xi}_n\cdot \vec{\sigma}\,)
\vec{\sigma}\,(\vec{k}_e \cdot \vec{\sigma}\,)\} \cdot {\rm
  tr}\{(\hat{k}_e + m_e \gamma^5 \hat{\zeta}_e) \gamma^0
\hat{k}_{\bar{\nu}} \vec{\gamma}\, (1 - \gamma^5)\} + g_A U_7 \nonumber\\
\hspace{-0.3in}&& \times\, {\rm tr}\{(1 + \vec{\xi}_n\cdot
\vec{\sigma}\,) (\vec{k}_e \cdot \vec{\sigma}\,) \vec{\sigma}\,\}
\cdot {\rm tr}\{(\hat{k}_e + m_e \gamma^5 \hat{\zeta}_e)
\vec{\gamma}\, \hat{k}_{\bar{\nu}} \gamma^0 (1 - \gamma^5)\} + g_A
U_8{\rm tr}\{(1 + \vec{\xi}_n\cdot \vec{\sigma}\,)\sigma^a (\vec{k}_e
\cdot \vec{\sigma}\,) \sigma^b\,\} {\rm tr}\{(\hat{k}_e + m_e \gamma^5
\hat{\zeta}_e) \nonumber\\
\hspace{-0.3in}&& \times \, \gamma^b \hat{k}_{\bar{\nu}}\gamma^a (1 -
\gamma^5)\} + g_A U_8{\rm tr}\{(1 + \vec{\xi}_n\cdot
\vec{\sigma}\,)\sigma^a (\vec{k}_e \cdot \vec{\sigma}\,) \sigma^b\,\}
    {\rm tr}\{(\hat{k}_e + m_e \gamma^5 \hat{\zeta}_e)\gamma^b
    \hat{k}_{\bar{\nu}}\gamma^a (1 - \gamma^5)\}.
\end{eqnarray}
Having calculated the traces over the nucleon degrees of freedom and
using the properties of the Dirac matrices \cite{Itzykson1980}
\begin{eqnarray}\label{eq:A.11}
  \hspace{-0.3in}
\gamma^{\alpha}\gamma^{\nu}\gamma^{\mu} =
\gamma^{\alpha}\eta^{\nu\mu} - \gamma^{\nu}\eta^{\mu\alpha} +
\gamma^{\mu}\eta^{\alpha\nu} +
i\,\varepsilon^{\alpha\nu\mu\beta}\,\gamma_{\beta}\gamma^5
\end{eqnarray}
and $\gamma^{\mu} \gamma^{\nu} + \gamma^{\nu} \gamma^{\mu} = 2
\eta^{\mu\nu}$, where $\eta^{\mu\nu}$ is the metric tensor of the
Minkowski space--time, $\varepsilon^{\alpha\nu\mu\beta}$ is the
Levi--Civita tensor defined by $\varepsilon^{0123} = 1$ and
$\varepsilon_{\alpha\nu\mu\beta}= - \varepsilon^{\alpha\nu\mu\beta}$
\cite{Itzykson1980}, we transcribe the right-hand-side (r.h.s.) of
Eq.(\ref{eq:A.10}) into the form \cite{Ivanov2013, Ivanov2017,
  Ivanov2019a}
\begin{eqnarray*}
\hspace{-0.3in}&&\sum_{\rm pol.}\frac{|M(n \to p e^-
  \bar{\nu}_e)|^2}{(1 + 3 g^2_A)|G_V|^2 64 m^2_n E_e E_{\bar{\nu}}} =
\frac{1}{4 E_e}\Big\{\Big[\Big(1 + \frac{2}{1 + 3 g^2_A}\, \big(U_1 +
  3 g^2_A U_2\big)\Big) + \Big(B_0 + \frac{2}{1 + 3 g^2_A}\big(g_A
  \big(U_1 + U_2) + 2 g^2_A\,U_2\big)\Big) \nonumber\\
\hspace{-0.3in}&& \times \, \frac{\vec{\xi}_n \cdot
  \vec{k}_{\bar{\nu}}}{E_{\bar{\nu}}} + \frac{2 E_e}{1 + 3
  g^2_A}\,(U_7 - g_A U_8)\, \frac{\vec{\xi}_n \cdot \vec{k}_e}{E_e} +
\frac{2 E_e}{1 + 3 g^2_A}\,\Big(g_A U_7 + (1 - 2 g_A)\,
U_8\Big)\,\frac{\vec{k}_e \cdot \vec{k}_{\bar{\nu}}}{E_e
  E_{\bar{\nu}}}\Big]\, {\rm tr}\{(\hat{k}_e + m_e \gamma^5
\hat{\zeta}_e) \gamma^0 (1 - \gamma^5)\} \nonumber\\
\hspace{-0.3in}&& + \Big[\Big(a_0 + \frac{2}{1 + 3 g^2_A}\big(U_1 -
  g^2_A U_2\big)\Big) \frac{\vec{k}_{\bar{\nu}}}{E_{\bar{\nu}}} +
  \Big(A_0 + \frac{2}{1 + 3 g^2_A}\big(g_A(U_1 + U_2) - 2 g^2_A
  U_2\big)\Big) \vec{\xi}_n + \frac{2 E_e}{1 + 3 g^2_A} (U_7 - g_A
  U_8) \frac{(\vec{\xi}_n \cdot \vec{k}_e)}{E_e}
  \frac{\vec{k}_{\bar{\nu}}}{E_{\bar{\nu}}}\nonumber\\
\hspace{-0.3in}&& + \frac{2 E_e}{1 + 3 g^2_A}\Big(g_A U_7 + (1 + 2
g_A) U_8\Big)\, \frac{\vec{k}_e}{E_e} + \frac{2 E_e}{1 + 3
  g^2_A}\Big(- g_A U_7 + (1 + 2 g_A) U_8\Big)\, \frac{(\vec{k}_e\cdot
  \vec{k}_{\bar{\nu}})}{E_e E_{\bar{\nu}}}\,\vec{\xi}_n + \frac{2
  E_e}{1 + 3 g^2_A}\Big( g_A U_7 - (1 - 2 g_A)\nonumber\\
\hspace{-0.3in}&& \times U_8\Big)\, \frac{(\vec{\xi}_n\cdot
  \vec{k}_{\bar{\nu}})}{E_{\bar{\nu}}}\,\frac{\vec{k}_e}{E_e}\Big]
\cdot \, {\rm tr}\{(\hat{k}_e + m_e \gamma^5 \hat{\zeta}_e)
\vec{\gamma}\, (1 - \gamma^5)\} + \Big[\frac{2}{1 + 3 g^2_A}\,\Big(U_3
  + 3 g^2_A U_4 + g_A E_{\bar{\nu}} U_6\Big) + \frac{2}{1 + 3
    g^2_A}\,\Big(g_A (U_3 + U_4)\nonumber\\
\hspace{-0.3in}&&+ 2 g^2_A U_4 + E_{\bar{\nu}} U_6\Big)\,
\frac{\vec{\xi}_n \cdot \vec{k}_{\bar{\nu}}}{E_{\bar{\nu}}} + \frac{2
  E_e}{1 + 3 g^2_A}\,U_5\,\frac{\vec{\xi}_n \cdot \vec{k}_e}{E_e} +
\frac{2 E_e}{1 + 3 g^2_A} \,g_A \, U_5\, \frac{\vec{k}_e \cdot
  \vec{k}_{\bar{\nu}}}{E_e E_{\bar{\nu}}}\Big]{\rm tr}\{(m_e +
\hat{k}_e \gamma^5 \hat{\zeta}_e)\} + \Big[- \frac{2 E_e}{1 + 3 g^2_A}\,
  g_A U_5 \nonumber\\  
\hspace{-0.3in}&& \times\, \frac{(\vec{\xi}_n \cdot \vec{k}_e)
  \vec{k}_{\bar{\nu}}}{E_e E_{\bar{\nu}}} - \frac{2 E_{\bar{\nu}}}{1 +
  3 g^2_A}\,(1 + g_A)\, U_6 \, \frac{(\vec{\xi}_n \cdot
  \vec{k}_{\bar{\nu}}) \vec{k}_{\bar{\nu}}}{ E^2_{\bar{\nu}}} +
\frac{2}{1 + 3 g^2_A}\,\Big(- g_A (U_3 + U_4) + 2 g^2_A U_4 + g_A
E_{\bar{\nu}} U_6\Big)\, \vec{\xi}_n + \frac{2}{1 + 3 g^2_A}
\nonumber\\
\hspace{-0.3in}&& \times\,\Big(- U_3 + g^2_A U_4 - g_A E_{\bar{\nu}}
U_6\Big)\, \frac{\vec{k}_{\bar{\nu}}}{E_{\bar{\nu}}} - \frac{2 E_e}{1
  + 3 g^2_A}\,g_A\, U_5\, \frac{\vec{k}_e}{E_e} + \frac{2 E_e}{1 + 3
  g^2_A}\, g_A\, U_5\, \frac{(\vec{\xi}_n\cdot
  \vec{k}_{\bar{\nu}})\vec{k}_e}{E_e E_{\bar{\nu}}} - \frac{2 E_e}{1 +
  3 g^2_A}\, g_A\, U_5\, \frac{(\vec{k}_e\cdot
  \vec{k}_{\bar{\nu}})\vec{\xi}_n}{E_e E_{\bar{\nu}}}\Big]\nonumber\\
 \end{eqnarray*}
\begin{eqnarray}\label{eq:A.12} 
\hspace{-0.3in}&& \cdot \, {\rm tr}\{( m_e + \hat{k}_e\gamma^5
\hat{\zeta}_e)\, \gamma^0 \vec{\gamma}\,\gamma^5\} + \Big[\frac{2}{1 +
    3 g^2_A}\,g_A \Big( U_3 - U_4 - E_{\bar{\nu}} U_6\Big)\, i\,
  \frac{\vec{\xi}_n \times \vec{k}_{\bar{\nu}}}{E_{\bar{\nu}}} -
  \frac{2 E_e}{1 + 3 g^2_A}\, g_A U_5\,i\, \frac{\vec{\xi}_n \times
    \vec{k}_e}{E_e} + \frac{2 E_e}{1 + 3 g^2_A}\, g_A U_5\nonumber\\
\hspace{-0.3in}&& \times \,i \,\frac{\vec{k}_e \times
  \vec{k}_{\bar{\nu}}}{E_e E_{\bar{\nu}}}\Big]\cdot {\rm tr}\{( m_e +
\hat{k}_e\gamma^5 \hat{\zeta}_e) \gamma^0 \vec{\gamma}\,\}\Big\}.
\end{eqnarray}
Having calcuated the traces over leptonic degrees of freedom we arrive
at the expression
\begin{eqnarray}\label{eq:A.13}
\hspace{-0.21in}&&\sum_{\rm pol.}\frac{|M(n \to p e^-
  \bar{\nu}_e)|^2}{(1 + 3 g^2_A)|G_V|^2 64 m^2_n E_e E_{\bar{\nu}}} =
\Big[\Big(1 + \frac{2}{1 + 3 g^2_A}\, \big(U_1 +
  3 g^2_A U_2\big)\Big) + \Big(B_0 + \frac{2}{1 + 3 g^2_A}\big(g_A
  \big(U_1 + U_2) + 2 g^2_A\,U_2\big)\Big) \nonumber\\
\hspace{-0.3in}&& \times \, \frac{\vec{\xi}_n \cdot
  \vec{k}_{\bar{\nu}}}{E_{\bar{\nu}}} + \frac{2 E_e}{1 + 3
  g^2_A}\,(U_7 - g_A U_8)\, \frac{\vec{\xi}_n \cdot \vec{k}_e}{E_e} +
\frac{2 E_e}{1 + 3 g^2_A}\,\Big(g_A U_7 + (1 - 2 g_A)\,
U_8\Big)\,\frac{\vec{k}_e \cdot \vec{k}_{\bar{\nu}}}{E_e
  E_{\bar{\nu}}}\Big]\, \Big(1 - \frac{m_e}{E_e}\,\zeta^0_e\Big)
+ \Big[\Big(a_0 + \frac{2}{1 + 3 g^2_A}\nonumber\\
\hspace{-0.3in}&&\times \big(U_1 - g^2_A U_2\big)\Big)
\frac{\vec{k}_{\bar{\nu}}}{E_{\bar{\nu}}} + \Big(A_0 + \frac{2}{1 + 3
  g^2_A}\big(g_A(U_1 + U_2) - 2 g^2_A U_2\big)\Big) \vec{\xi}_n +
\frac{2 E_e}{1 + 3 g^2_A} (U_7 - g_A U_8) \frac{(\vec{\xi}_n \cdot
  \vec{k}_e)}{E_e} \frac{\vec{k}_{\bar{\nu}}}{E_{\bar{\nu}}} + \frac{2
  E_e}{1 + 3 g^2_A}\nonumber\\
\hspace{-0.3in}&& \times \Big(g_A U_7 + (1 + 2 g_A) U_8\Big)\,
\frac{\vec{k}_e}{E_e} + \frac{2 E_e}{1 + 3 g^2_A}\Big(- g_A U_7 + (1 +
2 g_A) U_8\Big)\, \frac{(\vec{k}_e\cdot \vec{k}_{\bar{\nu}})}{E_e
  E_{\bar{\nu}}}\,\vec{\xi}_n + \frac{2 E_e}{1 + 3 g^2_A}\Big( g_A U_7
- (1 - 2 g_A) U_8\Big)\nonumber\\
\hspace{-0.3in}&& \times\, \frac{(\vec{\xi}_n\cdot
  \vec{k}_{\bar{\nu}})}{E_{\bar{\nu}}}\,\frac{\vec{k}_e}{E_e}\Big]
\cdot \, \Big(\frac{\vec{k}_e}{E_e} -
\frac{m_e}{E_e}\,\vec{\zeta}_e\Big) + \Big[\frac{2}{1 + 3
    g^2_A}\,\Big(U_3 + 3 g^2_A U_4 + g_A E_{\bar{\nu}} U_6\Big) +
  \frac{2}{1 + 3 g^2_A}\,\Big(g_A (U_3 + U_4) + 2 g^2_A U_4 +
  E_{\bar{\nu}} U_6\Big)\nonumber\\
\hspace{-0.3in}&& \times \, \frac{\vec{\xi}_n \cdot
  \vec{k}_{\bar{\nu}}}{E_{\bar{\nu}}} + \frac{2 E_e}{1 + 3
  g^2_A}\,U_5\,\frac{\vec{\xi}_n \cdot \vec{k}_e}{E_e} + \frac{2
  E_e}{1 + 3 g^2_A} \,g_A \, U_5\, \frac{\vec{k}_e \cdot
  \vec{k}_{\bar{\nu}}}{E_e E_{\bar{\nu}}}\Big]\, \frac{m_e}{E_e} +
\Big[- \frac{2 E_e}{1 + 3 g^2_A}\, g_A U_5\, \frac{(\vec{\xi}_n \cdot
    \vec{k}_e) \vec{k}_{\bar{\nu}}}{E_e E_{\bar{\nu}}} - \frac{2
    E_{\bar{\nu}}}{1 + 3 g^2_A}\,(1 + g_A)\, U_6 \nonumber\\
\hspace{-0.3in}&& \times \, \frac{(\vec{\xi}_n \cdot
  \vec{k}_{\bar{\nu}}) \vec{k}_{\bar{\nu}}}{ E^2_{\bar{\nu}}} +
\frac{2}{1 + 3 g^2_A} \Big(- g_A (U_3 + U_4) + 2 g^2_A U_4 + g_A
E_{\bar{\nu}} U_6\Big)\, \vec{\xi}_n + \frac{2}{1 + 3 g^2_A}\Big(- U_3
+ g^2_A U_4 - g_A E_{\bar{\nu}} U_6\Big)
\frac{\vec{k}_{\bar{\nu}}}{E_{\bar{\nu}}} - \frac{2 E_e}{1 + 3
  g^2_A}\nonumber\\
\hspace{-0.3in}&& \times \,g_A\, U_5\, \frac{\vec{k}_e}{E_e} + \frac{2
  E_e}{1 + 3 g^2_A}\, g_A\, U_5\, \frac{(\vec{\xi}_n\cdot
  \vec{k}_{\bar{\nu}})\vec{k}_e}{E_e E_{\bar{\nu}}} - \frac{2 E_e}{1 +
  3 g^2_A}\, g_A\, U_5\, \frac{(\vec{k}_e\cdot
  \vec{k}_{\bar{\nu}})\vec{\xi}_n}{E_e E_{\bar{\nu}}}\Big] \cdot
\Big(\vec{\zeta}_e - \frac{\vec{k}_e}{E_e}\,\zeta^0_e\Big) +
\Big[\frac{2}{1 + 3 g^2_A}\,g_A \Big( U_3 - U_4 - E_{\bar{\nu}}
  U_6\Big)\nonumber\\
\hspace{-0.3in}&& \times \, i\, \frac{\vec{\xi}_n \times
  \vec{k}_{\bar{\nu}}}{E_{\bar{\nu}}} - \frac{2 E_e}{1 + 3 g^2_A}\,
g_A U_5\,i\, \frac{\vec{\xi}_n \times \vec{k}_e}{E_e} + \frac{2 E_e}{1
  + 3 g^2_A}\, g_A U_5 \,i \,\frac{\vec{k}_e \times
  \vec{k}_{\bar{\nu}}}{E_e E_{\bar{\nu}}}\Big]\cdot i\,\frac{\vec{k}_e
  \times \vec{\xi}_e}{E_e}.
\end{eqnarray}
In terms of the irreducible correlation structures the r.h.s. of
Eq.(\ref{eq:A.13}) is given by
\begin{eqnarray}\label{eq:A.14}
\hspace{-0.3in}&&\sum_{\rm pol.}\frac{|M(n \to p e^-
  \bar{\nu}_e)|^2}{(1 + 3 g^2_A)|G_V|^2 64 m^2_n E_e E_{\bar{\nu}}} =
\zeta(E_e)_{\rm RC}\Big\{1 + b(E_e)_{\rm RC} \frac{m_e}{E_e} +
a(E_e)_{\rm RC} \frac{\vec{k}_e\cdot \vec{k}_{\bar{\nu}}}{E_e
  E_{\bar{\nu}}} + A(E_e)_{\rm RC} \frac{\vec{\xi}_n\cdot
  \vec{k}_e}{E_e} + B(E_e)_{\rm RC}\nonumber\\
\hspace{-0.3in}&& \times \frac{\vec{\xi}_n\cdot
  \vec{k}_{\bar{\nu}}}{E_{\bar{\nu}}} + K_n(E_e)_{\rm RC}
\frac{(\vec{\xi}_n\cdot \vec{k}_e)(\vec{k}_e\cdot
  \vec{k}_{\bar{\nu}})}{E^2_e E_{\bar{\nu}}} + G(E_e)_{\rm RC}
\frac{\vec{\xi}_e \cdot \vec{k}_e}{E_e} + H(E_e)_{\rm RC}
\frac{\vec{\xi}_e \cdot \vec{k}_{\bar{\nu}}}{E_{\bar{\nu}}} +
N(E_e)_{\rm RC} \vec{\xi}_n\cdot \vec{\xi}_e + Q_e(E_e)_{\rm RC}\nonumber\\
\hspace{-0.3in}&& \times \frac{(\vec{\xi}_n\cdot \vec{k}_e)(
  \vec{k}_e\cdot \vec{\xi}_e)}{(E_e + m_e) E_e} + K_e(E_e)_{\rm RC}
\frac{(\vec{\xi}_e\cdot \vec{k}_e)( \vec{k}_e\cdot
  \vec{k}_{\bar{\nu}})}{(E_e + m_e)E_e E_{\bar{\nu}}} + S(E_e)_{\rm
  RC} \frac{(\vec{\xi}_n\cdot \vec{\xi}_e)(\vec{k}_e \cdot
  \vec{k}_{\bar{\nu}})}{E_e E_{\bar{\nu}}} + T(E_e)_{\rm RC}
\frac{(\vec{\xi}_n \cdot \vec{k}_{\bar{\nu}})(\vec{\xi}_e \cdot
  \vec{k}_e)}{E_e E_{\bar{\nu}}} \nonumber\\
\hspace{-0.3in}&& +\, U(E_e)_{\rm RC} \frac{(\vec{\xi}_n\cdot
  \vec{k}_e)(\vec{\xi}_e \cdot \vec{k}_{\bar{\nu}})}{E_e
  E_{\bar{\nu}}} + \frac{2 E_e}{1 + 3 g^2_A}\,\Big(2g_A U_5 - (1 -
g_A)\, U_7 - (1 + g_A)\, U_8\Big)\, \frac{(\vec{\xi}_n \cdot
  \vec{k}_e)(\vec{\xi}_e \cdot \vec{k}_e)(\vec{k}_e \cdot
  \vec{k}_{\bar{\nu}})}{(E_e + m_e) E^2_e E_{\bar{\nu}}} \nonumber\\
\hspace{-0.3in}&& - \frac{2 E_e}{1 + 3 g^2_A}\, (1 + g_A)\, U_6
\Big[\Big(\frac{(\vec{\xi}_n \cdot \vec{k}_{\bar{\nu}})(\vec{\xi}_e
    \cdot \vec{k}_{\bar{\nu}})}{E^2_{\bar{\nu}}} -
  \frac{1}{3}\,\vec{\xi}_e \cdot \vec{\xi}_e\Big) + \Big(-
  \frac{(\vec{\xi}_n \cdot \vec{k}_{\bar{\nu}})(\vec{\xi}_e \cdot
    \vec{k}_e)(\vec{k}_e \cdot \vec{k}_{\bar{\nu}})}{(E_e + m_e) E_e
    E^2_{\bar{\nu}}} + \frac{1}{3}\,\frac{(\vec{\xi}_n \cdot
    \vec{k}_2)(\vec{\xi}_e \cdot \vec{k}_e)}{(E_e + m_e)
    E_e}\Big)\Big] \Big\}, \nonumber\\
\hspace{-0.3in}&&
\end{eqnarray}
where the index RC means that these corrections are defined by the
outer radiative corrections of order $O(\alpha/\pi)$, calculated to LO
in the large nucleon mass expansion \cite{Sirlin1967, Shann1971,
  Ivanov2017, Ivanov2019a}, and the inner radiative corrections of
order $O(\alpha E_e/m_N)$ \cite{Ivanov2019b, Ivanov2020a}. The
correlation function $\zeta(E_e)_{\rm RC}$ and the correlation
coefficients $\zeta(E_e)_{\rm RC} X(E_e)_{\rm RC}$ for $X =
b,a,A,B,\ldots,T$ and $U$ are equal to
\begin{eqnarray*}
\hspace{-0.15in}\zeta(E_e)_{\rm RC} &=& \Big(1 + \frac{2}{1 + 3
  g^2_A}\,\big(U_1 + 3 g^2_A U_2\big)\Big) + \frac{2}{1 + 3
  g^2_A}\,\Big(U_3 + 3g^2_A U_4 + g_A E_{\bar{\nu}}
U_6\Big)\frac{m_e}{E_e} \nonumber\\
\hspace{-0.15in}&&+ \frac{2E_e}{1 + 3 g^2_A}\,\Big(g_A U_7 + (1 + 2
g_A) U_8\Big)\,\beta^2, \nonumber\\
\hspace{-0.15in}\zeta(E_e)_{\rm RC} b(E_e)_{\rm RC} &=& 0, \nonumber\\
\end{eqnarray*}
\begin{eqnarray}\label{eq:A.15}
\hspace{-0.15in}\zeta(E_e)_{\rm RC} a(E_e)_{\rm RC} &=& \Big(a_0 +
\frac{2}{1 + 3 g^2_A}\,\big(U_1 - g^2_A U_2\big)\Big) + \frac{2 E_e}{1
  + 3 g^2_A}\,g_A\, U_5 \, \frac{m_e}{E_e} + \frac{2 E_e}{1 + 3
  g^2_A}\, \Big(g_A U_7 + (1 - 2 g_A) U_8\Big), \nonumber\\
\hspace{-0.15in}\zeta(E_e)_{\rm RC} A(E_e)_{\rm RC} &=& \Big(A_0 +
\frac{2}{1 + 3 g^2_A}\,\big(g_A(U_1 + U_2) - 2 g^2_A U_2\big)\Big) +
\frac{2 E_e}{1 + 3 g^2_A}\,\frac{m_e}{E_e}\, U_5 + \frac{2 E_e}{1 + 3
  g^2_A}\, \Big(U_7 - g_A U_8\Big),\nonumber\\
\hspace{-0.15in} \zeta(E_e)_{\rm RC} B(E_e)_{\rm RC} &=& \Big(B_0 +
\frac{2}{1 + 3 g^2_A}\,\big(g_A(U_1 + U_2) + 2 g^2_A U_2\big)\Big) +
\frac{2}{1 + 3 g^2_A}\,\Big(g_A (U_3 + U_4) + 2 g^2_A U_4 +
E_{\bar{\nu}} U_6\Big)\, \frac{m_e}{E_e} \nonumber\\
\hspace{-0.15in}&&+ \frac{2 E_e}{1 + 3 g^2_A}\, \Big(g_A U_7 - (1 - 2
g_A)\, U_8\Big)\, \beta^2, \nonumber\\
\hspace{-0.3in} \zeta(E_e)_{\rm RC} K_n(E_e)_{\rm RC} &=&\frac{2
  E_e}{1 + 3 g^2_A}\, \Big((1 - g_A)\, U_7 + (1 + g_A)\, U_8\Big),
\nonumber\\
\hspace{-0.15in} \zeta(E_e)_{\rm RC} Q_n(E_e)_{\rm RC} &=& 0,
\nonumber\\
\hspace{-0.15in} \zeta(E_e)_{\rm RC} G(E_e)_{\rm RC} &=& - \Big(1 +
\frac{2}{1 + 3 g^2_A}\,\big(U_1 + 3 g^2_A U_2\big)\Big) - \frac{2
  E_e}{1 + 3 g^2_A}\,g_A\, U_5\, \frac{m_e}{E_e} - \frac{2 E_e}{1 + 3
  g^2_A}\,\Big(g_A U_7 + (1 + 2 g_A)\, U_8\Big),
\nonumber\\
\hspace{-0.15in} \zeta(E_e)_{\rm RC} H(E_e)_{\rm RC} &=& -
\frac{m_e}{E_e}\,\Big( a_0 + \frac{2}{1 + 3 g^2_A}\,\big(U_1 - g^2_A
U_2\big)\Big) + \frac{2}{1 + 3 g^2_A}\,\big(- U_3 + g^2_A U_4 - g_A
E_{\bar{\nu}} U_6\big) + \frac{2 E_e}{1 + 3 g^2_A}\,g_A\, U_5\,
\beta^2, \nonumber\\
\hspace{-0.15in} \zeta(E_e)_{\rm RC} N(E_e)_{\rm RC} &=& -
\frac{m_e}{E_e}\, \Big(A_0 + \frac{2}{1 + 3 g^2_A}\,\big(g_A(U_1 +
U_2) - 2 g^2_A U_2\big)\Big) + \frac{2}{1 + 3 g^2_A}\,\Big(- g_A (U_3
+ U_4) + 2 g^2_A)\, U_4 + g_A E_{\bar{\nu}} U_6\Big)  \nonumber\\
\hspace{-0.15in} && + \frac{2 E_e}{1
  + 3 g^2_A}\,g_A\, U_5\, \beta^2 - \frac{1}{3}\, \frac{2
  E_{\bar{\nu}}}{1 + 3 g^2_A}\,(1 + g_A)\, U_6,\nonumber\\
\hspace{-0.15in} \zeta(E_e)_{\rm RC} Q_e(E_e)_{\rm RC} &=& - \Big(A_0 +
\frac{2}{1 + 3 g^2_A}\,\big(g_A(U_1 + U_2) - 2 g^2_A U_2\big)\Big) -
\frac{2}{1 + 3 g^2_A}\, \Big( - g_A (U_3 + U_4) + 2 g^2_A \, U_4 + g_A
E_{\bar{\nu}} U_6\big)  \nonumber\\
\hspace{-0.15in} && + \frac{2}{1 + 3 g^2_A}\,\Big(1 +
\frac{m_e}{E_e}\Big)\, \Big(g_A U_5 - U_7 + g_A U_8\Big) +
\frac{1}{3}\, \frac{2 E_{\bar{\nu}}}{1 + 3 g^2_A}\,(1 + g_A)\,
U_6,\nonumber\\
\hspace{-0.15in} \zeta(E_e)_{\rm RC} K_e(E_e)_{\rm RC} &=& - \Big(a_0 +
\frac{2}{1 + 3 g^2_A}\,\big(U_1 - g^2_A U_2\big)\Big) - \frac{2}{1 + 3
  g^2_A}\,\Big(- U_3 + g^2_A U_4 - g_A E_{\bar{\nu}} U_6\Big) -
\frac{2 E_e}{1 + 3 g^2_A}\,\Big(g_A U_5 + g_A U_7 \nonumber\\
\hspace{-0.3in}&&+ (1 - 2 g_A)\, U_8\Big)\, \Big(1 +
\frac{m_e}{E_e}\Big),\nonumber\\
\hspace{-0.15in} \zeta(E_e)_{\rm RC} S(E_e)_{\rm RC} &=& - \frac{2}{1 +
  3 g^2_A}\,g_A\Big(U_3 - U_4 - E_{\bar{\nu}} U_6\Big) - \frac{2 E_e}{1 +
  3 g^2_A}\,g_A U_5 - \frac{2 E_e}{1 + 3 g^2_A}\, \Big(- g_A U_7 + (1
+ 2 g_A)\, U_8\Big) \frac{m_e}{E_e},\nonumber\\
\hspace{-0.15in} \zeta(E_e)_{\rm RC} T(E_e)_{\rm RC} &=& - \Big(B_0 +
\frac{2}{1 + 3 g^2_A}\,\big(g_A(U_1 + U_2) + 2 g^2_A U_2\big)\Big) +
\frac{2 E_e}{1 + 3 g^2_A}\, \Big(g_A\, U_5 - g_A U_7 + (1 - 2 g_A)\,
U_8\Big),\nonumber\\
\hspace{-0.15in} \zeta(E_e)_{\rm RC} U(E_e)_{\rm RC} &=& \frac{2}{1 +
  3 g^2_A}\, g_A \Big(U_3 - U_4 - E_{\bar{\nu}} U_6\Big) - \frac{2
  E_e}{1 + 3 g^2_A}\,g_A\, U_5 - \frac{2 E_e}{1 + 3 g^2_A}\, \Big(U_7 -
  g_A U_8\Big)\, \frac{m_e}{E_e},
\end{eqnarray}
where the correlation coefficients $a_0, A_0$ and $B_0$ depend only on
the axial coupling constant $g_A$ \cite{Abele2008, Nico2009} (see also
\cite{Ivanov2021a})
\begin{eqnarray}\label{eq:A.16}
\hspace{-0.3in} a_0 = \frac{1 - g^2_A}{1 + 3 g^2_A}\quad,\quad A_0 = 2
\frac{g_A(1 - g_A)}{1 + 3 g^2_A}\quad,\quad B_0 = 2 \frac{g_A(1 +
  g_A)}{1 + 3 g^2_A}.
\end{eqnarray}
Using the definitions of the functions $U_j$ for $j = 1,2, \ldots,8$
in Eq.(\ref{eq:A.4}) and Eq.(\ref{eq:A.7}) we transcribe
Eq.(\ref{eq:A.15}) as follows
\begin{eqnarray*}
  \hspace{-0.3in}&&\zeta(E_e)_{\rm RC} = 1 +
  \frac{\alpha}{\pi}\,\Big(\bar{g}_n(E_e) -
  g^{(1)}_{\beta\gamma}(E_e,\mu)\Big) + \frac{1}{1 + 3 g^2_A}\,
  \frac{\alpha}{\pi}\, \frac{E_e}{m_N} \Big( f_V(E_e) + \sqrt{1 -
    \beta^2}\,f_S(E_e) + g_A g_V(E_e)\, \beta^2  \nonumber\\
\hspace{-0.3in}&&+ (1 + 2 g_A)\,h_A(E_e)\, \beta^2 + g_A \frac{E_0 -
  E_e}{E_e}\,\sqrt{1 - \beta^2}\, h_S(E_e) + 3 g^2_A f_A(E_e) + 3
g^2_A \sqrt{1 - \beta^2}\, f_T(E_e)\Big) + \frac{3 g^2_A}{1 + 3 g^2_A}
\nonumber\\
\hspace{-0.3in}&& \times \, \frac{\alpha}{\pi}\, \frac{5}{2}\,
\frac{m^2_N}{M^2_W}\,{\ell n}\frac{M^2_W}{m^2_N} + \frac{1}{1 + 3
  g^2_A}\,\frac{\alpha}{\pi}\, \Big(\bar{g}_{\rm st}(E_e) + 3 g_A
\bar{f}_{\rm st}(E_e)\Big), \nonumber\\
\hspace{-0.3in}&&\zeta(E_e)_{\rm RC} b(E_e)_{\rm RC} = 0,
\nonumber\\
\hspace{-0.3in}&&\zeta(E_e)_{\rm RC} a(E_e)_{\rm RC} = a_0 \,\Big(1 +
\frac{\alpha}{\pi}\,\Big(\bar{g}_n(E_e) + \frac{1 -
  \beta^2}{2\beta}\,{\ell n}\Big(\frac{1 + \beta}{1 - \beta}\Big) -
g^{(1)}_{\beta\gamma}(E_e,\mu)\Big)\Big) + \frac{1}{1 + 3
  g^2_A}\,\frac{\alpha}{\pi}\,\frac{E_e}{m_N}  \Big(f_V(E_e)\nonumber\\
\end{eqnarray*}
\begin{eqnarray*}
\hspace{-0.3in}&& + g_A \sqrt{1 - \beta^2}\,g_S(E_e) + g_A\, g_V(E_e)
+ (1 - 2 g_A)\, h_A(E_e) - g^2_A f_A(E_e)\Big) - \frac{g^2_A}{1 + 3
  g^2_A}\,\frac{\alpha}{\pi}\,\frac{5}{2}\, \frac{m^2_N}{M^2_W}\,{\ell
  n}\frac{M^2_W}{m^2_N} + \frac{1}{1 + 3 g^2_A}\nonumber\\
\hspace{-0.3in}&& \times \,\frac{\alpha}{\pi}\, \Big(\bar{g}_{\rm
  st}(E_e) - g_A \bar{f}_{\rm st}(E_e)\Big), \nonumber\\
\hspace{-0.3in}&&\zeta(E_e)_{\rm RC} A(E_e)_{\rm RC} = A_0 \,\Big(1 +
\frac{\alpha}{\pi}\,\Big(\bar{g}_n(E_e) + \frac{1 -
  \beta^2}{2\beta}\,{\ell n}\Big(\frac{1 + \beta}{1 - \beta}\Big) -
g^{(1)}_{\beta\gamma}(E_e,\mu)\Big)\Big) + \frac{1}{1 + 3 g^2_A}\,
\frac{\alpha}{\pi}\, \frac{E_e}{m_N}\, \Big(g_A f_V(E_e) \nonumber\\
\hspace{-0.3in}&& + \sqrt{1 - \beta^2}\, g_S(E_e) + g_V(E_e) - g_A
h_A(E_e) + g_A(1 - 2 g_A) \, f_A(E_e)\Big) + \frac{1}{1 + 3 g^2_A}\,
g_A(1 - 2 g_A)\,\frac{\alpha}{\pi}\, \frac{5}{2}\,
\frac{m^2_N}{M^2_W}\,{\ell n}\frac{M^2_W}{m^2_N} \nonumber\\
\hspace{-0.3in}&&+ \frac{1}{1 + 3 g^2_A}\,\frac{\alpha}{\pi}\,
\Big(g_A \bar{g}_{\rm st}(E_e) + (1 - 2 g_A)\, \bar{f}_{\rm
  st}(E_e)\Big),\nonumber\\
\hspace{-0.3in}&& \zeta(E_e)_{\rm RC} B(E_e)_{\rm RC} = B_0 \,\Big(1 +
\frac{\alpha}{\pi}\,\Big(\bar{g}_n(E_e) -
g^{(1)}_{\beta\gamma}(E_e,\mu)\Big)\Big) + \frac{1}{1 + 3 g^2_A}\,
\frac{\alpha}{\pi}\, \frac{E_e}{m_N}\,\Big(g_A f_V(E_e) + g_A \sqrt{1
  - \beta^2}\, f_S(E_e) \nonumber\\
\hspace{-0.3in}&&+ \frac{E_0 - E_e}{E_e}\, \sqrt{1 - \beta^2}\,
h_S(E_e) + g_A g_V(E_e)\, \beta^2 - (1 - 2 g_A)\, h_A(E_e)\, \beta^2 +
g_A(1 + 2 g_A) \, f_A(E_e) + g_A(1 + 2 g_A) \sqrt{1 - \beta^2}\nonumber\\
\hspace{-0.3in}&& \times \, f_T(E_e) \Big) + \frac{1}{1 + 3
  g^2_A}\,g_A(1 + 2 g_A)\,\frac{\alpha}{\pi}\, \frac{5}{2}\,
\frac{m^2_N}{M^2_W}\,{\ell n}\frac{M^2_W}{m^2_N} + \frac{1}{1 + 3
  g^2_A}\,\frac{\alpha}{\pi}\, \Big(g_A \bar{g}_{\rm st}(E_e) + (1 + 2
g_A)\, \bar{f}_{\rm st}(E_e)\Big), \nonumber\\
\hspace{-0.3in}&& \zeta(E_e)_{\rm RC} K_n(E_e)_{\rm RC} = \frac{1}{1 +
  3 g^2_A}\, \frac{\alpha}{\pi}\, \frac{E_e}{m_N}\,\Big((1 -
g_A)\,g_V(E_e) + (1 + 2 g_A)\,
h_A(E_e)\Big),\nonumber\\ \hspace{-0.3in}&& \zeta(E_e)_{\rm RC}
Q_n(E_e)_{\rm RC} = 0,\nonumber\\ 
\hspace{-0.3in} &&\zeta(E_e)_{\rm RC} G(E_e)_{\rm RC} = - \Big(1 +
\frac{\alpha}{\pi}\,\Big(\bar{g}_n(E_e) + \frac{1 - \beta^2}{2\beta}\,
     {\ell n}\Big(\frac{1 + \beta}{1 - \beta}\Big) -
     g^{(1)}_{\beta\gamma}(E_e,\mu)\Big)\Big) - \frac{1}{1 + 3 g^2_A}\,
     \frac{\alpha}{\pi}\, \frac{E_e}{m_N}\, \Big(f_V(E_e) +
     g_A\nonumber\\ \hspace{-0.3in}&&\times \, \sqrt{1 - \beta^2}\,
     g_S(E_e) + g_A g_V(E_e) + (1 + 2 g_A)\, h_A(E_e) + 3 g^2_A
     f_A(E_e) \Big) - \frac{3 g^2_A}{1 + 3 g^2_A}\,
     \frac{\alpha}{\pi}\, \frac{5}{2}\, \frac{m^2_N}{M^2_W}\,{\ell
       n}\frac{M^2_W}{m^2_N}\nonumber\\   
\hspace{-0.3in} && - \frac{1}{1 + 3 g^2_A}\,\frac{\alpha}{\pi}\,
\Big(\bar{g}_{\rm st}(E_e) + 3 g_A \bar{f}_{\rm st}(E_e)\Big),
\nonumber\\
\hspace{-0.3in}&& \zeta(E_e)_{\rm RC} H(E_e)_{\rm RC} = -
\frac{m_e}{E_e}\,a_0 \Big(1 + \frac{\alpha}{\pi}\,\Big(\bar{g}_n(E_e)
- \frac{\beta}{2}\,{\ell n}\Big(\frac{1 + \beta}{1 - \beta}\Big) -
g^{(1)}_{\beta\gamma}(E_e,\mu)\Big)\Big) - \frac{1}{1 + 3 g^2_A}\,
\frac{\alpha}{\pi}\, \frac{E_e}{m_N}\Big( \sqrt{1 - \beta^2}\,
f_V(E_e)\nonumber\\
\hspace{-0.3in}&& + f_S(E_e) + g_A g_S(E_e)\, \beta^2 + g_A\,
\frac{E_0 - E_e}{E_e}\, h_S(E_e) - g^2_A \sqrt{1 - \beta^2}\, f_A(E_e)
- g^2_A f_T(E_e) \Big) + \sqrt{1 - \beta^2}\, \frac{g^2_A}{1 + 3 g^2_A}\,
\frac{\alpha}{\pi}\, \frac{5}{2}\, \frac{m^2_N}{M^2_W}\nonumber\\
\hspace{-0.3in}&& \times \,{\ell n}\frac{M^2_W}{m^2_N} - \frac{1}{1 +
  3 g^2_A}\,\frac{\alpha}{\pi}\,\sqrt{1 - \beta^2} \Big(\bar{g}_{\rm
  st}(E_e) - g_A \bar{f}_{\rm st}(E_e)\Big),\nonumber\\
\hspace{-0.3in}&& \zeta(E_e)_{\rm RC} N(E_e)_{\rm RC} = -
\frac{m_e}{E_e}\,A_0 \Big(1 + \frac{\alpha}{\pi}\,\Big(\bar{g}_n(E_e)
- \frac{\beta}{2}\,{\ell n}\Big(\frac{1 + \beta}{1 - \beta}\Big) -
g^{(1)}_{\beta\gamma}(E_e,\mu)\Big)\Big) - \frac{1}{1 + 3 g^2_A}\,
\frac{\alpha}{\pi}\,\frac{E_e}{m_N}\Big(g_A f_V(E_e)\nonumber\\
\hspace{-0.3in}&& \times\, \sqrt{1 - \beta^2} + g_A f_S(E_e) - g_A
g_S(E_e)\,\beta^2 + \frac{1}{3}\,(1 - 2 g_A)\,\frac{E_0 - E_e}{E_e}\,
h_S(E_e) + g_A (1 - 2 g_A)\, \sqrt{1 - \beta^2}\, f_A(E_e) 
\nonumber\\
\hspace{-0.3in}&& + g_A (1 - 2 g_A)\, f_T(E_e)\Big) - \sqrt{1 -
  \beta^2}\, \frac{1}{1 + 3 g^2_A}\,g_A(1 - 2 g_A)\,
\frac{\alpha}{\pi}\, \frac{5}{2}\, \frac{m^2_N}{M^2_W}\,{\ell
  n}\frac{M^2_W}{m^2_N} \nonumber\\
\hspace{-0.3in}&& - \frac{1}{1 + 3 g^2_A}\,\frac{\alpha}{\pi}\,\sqrt{1
  - \beta^2} \Big(g_A \bar{g}_{\rm st}(E_e) + (1 - 2 g_A)\,
\bar{f}_{\rm st}(E_e)\Big),\nonumber\\
\hspace{-0.3in}&& \zeta(E_e)_{\rm RC} Q_e(E_e)_{\rm RC} = - A_0
\,\Big(1 + \frac{\alpha}{\pi}\,\Big(\bar{g}_n(E_e) + \big(1 + \sqrt{1
  - \beta^2}\,\big) \frac{\sqrt{1 - \beta^2}}{2\beta}\,{\ell
  n}\Big(\frac{1 + \beta}{1 - \beta}\Big) -
g^{(1)}_{\beta\gamma}(E_e,\mu)\Big)\Big) \nonumber\\
\hspace{-0.3in}&& - \frac{1}{1 + 3 g^2_A}\, \frac{\alpha}{\pi}\,
\frac{E_e}{m_N}\, \Big(g_A f_V(E_e) - g_A f_S(E_e) - g_A (1 + \sqrt{1
  - \beta^2}\,g_S(E_e) + (1 + \sqrt{1 - \beta^2}\,)\,g_V(E_e) -
\frac{1}{3}\,(1 - 2 g_A)\nonumber\\
\hspace{-0.3in}&& \times \,\frac{E_0 - E_e}{E_e}\, h_S(E_e) + g_A (1 -
2 g_A)\, f_A(E_e) - g_A (1 - 2 g_A) \, f_T(E_e)\Big) - \frac{1}{1 + 3
  g^2_A}\,g_A(1 - 2 g_A)\, \frac{\alpha}{\pi}\, \frac{5}{2}\,
\frac{m^2_N}{M^2_W}\,{\ell n}\frac{M^2_W}{m^2_N}\nonumber\\
\hspace{-0.3in}&& - \frac{1}{1 + 3 g^2_A}\,\frac{\alpha}{\pi} \Big(g_A
\bar{g}_{\rm st}(E_e) + (1 - 2 g_A)\, \bar{f}_{\rm
  st}(E_e)\Big),\nonumber\\
  \hspace{-0.3in}&& \zeta(E_e)_{\rm RC} K_e(E_e)_{\rm RC} = - a_0
  \Big(1 + \frac{\alpha}{\pi}\Big(\bar{g}_n(E_e) + (1 + \sqrt{1 -
    \beta^2}\,)\frac{\sqrt{1 - \beta^2}}{2 \beta}\,{\ell
    n}\Big(\frac{1 + \beta}{1 - \beta}\Big) -
  g^{(1)}_{\beta\gamma}(E_e,\mu)\Big)\Big)\nonumber\\
\hspace{-0.3in}&& - \frac{1}{1 + 3 g^2_A}\, \frac{\alpha}{\pi}\,
\frac{E_e}{m_N}\Big(f_V(E_e) - f_S(E_e) + g_A (1 + \sqrt{1 -
  \beta^2}\,) \,g_S(E_e) + g_A (1 + \sqrt{1 - \beta^2}\,)\,g_V(E_e) +
(1 - 2 g_A) \nonumber\\
\hspace{-0.3in}&&\times\, (1 + \sqrt{1 - \beta^2}\,)\,h_A(E_e) -
g_A\,\frac{E_0 - E_e}{E_e}\,h_S(E_e) - g^2_A f_A(E_e) + g^2_A
f_T(E_e)\Big) + \frac{g^2_A}{1 + 3 g^2_A}\, \frac{\alpha}{\pi}\,
\frac{5}{2}\, \frac{m^2_N}{M^2_W}\,{\ell n}\frac{M^2_W}{m^2_N}\nonumber\\
\hspace{-0.3in}&& - \frac{1}{1 + 3 g^2_A}\,\frac{\alpha}{\pi}
\Big(\bar{g}_{\rm st}(E_e) - g_A\, \bar{f}_{\rm
  st}(E_e)\Big),\nonumber\\
\end{eqnarray*}
\begin{eqnarray}\label{eq:A.17}
\hspace{-0.3in}&& \zeta(E_e)_{\rm RC} S(E_e)_{\rm RC} = - \frac{1}{1 +
  3 g^2_A}\,\frac{\alpha}{\pi}\, \frac{E_e}{m_N}\Big(g_A f_S(E_e) -
g_A f_T(E_e) + g_A g_S(E_e) - g_A \frac{E_0 - E_e}{E_e}\, h_S(E_e)\nonumber\\
\hspace{-0.3in}&& - g_A \sqrt{1 - \beta^2}\, g_V(E_e) + (1 + 2 g_A)\,
\sqrt{1 - \beta^2}\,h_A(E_e)\Big),\nonumber\\
\hspace{-0.3in}&& \zeta(E_e)_{\rm RC} T(E_e)_{\rm RC} = - B_0\Big(1 +
\frac{\alpha}{\pi}\,\Big(\bar{g}_n(E_e) + \frac{1 -
  \beta^2}{2\beta}\,{\ell n}\Big(\frac{1 + \beta}{1 - \beta}\Big) -
g^{(1)}_{\beta\gamma}(E_e,\mu)\Big)\Big) - \frac{1}{1 + 3
  g^2_A}\,\frac{\alpha}{\pi}\, \frac{E_e}{m_N}\nonumber\\
\hspace{-0.3in}&& \times \Big(g_A f_V(E_e) - g_A g_S(E_e) + g_A
g_V(E_e) - (1 - 2 g_A)\, h_A(E_e) + g_A(1 + 2 g_A)\, f_A(E_e)\Big) -
\frac{1}{1 + 3 g^2_A}\,g_A(1 + 2 g_A)\nonumber\\
\hspace{-0.3in}&& \times\, \frac{\alpha}{\pi}\, \frac{5}{2}\,
\frac{m^2_N}{M^2_W}\,{\ell n}\frac{M^2_W}{m^2_N} - \frac{1}{1 + 3
  g^2_A}\,\frac{\alpha}{\pi} \Big(g_A \bar{g}_{\rm st}(E_e) + (1 + 2
g_A)\, \bar{f}_{\rm st}(E_e)\Big),\nonumber\\
\hspace{-0.3in} &&\zeta(E_e)_{\rm RC} U(E_e)_{\rm RC} = \frac{1}{1 +
  3 g^2_A}\,\frac{\alpha}{\pi}\, \frac{E_e}{m_N}\Big(g_A f_S(E_e) -
g_A f_T(E_e) - g_A g_S(E_e) - g_A \frac{E_0 - E_e}{E_e}\, h_S(E_e)\nonumber\\
\hspace{-0.3in}&& - \sqrt{1 - \beta^2}\, g_V(E_e) + g_A\, \sqrt{1 -
  \beta^2}\,h_A(E_e)\Big).
\end{eqnarray}
Taking into account the contribution of the neutron radiative beta
decay \cite{Ivanov2013, Ivanov2017, Ivanov2019a, Ivanov2021a,
  Ivanov2021b} we obtain the correlation function $\zeta(E_e)_{\rm
  RC}$ and correlation coefficients $\zeta(E_e)_{\rm RC} X(E_e)_{\rm
  RC}$ for $X = b,a,A,B,\ldots,T$ and $U$ in Eq.(\ref{eq:A.17}) in the
form
\begin{eqnarray*}
  \hspace{-0.3in}&&\zeta(E_e)_{\rm RC} = 1 +
  \frac{\alpha}{\pi}\,\bar{g}_n(E_e) + \frac{1}{1 + 3 g^2_A}\,
  \frac{\alpha}{\pi}\, \frac{E_e}{m_N} \Big( f_V(E_e) + \sqrt{1 -
    \beta^2}\,f_S(E_e) + g_A g_V(E_e)\, \beta^2 + (1 + 2
  g_A)\,h_A(E_e)\, \beta^2 \nonumber\\
\hspace{-0.3in}&& + g_A \frac{E_0 - E_e}{E_e}\,\sqrt{1 - \beta^2}\,
h_S(E_e) + 3 g^2_A f_A(E_e) + 3 g^2_A \sqrt{1 - \beta^2}\,
f_T(E_e)\Big) + \frac{3 g^2_A}{1 + 3 g^2_A} \, \frac{\alpha}{\pi}\,
\frac{5}{2}\, \frac{m^2_N}{M^2_W}\,{\ell n}\frac{M^2_W}{m^2_N}\nonumber\\
\hspace{-0.3in}&& + \frac{1}{1 + 3 g^2_A}\,\frac{\alpha}{\pi}\,
\Big(\bar{g}_{\rm st}(E_e) + 3 g_A \bar{f}_{\rm st}(E_e)\Big),
\nonumber\\
\hspace{-0.3in}&&\zeta(E_e)_{\rm RC} b(E_e)_{\rm RC} = 0,
\nonumber\\
\hspace{-0.3in}&&\zeta(E_e)_{\rm RC} a(E_e)_{\rm RC} = a_0 \,\Big(1 +
\frac{\alpha}{\pi}\,\bar{g}_n(E_e) + \frac{\alpha}{\pi}\,f_n(E_e)
\Big) + \frac{1}{1 + 3 g^2_A}\,\frac{\alpha}{\pi}\,\frac{E_e}{m_N}
\Big(f_V(E_e) + g_A \sqrt{1 - \beta^2}\,g_S(E_e)\nonumber\\
\hspace{-0.3in}&& + g_A\, g_V(E_e) + (1 - 2 g_A)\, h_A(E_e) - g^2_A
f_A(E_e)\Big) - \frac{g^2_A}{1 + 3
  g^2_A}\,\frac{\alpha}{\pi}\,\frac{5}{2}\, \frac{m^2_N}{M^2_W}\,{\ell
  n}\frac{M^2_W}{m^2_N}\nonumber\\
\hspace{-0.3in}&& + \frac{1}{1 + 3 g^2_A}\,\frac{\alpha}{\pi}\,
\Big(\bar{g}_{\rm st}(E_e) - g_A \bar{f}_{\rm st}(E_e)\Big),
\nonumber\\
\hspace{-0.3in}&&\zeta(E_e)_{\rm RC} A(E_e)_{\rm RC} = A_0 \,\Big(1 +
\frac{\alpha}{\pi}\,\bar{g}_n(E_e) + \frac{\alpha}{\pi}\,f_n(E_e)\Big)
+ \frac{1}{1 + 3 g^2_A}\, \frac{\alpha}{\pi}\, \frac{E_e}{m_N}\,
\Big(g_A f_V(E_e) + \sqrt{1 - \beta^2}\, g_S(E_e) \nonumber\\
\hspace{-0.3in}&& + g_V(E_e) - g_A h_A(E_e) + g_A(1 - 2 g_A) \,
f_A(E_e)\Big) + \frac{1}{1 + 3 g^2_A}\, g_A(1 - 2
g_A)\,\frac{\alpha}{\pi}\, \frac{5}{2}\, \frac{m^2_N}{M^2_W}\,{\ell
  n}\frac{M^2_W}{m^2_N} \nonumber\\
\hspace{-0.3in}&&+ \frac{1}{1 + 3 g^2_A}\,\frac{\alpha}{\pi}\,
\Big(g_A \bar{g}_{\rm st}(E_e) + (1 - 2 g_A)\, \bar{f}_{\rm
  st}(E_e)\Big),\nonumber\\
\hspace{-0.3in}&& \zeta(E_e)_{\rm RC} B(E_e)_{\rm RC} = B_0 \,\Big(1 +
\frac{\alpha}{\pi}\,\bar{g}_n(E_e)\Big) + \frac{1}{1 + 3 g^2_A}\,
\frac{\alpha}{\pi}\, \frac{E_e}{m_N}\,\Big(g_A f_V(E_e) + g_A \sqrt{1
  - \beta^2}\, f_S(E_e) + \frac{E_0 - E_e}{E_e}\nonumber\\
\hspace{-0.3in}&& \times \, \sqrt{1 - \beta^2}\, h_S(E_e) + g_A
g_V(E_e)\, \beta^2 - (1 - 2 g_A)\, h_A(E_e)\, \beta^2 + g_A(1 + 2 g_A)
\, f_A(E_e) + g_A(1 + 2 g_A) \sqrt{1 - \beta^2}\nonumber\\
\hspace{-0.3in}&& \times \, f_T(E_e) \Big) + \frac{1}{1 + 3
  g^2_A}\,g_A(1 + 2 g_A)\,\frac{\alpha}{\pi}\, \frac{5}{2}\,
\frac{m^2_N}{M^2_W}\,{\ell n}\frac{M^2_W}{m^2_N} + \frac{1}{1 + 3
  g^2_A}\,\frac{\alpha}{\pi}\, \Big(g_A \bar{g}_{\rm st}(E_e) + (1 + 2
g_A)\, \bar{f}_{\rm st}(E_e)\Big), \nonumber\\
\hspace{-0.3in}&& \zeta(E_e)_{\rm RC} K_n(E_e)_{\rm RC} = \frac{1}{1 +
  3 g^2_A}\, \frac{\alpha}{\pi}\, \frac{E_e}{m_N}\,\Big((1 -
g_A)\,g_V(E_e) + (1 + 2 g_A)\,
h_A(E_e)\Big),\nonumber\\ \hspace{-0.3in}&& \zeta(E_e)_{\rm RC}
Q_n(E_e)_{\rm RC} = 0,\nonumber\\
\hspace{-0.3in} &&\zeta(E_e)_{\rm RC} G(E_e)_{\rm RC} = - \Big(1 +
\frac{\alpha}{\pi}\,\bar{g}_n(E_e) + \frac{\alpha}{\pi}\,f_n(E_e)\Big)
- \frac{1}{1 + 3 g^2_A}\, \frac{\alpha}{\pi}\, \frac{E_e}{m_N}\,
\Big(f_V(E_e) + g_A \sqrt{1 - \beta^2}\, g_S(E_e) \nonumber\\
\hspace{-0.3in}&&+ g_A g_V(E_e) + (1
+ 2 g_A)\, h_A(E_e) + 3 g^2_A f_A(E_e) \Big) - \frac{3 g^2_A}{1 + 3
  g^2_A}\, \frac{\alpha}{\pi}\, \frac{5}{2}\,
\frac{m^2_N}{M^2_W}\,{\ell n}\frac{M^2_W}{m^2_N}\nonumber\\
\end{eqnarray*}
\begin{eqnarray}\label{eq:A.18}
\hspace{-0.3in} && - \frac{1}{1 + 3 g^2_A}\,\frac{\alpha}{\pi}\,
\Big(\bar{g}_{\rm st}(E_e) + 3 g_A \bar{f}_{\rm st}(E_e)\Big),
\nonumber\\
\hspace{-0.3in}&& \zeta(E_e)_{\rm RC} H(E_e)_{\rm RC} = -
\frac{m_e}{E_e}\,a_0 \Big(1 + \frac{\alpha}{\pi}\,\bar{g}_n(E_e) +
\frac{\alpha}{\pi}\,h^{(1)}_n(E_e)\Big) - \frac{1}{1 + 3 g^2_A}\,
\frac{\alpha}{\pi}\, \frac{E_e}{m_N}\Big( \sqrt{1 - \beta^2}\,
f_V(E_e) + f_S(E_e) \nonumber\\
\hspace{-0.3in}&& + g_A g_S(E_e)\, \beta^2 + g_A\,
\frac{E_0 - E_e}{E_e}\, h_S(E_e) - g^2_A \sqrt{1 - \beta^2}\, f_A(E_e)
- g^2_A f_T(E_e) \Big) + \sqrt{1 - \beta^2}\, \frac{g^2_A}{1 + 3 g^2_A}\,
\frac{\alpha}{\pi}\, \frac{5}{2}\, \frac{m^2_N}{M^2_W}\nonumber\\
\hspace{-0.3in}&& \times \,{\ell n}\frac{M^2_W}{m^2_N} - \frac{1}{1 +
  3 g^2_A}\,\frac{\alpha}{\pi}\,\sqrt{1 - \beta^2} \Big(\bar{g}_{\rm
  st}(E_e) - g_A \bar{f}_{\rm st}(E_e)\Big),\nonumber\\
\hspace{-0.3in}&& \zeta(E_e)_{\rm RC} N(E_e)_{\rm RC} = -
\frac{m_e}{E_e}\,A_0 \Big( 1 + \frac{\alpha}{\pi}\,\bar{g}_n(E_e) +
\frac{\alpha}{\pi}\,h^{(1)}_n(E_e)\Big) - \frac{1}{1 + 3 g^2_A}\,
\frac{\alpha}{\pi}\,\frac{E_e}{m_N}\Big(g_A f_V(E_e) \sqrt{1 -
  \beta^2} + g_A f_S(E_e)\nonumber\\
\hspace{-0.3in}&&  - g_A g_S(E_e)\,\beta^2 +
\frac{1}{3}\,(1 - 2 g_A)\,\frac{E_0 - E_e}{E_e}\, h_S(E_e) + g_A (1 -
2 g_A)\, \sqrt{1 - \beta^2}\, f_A(E_e) + g_A (1 - 2 g_A)\,
f_T(E_e)\Big)\nonumber\\
\hspace{-0.3in}&& - \sqrt{1 - \beta^2}\, \frac{1}{1 + 3 g^2_A}\,g_A(1
- 2 g_A)\, \frac{\alpha}{\pi}\, \frac{5}{2}\,
\frac{m^2_N}{M^2_W}\,{\ell n}\frac{M^2_W}{m^2_N} - \frac{1}{1 + 3
  g^2_A}\,\frac{\alpha}{\pi}\,\sqrt{1 - \beta^2} \Big(g_A \bar{g}_{\rm
  st}(E_e) + (1 - 2 g_A)\, \bar{f}_{\rm st}(E_e)\Big),\nonumber\\
\hspace{-0.3in}&& \zeta(E_e)_{\rm RC} Q_e(E_e)_{\rm RC} = - A_0
\,\Big(1 + \frac{\alpha}{\pi}\,\bar{g}_n(E_e) +
\frac{\alpha}{\pi}\,h^{(2)}_n(E_e)\Big) - \frac{1}{1 + 3 g^2_A}\,
\frac{\alpha}{\pi}\, \frac{E_e}{m_N}\, \Big(g_A f_V(E_e) - g_A
f_S(E_e) \nonumber\\
\hspace{-0.3in}&& - g_A (1 + \sqrt{1 - \beta^2}\,g_S(E_e) + (1 +
\sqrt{1 - \beta^2}\,)\,g_V(E_e) - \frac{1}{3}\,(1 - 2 g_A)\,\frac{E_0
  - E_e}{E_e}\, h_S(E_e) + g_A (1 - 2 g_A)\, f_A(E_e)\nonumber\\
\hspace{-0.3in}&& - g_A (1 - 2 g_A) \, f_T(E_e)\Big) - \frac{1}{1 + 3
  g^2_A}\,g_A(1 - 2 g_A)\, \frac{\alpha}{\pi}\, \frac{5}{2}\,
\frac{m^2_N}{M^2_W}\,{\ell n}\frac{M^2_W}{m^2_N} - \frac{1}{1 + 3
  g^2_A}\,\frac{\alpha}{\pi} \Big(g_A \bar{g}_{\rm st}(E_e) + (1 - 2
g_A)\, \bar{f}_{\rm st}(E_e)\Big),\nonumber\\
  \hspace{-0.3in}&& \zeta(E_e)_{\rm RC} K_e(E_e)_{\rm RC} = - a_0
  \Big(1 + \frac{\alpha}{\pi}\,\bar{g}_n(E_e) + \frac{\alpha}{\pi}\,
  h^{(2)}_n(E_e)\Big) - \frac{1}{1 + 3 g^2_A}\, \frac{\alpha}{\pi}\,
  \frac{E_e}{m_N}\Big(f_V(E_e) - f_S(E_e) \nonumber\\
\hspace{-0.3in}&& + g_A (1 + \sqrt{1 - \beta^2}\,) \,g_S(E_e) + g_A (1
+ \sqrt{1 - \beta^2}\,)\,g_V(E_e) + (1 - 2 g_A) \, (1 + \sqrt{1 -
  \beta^2}\,)\,h_A(E_e) \nonumber\\
\hspace{-0.3in}&&- g_A\,\frac{E_0 - E_e}{E_e}\,h_S(E_e) - g^2_A
f_A(E_e) + g^2_A f_T(E_e)\Big) + \frac{g^2_A}{1 + 3 g^2_A}\,
\frac{\alpha}{\pi}\, \frac{5}{2}\, \frac{m^2_N}{M^2_W}\,{\ell
  n}\frac{M^2_W}{m^2_N}\nonumber\\
\hspace{-0.3in}&& - \frac{1}{1 + 3 g^2_A}\,\frac{\alpha}{\pi}
\Big(\bar{g}_{\rm st}(E_e) - g_A\, \bar{f}_{\rm
  st}(E_e)\Big),\nonumber\\
\hspace{-0.3in}&& \zeta(E_e)_{\rm RC} S(E_e)_{\rm RC} = - \frac{1}{1 +
  3 g^2_A}\,\frac{\alpha}{\pi}\, \frac{E_e}{m_N}\Big(g_A f_S(E_e) -
g_A f_T(E_e) + g_A g_S(E_e) - g_A \frac{E_0 - E_e}{E_e}\, h_S(E_e)\nonumber\\
\hspace{-0.3in}&& - g_A \sqrt{1 - \beta^2}\, g_V(E_e) + (1 + 2 g_A)\,
\sqrt{1 - \beta^2}\,h_A(E_e)\Big),\nonumber\\
\hspace{-0.3in}&& \zeta(E_e)_{\rm RC} T(E_e)_{\rm RC} = - B_0\Big(1 +
\frac{\alpha}{\pi}\,\bar{g}_n(E_e) + \frac{\alpha}{\pi}\,f_n(E_e)
\Big) - \frac{1}{1 + 3 g^2_A}\,\frac{\alpha}{\pi}\, \frac{E_e}{m_N}
\Big(g_A f_V(E_e) - g_A g_S(E_e) \nonumber\\
\hspace{-0.3in}&& + g_A g_V(E_e) - (1 - 2 g_A)\, h_A(E_e) + g_A(1 + 2
g_A)\, f_A(E_e)\Big) - \frac{1}{1 + 3 g^2_A}\,g_A(1 + 2 g_A)\,
\frac{\alpha}{\pi}\, \frac{5}{2}\, \frac{m^2_N}{M^2_W}\,{\ell
  n}\frac{M^2_W}{m^2_N} \nonumber\\
\hspace{-0.3in}&& - \frac{1}{1 + 3 g^2_A}\,\frac{\alpha}{\pi} \Big(g_A
\bar{g}_{\rm st}(E_e) + (1 + 2 g_A)\, \bar{f}_{\rm
  st}(E_e)\Big),\nonumber\\
\hspace{-0.3in} &&\zeta(E_e)_{\rm RC} U(E_e)_{\rm RC} = \frac{1}{1 +
  3 g^2_A}\,\frac{\alpha}{\pi}\, \frac{E_e}{m_N}\Big(g_A f_S(E_e) -
g_A f_T(E_e) - g_A g_S(E_e) - g_A \frac{E_0 - E_e}{E_e}\, h_S(E_e)\nonumber\\
\hspace{-0.3in}&& - \sqrt{1 - \beta^2}\, g_V(E_e) + g_A\, \sqrt{1 -
  \beta^2}\,h_A(E_e)\Big).
\end{eqnarray}
The functions $\bar{g}_n(E_e)$ and $f_n(E_e)$ have been calculated by
Sirlin \cite{Sirlin1967} and Shann \cite{Shann1971} (see also
\cite{Ivanov2013, Ivanov2017, Ivanov2021a}), whereas the function
$h^{(1)}_n(E_e)$ and $h^{(2)}_n(E_e)$ have been calculated in
\cite{Ivanov2017, Ivanov2019a}. They are equal to
\begin{eqnarray*}
\hspace{-0.3in}&&\bar{g}_n(E_e) = \frac{3}{4}\,{\ell
  n}\Big(\frac{m^2_N}{m^2_e}\Big) - \frac{3}{8} +
\Big[\frac{1}{\beta} \,{\ell n}\Big(\frac{1 + \beta}{1 -
    \beta}\Big) - 2\Big]\Big[{\ell n}\Big(\frac{2(E_0 -
    E_e)}{m_e}\Big) - \frac{3} {2} + \frac{1}{3}\,\frac{E_0 -
    E_e}{E_e}\Big]- \frac{2}{\beta}\, {\rm Li}_2\Big(\frac{2\beta}{1 +
  \beta}\Big) \nonumber\\
\hspace{-0.3in} && + \frac{1}{2\beta}{\ell n}\Big(\frac{1 + \beta}{1 -
  \beta}\Big)\,\Big[(1+\beta^2) + \frac{1}{12} \frac{(E_0 -
    E_e)^2}{E^2_e} - {\ell n}\Big(\frac{1 + \beta}{1 -
    \beta}\Big)\Big] , \nonumber\\
\end{eqnarray*}
\begin{eqnarray}\label{eq:A.19}
\hspace{-0.3in}&&f_n(E_e) = \frac{1}{3}\,\frac{E_0 - E_e}{E_e}\Big(1 +
\frac{1}{8}\frac{E_0 - E_e}{E_e}\Big)\,\frac{1 - \beta^2}{\beta^2}\,
\Big[\frac{1}{\beta}\,{\ell n}\Big(\frac{1 + \beta}{1 - \beta}\Big) -
  2\Big]- \frac{1}{12}\frac{(E_0 - E_e)^2}{E^2_e} + \frac{1 -
  \beta^2}{2\beta}\,{\ell n}\Big(\frac{1 + \beta}{1 -
  \beta}\Big),\nonumber\\
\hspace{-0.3in}&& h^{(1)}_n(E_e) =  - \frac{1}{3}\,\frac{E_0 -
  E_e}{E_e}\Big\{\Big(1 + \frac{1 + \beta^2}{8\beta^2}\,\frac{E_0 -
  E_e}{E_e}\Big) \Big[\frac{1}{\beta}\,{\ell n}\Big(\frac{1 + \beta}{1
    - \beta}\Big) - 2\Big] + \frac{1}{4}\, \frac{E_0 - E_e}{E_e}\Big\}
- \frac{\beta}{2}\,{\ell n}\Big(\frac{1 + \beta}{1 - \beta}\Big),
\nonumber\\
\hspace{-0.3in}&& h^{(2)}_n(E_e) = - \frac{1}{3}\,\frac{E_0 -
  E_e}{E_e}\Big\{\Big(1 + \frac{1 + \beta^2}{8\beta^2}\,\frac{E_0 -
  E_e}{E_e}\Big) \Big[\frac{1}{\beta}\,{\ell n}\Big(\frac{1 + \beta}{1
    - \beta}\Big) - 2\Big] + \frac{1}{4}\, \frac{E_0 -
  E_e}{E_e}\Big\}+ \big(1 + \sqrt{1 - \beta^2}\big)\nonumber\\
\hspace{-0.3in}&&\times \, \Big\{\frac{1}{3}\, \frac{E_0 -
  E_e}{\beta^2 E_e}\Big[\frac{1}{\beta}\,{\ell n}\Big(\frac{1 +
    \beta}{1 - \beta}\Big) - 2\Big] + \frac{1}{24}\, \frac{(E_0 -
  E_e)^2}{\beta^2 E^2_e} \Big(\frac{3 -
  \beta^2}{\beta^2}\Big[\frac{1}{\beta}\,{\ell n}\Big(\frac{1 +
    \beta}{1 - \beta}\Big) - 2\Big] - 2\Big) + \frac{\sqrt{1 -
    \beta^2}}{2 \beta}\, {\ell n}\Big(\frac{1 + \beta}{1 -
  \beta}\Big)\Big\}.\nonumber\\
\hspace{-0.3in}&&
\end{eqnarray}
The radiative corrections of order $O(\alpha/\pi)$ and $O(\alpha
E_e/m_N)$ to the neutron lifetime and the correlation coefficients of
the electron-energy and angular distribution Eq.(\ref{eq:1}) are given
by
\begin{eqnarray*}
  \hspace{-0.3in}&&\zeta(E_e)_{\rm RC} = 1 +
  \frac{\alpha}{\pi}\,\bar{g}_n(E_e) + \zeta(E_e)_{\rm RC-NLO},\nonumber\\
 \hspace{-0.3in}&&\zeta(E_e)_{\rm RC-NLO} = \frac{1}{1 + 3 g^2_A}\,
 \frac{\alpha}{\pi}\, \frac{E_e}{m_N} \Big( f_V(E_e) + \sqrt{1 -
   \beta^2}\,f_S(E_e) + g_A g_V(E_e)\, \beta^2 + (1 + 2 g_A)
 \,h_A(E_e)\, \beta^2 \nonumber\\
\hspace{-0.3in}&&+ g_A \frac{E_0 - E_e}{E_e}\,\sqrt{1 - \beta^2}\,
h_S(E_e) + 3 g^2_A f_A(E_e) + 3 g^2_A \sqrt{1 - \beta^2}\,
f_T(E_e)\Big) + \frac{3 g^2_A}{1 + 3 g^2_A} \, \frac{\alpha}{\pi}\,
\frac{5}{2}\, \frac{m^2_N}{M^2_W}\,{\ell n}\frac{M^2_W}{m^2_N} \nonumber\\
\hspace{-0.3in}&& + \frac{1}{1 + 3 g^2_A}\,\frac{\alpha}{\pi}\,
\Big(\bar{g}_{\rm st}(E_e) + 3 g_A \bar{f}_{\rm st}(E_e)\Big),
\nonumber\\
\hspace{-0.3in}&&a(E_e)_{\rm RC} = a_0 \,\Big(1 +
\frac{\alpha}{\pi}\,f_n(E_e) \Big) + a(E_e)_{\rm RC-NLO},\nonumber\\
\hspace{-0.3in}&&a(E_e)_{\rm RC-NLO} = \frac{1}{1 + 3
  g^2_A}\,\frac{\alpha}{\pi}\,\frac{E_e}{m_N} \Big(f_V(E_e) + g_A
\sqrt{1 - \beta^2}\,g_S(E_e) + g_A\, g_V(E_e)+ (1 - 2 g_A)\, h_A(E_e)
\nonumber\\
\hspace{-0.3in} && - g^2_A f_A(E_e)\Big) - \frac{g^2_A}{1 + 3
  g^2_A}\,\frac{\alpha}{\pi}\,\frac{5}{2}\, \frac{m^2_N}{M^2_W}\,{\ell
  n}\frac{M^2_W}{m^2_N} + \frac{1}{1 + 3 g^2_A}\,\frac{\alpha}{\pi}\,
\Big(\bar{g}_{\rm st}(E_e) - g_A \bar{f}_{\rm st}(E_e)\Big) - a_0
\zeta(E_e)_{\rm RC-NLO}, \nonumber\\
\hspace{-0.3in}&&A(E_e)_{\rm RC} = A_0 \,\Big(1 +
\frac{\alpha}{\pi}\,f_n(E_e)\Big) + A(E_e)_{\rm RC-NLO}, \nonumber\\
\hspace{-0.3in}&&A(E_e)_{\rm RC-NLO} = \frac{1}{1 + 3 g^2_A}\,
\frac{\alpha}{\pi}\, \frac{E_e}{m_N}\, \Big(g_A f_V(E_e) + \sqrt{1 -
  \beta^2}\, g_S(E_e) + g_V(E_e) - g_A h_A(E_e) \nonumber\\
\hspace{-0.3in}&& + g_A(1 - 2 g_A) \, f_A(E_e)\Big) + \frac{1}{1 + 3
  g^2_A}\, g_A(1 - 2 g_A)\,\frac{\alpha}{\pi}\, \frac{5}{2}\,
\frac{m^2_N}{M^2_W}\,{\ell n}\frac{M^2_W}{m^2_N} + \frac{1}{1 + 3
  g^2_A}\,\frac{\alpha}{\pi}\, \Big(g_A \bar{g}_{\rm st}(E_e) + (1 - 2
g_A)\, \bar{f}_{\rm st}(E_e)\Big) \nonumber\\
\hspace{-0.3in}&& - A_0\,\zeta(E_e)_{\rm RC-NLO}, \nonumber\\
\hspace{-0.3in}&&B(E_e)_{\rm RC} = B_0 \,\Big(1 +
\frac{\alpha}{\pi}\,\bar{g}_n(E_e)\Big) + B(E_e)_{\rm RC-NLO}, \nonumber\\
\hspace{-0.3in}&&B(E_e)_{\rm RC-NLO} = \frac{1}{1 + 3 g^2_A}\,
\frac{\alpha}{\pi}\, \frac{E_e}{m_N}\,\Big(g_A f_V(E_e) + g_A \sqrt{1
  - \beta^2}\, f_S(E_e) + \frac{E_0 - E_e}{E_e}\, \sqrt{1 - \beta^2}\,
h_S(E_e) \nonumber\\
\hspace{-0.3in}&& + g_A g_V(E_e)\, \beta^2 - (1 - 2 g_A)\, h_A(E_e)\,
\beta^2 + g_A(1 + 2 g_A) \, f_A(E_e) + g_A(1 + 2 g_A) \sqrt{1 -
  \beta^2}\, f_T(E_e) \Big) \nonumber\\
\hspace{-0.3in}&&+ \frac{1}{1 + 3 g^2_A}\,g_A(1 + 2
g_A)\,\frac{\alpha}{\pi}\, \frac{5}{2}\, \frac{m^2_N}{M^2_W}\,{\ell
  n}\frac{M^2_W}{m^2_N} + \frac{1}{1 + 3 g^2_A}\,\frac{\alpha}{\pi}\,
\Big(g_A \bar{g}_{\rm st}(E_e) + (1 + 2 g_A)\, \bar{f}_{\rm
  st}(E_e)\Big) \nonumber\\
\hspace{-0.3in}&& - B_0\,\zeta(E_e)_{\rm RC-NLO}, \nonumber\\
\hspace{-0.3in}&&K_n(E_e)_{\rm RC} = K_n(E_e)_{\rm RC-NLO} =
\frac{1}{1 + 3 g^2_A}\, \frac{\alpha}{\pi}\, \frac{E_e}{m_N}\,\Big((1
- g_A)\,g_V(E_e) + (1 + 2 g_A)\,
h_A(E_e)\Big),\nonumber\\ \hspace{-0.3in}&& Q_n(E_e)_{\rm RC} =
0,\nonumber\\
\hspace{-0.3in} && G(E_e)_{\rm RC} = - \Big(1 +
\frac{\alpha}{\pi}\,f_n(E_e)\Big) +  G(E_e)_{\rm RC-NLO},\nonumber\\
\hspace{-0.3in} && G(E_e)_{\rm RC-NLO} = - \frac{1}{1 + 3 g^2_A}\,
\frac{\alpha}{\pi}\, \frac{E_e}{m_N}\, \Big(g_A \, \sqrt{1 -
  \beta^2}\, g_S(E_e) + g_A (1 - \beta^2)\, g_V(E_e) + (1 + 2 g_A)\,(1
- \beta^2)\nonumber\\
\hspace{-0.3in} && \times \, h_A(E_e) - \sqrt{1 - \beta^2}\,f_S(E_e) -
g_A \frac{E_0 - E_e}{E_e}\,\sqrt{1 - \beta^2}\,h_S(E_e) - 3 g^2_A \sqrt{1 -
  \beta^2}\, f_T(E_e)\Big),\nonumber\\
\end{eqnarray*}
\begin{eqnarray}\label{eq:A.20}
\hspace{-0.3in}&& H(E_e)_{\rm RC} = - \frac{m_e}{E_e}\,a_0 \Big(1 +
\frac{\alpha}{\pi}\,h^{(1)}_n(E_e)\Big) +  H(E_e)_{\rm RC-NLO},\nonumber\\
\hspace{-0.3in}&& H(E_e)_{\rm RC-NLO} = - \frac{1}{1 + 3 g^2_A}\,
\frac{\alpha}{\pi}\, \frac{E_e}{m_N}\Big( \sqrt{1 - \beta^2}\,
f_V(E_e) + f_S(E_e) + g_A g_S(E_e)\, \beta^2 \nonumber\\
\hspace{-0.3in} && + g_A\, \frac{E_0 -
  E_e}{E_e}\, h_S(E_e) - g^2_A \sqrt{1 - \beta^2}\, f_A(E_e) - g^2_A
f_T(E_e) \Big) + \sqrt{1 - \beta^2}\, \frac{g^2_A}{1 + 3 g^2_A}\,
\frac{\alpha}{\pi}\, \frac{5}{2}\, \frac{m^2_N}{M^2_W}\,{\ell
  n}\frac{M^2_W}{m^2_N} \nonumber\\
\hspace{-0.3in} && - \frac{1}{1 + 3
  g^2_A}\,\frac{\alpha}{\pi}\,\sqrt{1 - \beta^2} \Big(\bar{g}_{\rm
  st}(E_e) - g_A \bar{f}_{\rm st}(E_e)\Big) +
\frac{m_e}{E_e}\,a_0\,\zeta(E_e)_{\rm RC-NLO}, \nonumber\\
\hspace{-0.3in}&& N(E_e)_{\rm RC} = - \frac{m_e}{E_e}\,A_0 \Big( 1 +
\frac{\alpha}{\pi}\,h^{(1)}_n(E_e)\Big) +  N(E_e)_{\rm RC-NLO},
\nonumber\\
\hspace{-0.3in}&& N(E_e)_{\rm RC-NLO} = - \frac{1}{1 + 3 g^2_A}\,
\frac{\alpha}{\pi}\,\frac{E_e}{m_N}\Big(g_A f_V(E_e) \sqrt{1 -
  \beta^2} + g_A f_S(E_e) - g_A g_S(E_e)\,\beta^2 \nonumber\\
\hspace{-0.3in}&& + \frac{1}{3}\,(1 - 2 g_A)\,\frac{E_0 - E_e}{E_e}\,
h_S(E_e) + g_A (1 - 2 g_A)\, \sqrt{1 - \beta^2}\, f_A(E_e) + g_A (1 -
2 g_A)\, f_T(E_e)\Big) - \sqrt{1 - \beta^2}\, \frac{1}{1 + 3
  g^2_A} \nonumber\\
\hspace{-0.3in}&& \times \,g_A(1 - 2 g_A)\, \frac{\alpha}{\pi}\,
\frac{5}{2}\, \frac{m^2_N}{M^2_W}\,{\ell n}\frac{M^2_W}{m^2_N} -
\frac{1}{1 + 3 g^2_A}\,\frac{\alpha}{\pi}\,\sqrt{1 - \beta^2} \Big(g_A
\bar{g}_{\rm st}(E_e) + (1 - 2 g_A)\, \bar{f}_{\rm st}(E_e)\Big) +
\frac{m_e}{E_e}\, A_0\,\zeta(E_e)_{\rm RC-NLO}, \nonumber\\
\hspace{-0.3in}&& Q_e(E_e)_{\rm RC} = - A_0 \,\Big(1 +
\frac{\alpha}{\pi}\,h^{(2)}_n(E_e)\Big) +  Q_e(E_e)_{\rm RC-NLO},
\nonumber\\
\hspace{-0.3in}&& Q_e(E_e)_{\rm RC-NLO} = - \frac{1}{1 + 3 g^2_A}\,
\frac{\alpha}{\pi}\, \frac{E_e}{m_N}\, \Big(g_A f_V(E_e) - g_A
f_S(E_e) - g_A (1 + \sqrt{1 - \beta^2}\,)\,g_S(E_e) \nonumber\\
\hspace{-0.3in} && + (1 + \sqrt{1 - \beta^2}\,)\,g_V(E_e) -
\frac{1}{3}\,(1 - 2 g_A)\,\frac{E_0 - E_e}{E_e}\, h_S(E_e) + g_A (1 -
2 g_A)\, f_A(E_e) - g_A (1 - 2 g_A) \, f_T(E_e)\Big) \nonumber\\
\hspace{-0.3in} && - \frac{1}{1 + 3 g^2_A}\,g_A(1 - 2 g_A)\,
\frac{\alpha}{\pi}\, \frac{5}{2}\, \frac{m^2_N}{M^2_W}\,{\ell
  n}\frac{M^2_W}{m^2_N} - \frac{1}{1 + 3 g^2_A}\,\frac{\alpha}{\pi}
\Big(g_A \bar{g}_{\rm st}(E_e) + (1 - 2 g_A)\, \bar{f}_{\rm
  st}(E_e)\Big) + A_0\,\zeta(E_e)_{\rm RC-NLO}, \nonumber\\
  \hspace{-0.3in}&&K_e(E_e)_{\rm RC} = - a_0 \Big(1 +
  \frac{\alpha}{\pi}\, h^{(2)}_n(E_e)\Big) + K_e(E_e)_{\rm RC-NLO}, \nonumber\\
  \hspace{-0.3in}&&K_e(E_e)_{\rm RC-NLO} = - \frac{1}{1 + 3 g^2_A}\,
  \frac{\alpha}{\pi}\, \frac{E_e}{m_N}\Big(f_V(E_e) - f_S(E_e) + g_A
  (1 + \sqrt{1 - \beta^2}\,) \,g_S(E_e) \nonumber\\
  \hspace{-0.3in}&& + g_A (1 + \sqrt{1 - \beta^2}\,)\,g_V(E_e) + (1 - 2
  g_A) \, (1 + \sqrt{1 - \beta^2}\,)\,h_A(E_e) - g_A\,\frac{E_0 -
    E_e}{E_e}\,h_S(E_e) - g^2_A f_A(E_e) + g^2_A f_T(E_e)\Big)\nonumber\\
  \hspace{-0.3in}&& + \frac{g^2_A}{1 + 3 g^2_A}\, \frac{\alpha}{\pi}\,
  \frac{5}{2}\, \frac{m^2_N}{M^2_W}\,{\ell n}\frac{M^2_W}{m^2_N} -
  \frac{1}{1 + 3 g^2_A}\,\frac{\alpha}{\pi} \Big(\bar{g}_{\rm st}(E_e)
  - g_A\, \bar{f}_{\rm st}(E_e)\Big) + a_0\,\zeta(E_e)_{\rm RC-NLO},
  \nonumber\\
\hspace{-0.3in}&& S(E_e)_{\rm RC} = S(E_e)_{\rm RC-NLO} = - \frac{1}{1 + 3
  g^2_A}\,\frac{\alpha}{\pi}\, \frac{E_e}{m_N}\Big(g_A f_S(E_e) - g_A
f_T(E_e) + g_A g_S(E_e) - g_A \frac{E_0 - E_e}{E_e}\, h_S(E_e)\nonumber\\
\hspace{-0.3in}&&- g_A
\sqrt{1 - \beta^2}\, g_V(E_e)  + (1 + 2 g_A)\, \sqrt{1 -
  \beta^2}\,h_A(E_e)\Big),\nonumber\\
\hspace{-0.3in}&& T(E_e)_{\rm RC} = - B_0\Big(1 +
\frac{\alpha}{\pi}\,f_n(E_e)\Big) +  T(E_e)_{\rm RC-NLO},\nonumber\\
\hspace{-0.3in}&& T(E_e)_{\rm RC-NLO} = - \frac{1}{1 + 3
  g^2_A}\,\frac{\alpha}{\pi}\, \frac{E_e}{m_N}\, \Big(g_A f_V(E_e) -
g_A g_S(E_e) + g_A g_V(E_e) - (1 - 2 g_A)\, h_A(E_e) \nonumber\\
\hspace{-0.3in}&& + g_A(1 + 2 g_A)\, f_A(E_e)\Big) - \frac{1}{1 + 3
  g^2_A}\,g_A(1 + 2 g_A)\, \frac{\alpha}{\pi}\, \frac{5}{2}\,
\frac{m^2_N}{M^2_W}\,{\ell n}\frac{M^2_W}{m^2_N} - \frac{1}{1 + 3
  g^2_A}\,\frac{\alpha}{\pi} \Big(g_A \bar{g}_{\rm st}(E_e) + (1 + 2
g_A)\, \bar{f}_{\rm st}(E_e)\Big) \nonumber\\
\hspace{-0.3in}&& + B_0\,\zeta(E_e)_{\rm RC-NLO}, \nonumber\\
\hspace{-0.3in} &&U(E_e)_{\rm RC} = U(E_e)_{\rm RC-NLO} = \frac{1}{1 + 3
  g^2_A}\,\frac{\alpha}{\pi}\, \frac{E_e}{m_N}\Big(g_A f_S(E_e) - g_A
f_T(E_e) - g_A g_S(E_e) - g_A \frac{E_0 - E_e}{E_e}\, h_S(E_e) \nonumber\\
\hspace{-0.3in}&&- \sqrt{1 - \beta^2}\, g_V(E_e) + g_A\, \sqrt{1 -
  \beta^2}\,h_A(E_e)\Big).
\end{eqnarray}
The correlation function $\zeta(E_e)_{\rm RC}$ and correlation
coefficients $X(E_e)_{\rm RC}$ contain a complete set of outer
radiative corrections of order $O(\alpha/\pi)$ \cite{Sirlin1967,
  Shann1971, Ivanov2017, Ivanov2019a}, calculated to LO in the large
nucleon mass $m_N$ expansion, and radiative corrections of order
$O(\alpha E_e/m_N)$ \cite{Ivanov2019b, Ivanov2020a}, obtained as NLO
corrections in the large nucleon mass $m_N$ expansion to Sirlin's
outer and inner radiative corrections, calculated to LO in the large
nucleon mass $m_N$ expansion. For $\alpha = 0$ the correlation
function $\zeta(E_e)_{\rm RC}$ and the correlation coefficients
$X(E_e)_{\rm RC}$ acquire their expressions, calculated to LO in the
large nucleon mass $m_N$ expansion \cite{Abele2008, Nico2009}(see also
\cite{Ivanov2021a}). We have plotted the $Y(E_e)_{\rm RC-NLO}$
corrections for $Y = \zeta, a, A, B, \ldots U$ and $A^{(\beta)}(E_e)$
(see Appendix E) in \cite{MathW}.

\section*{Appendix B: The corrections of order $O(E_e/m_N)$ and
  $O(E^2_e/m^2_N)$, caused by weak magnetism and proton recoil, to
  next-to-leading and next-to-next-to-leading order in the large
  nucleon mass $m_N$ expansion }
\renewcommand{\theequation}{B-\arabic{equation}}
\setcounter{equation}{0}

The corrections to the structure function $\zeta(E_e)_{\rm WP}$ and
the correlation coefficients $X(E_e)_{\rm WP}$ for $X = a, A, B,
\ldots,T$ and $U$, caused by weak magnetism and proton recoil, we
define as follows
\begin{eqnarray}\label{eq:B.1}
  \hspace{-0.3in}\zeta(E_e)_{\rm WP} &=& \zeta(E_e)_{\rm NLO} +
  \zeta(E_e)_{\rm N^2LO},\nonumber\\ X(E_e)_{\rm WP} &=& X(E_e)_{\rm
    NLO} + X(E_e)_{\rm N^2LO}.
\end{eqnarray}
The corrections
$\zeta(E_e)_{\rm NLO}$ and $X(E_e)_{\rm NLO}$, which are in principle
of order $10^{-3}$ \cite{Ivanov2013, Ivanov2017, Ivanov2019a} have
been calculated in \cite{Bilenky1959} and \cite{Wilkinson1982,
  Ando2004, Gudkov2006, Ivanov2013, Ivanov2017, Ivanov2019a,
  Ivanov2021a, Ivanov2021b} (see also \cite{Ivanov2020b}). The
analytical expressions of these corrections are given by
\begin{eqnarray}\label{eq:B.2}
\hspace{-0.3in}\zeta(E_e)_{\rm NLO} &=&\frac{1}{1 + 3
  g^2_A}\,\frac{E_0}{m_N}\, \Big[ - 2\,g_A\Big(g_A + (\kappa +
  1)\Big) + \Big(10 g^2_A + 4(\kappa + 1)\, g_A + 2\Big)\,\frac{E_e}{E_0}
  \nonumber\\
\hspace{-0.3in}&-& 2 g_A\,\Big(g_A + (\kappa +
1)\Big)\,\frac{m^2_e}{E^2_0}\,\frac{E_0}{E_e}\Big],\nonumber\\
\hspace{-0.3in}a(E_e)_{\rm NLO} &=& \frac{1}{1 + 3 g^2_A}\,
\frac{E_0}{m_N}\,\Big[2 g_A\,\Big(g_A + (\kappa + 1)\Big) - 4 g_A
  \Big(3 g_A + (\kappa + 1)\Big)\,\frac{E_e}{E_0}\Big]
- a_0\,\zeta(E_e)_{\rm NLO},\nonumber\\
\hspace{-0.3in} A(E_e)_{\rm NLO} &=&\frac{1}{1 + 3 g^2_A}\,
\frac{E_0}{m_N}\,\Big[\big(g^2_A + \kappa\, g_A - (\kappa + 1)\big) -
  \big(5 g^2_A + (3\kappa - 4)\,g_A - (\kappa +
  1)\big)\,\frac{E_e}{E_0}\Big] - A_0\zeta(E_e)_{\rm NLO},\nonumber\\
\hspace{-0.3in} B(E_e)_{\rm NLO} &=& \frac{1}{1 + 3 g^2_A}\,
\frac{E_0}{m_N}\,\Big[- 2\,g_A\big(g_A + (\kappa + 1)\big) + \big(7
  g^2_A + (3 \kappa + 8)\, g_A + (\kappa + 1)\big)\,\frac{E_e}{E_0}
  \nonumber\\
\hspace{-0.3in}&-& \big(g^2_A + (\kappa + 2)\, g_A + (\kappa +
1)\big)\,\frac{m^2_e}{E^2_0}\,\frac{E_0}{E_e}\Big] - B_0\zeta(E_e)_{\rm
  NLO},\nonumber\\
\hspace{-0.3in} K_n(E_e)_{\rm NLO} &=& \frac{1}{1 + 3
  g^2_A}\,\frac{E_0}{m_N}\,\Big(5 g^2_A + (\kappa - 4)\, g_A - (\kappa +
1)\Big)\,\frac{E_e}{E_0},\nonumber\\
\hspace{-0.3in} Q_n(E_e)_{\rm NLO} &=& \frac{1}{1 + 3
  g^2_A}\,\frac{E_0}{m_N}\,\Big[ \Big(g^2_A + (\kappa + 2)\, g_A +
  (\kappa + 1)\Big) - \Big(7 g^2_A + (\kappa + 8)\,g_A + (\kappa +
  1)\Big)\, \frac{E_e}{E_0}\Big],\nonumber\\
\hspace{-0.3in} G(E_e)_{\rm NLO} &=& \frac{1}{1 + 3
  g^2_A}\,\frac{E_0}{m_N}\,\Big[- 2 g_A\,\Big(g_A + (\kappa +
  1)\Big)\,\frac{m^2_e}{E^2_0}\,\frac{E_0}{E_e}\Big],\nonumber\\
\hspace{-0.3in} H(E_e)_{\rm NLO} &=& \frac{1}{1 + 3
  g^2_A}\,\frac{E_0}{m_N}\,\frac{m_e}{E_e}\,\Big[ - 2\, g_A \Big(g_A +
  (\kappa + 1)\Big) + \Big(4 g^2_A + 2 (\kappa + 1)\, g_A -
  2\Big)\,\frac{E_e}{E_0}\Big] + a_0\,\frac{m_e}{E_e}\,\zeta(E_e)_{\rm
  NLO},\nonumber\\
\hspace{-0.3in} N(E_e)_{\rm NLO} &=& \frac{1}{1 + 3
  g^2_A}\,\frac{E_0}{m_N}\,\frac{m_e}{E_e}\,\Big[-
  \Big(\frac{4}{3}\,g^2_A + \Big(\frac{4}{3} \kappa -
  \frac{1}{3}\Big)\,g_A - \frac{2}{3} (\kappa + 1)\Big)
  \nonumber\\
\hspace{-0.3in}&+& \Big(\frac{16}{3}\, g^2_A + \Big(\frac{4}{3} \kappa
- \frac{16}{3}\Big)\,g_A - \frac{2}{3} (\kappa + 1)\Big)\,
\frac{E_e}{E_0}\Big] + A_0\, \frac{m_e}{E_e}\,\zeta(E_e)_{\rm
  NLO},\nonumber\\
\hspace{-0.3in}Q_e(E_e)_{\rm NLO} &=&\frac{1}{1 + 3
  g^2_A}\,\frac{E_0}{m_N}\,\Big[- \Big(\frac{4}{3}\,g^2_A +
  \Big(\frac{4}{3} \kappa - \frac{1}{3}\Big)\,g_A - \frac{2}{3}
  (\kappa + 1)\Big) + 2\,g_A\Big(g_A + (\kappa +
  1)\Big)\,\frac{m_e}{E_0} \nonumber\\
\hspace{-0.3in}&+& \Big(\frac{22}{3}\, g^2_A + \Big(\frac{10}{3}\kappa
- \frac{10}{3}\Big)\,g_A - \frac{2}{3} (\kappa + 1)\Big)\,
\frac{E_e}{E_0}\Big] + A_0 \,\zeta(E_e)_{\rm NLO}, \nonumber\\
\hspace{-0.3in} K_e(E_e)_{\rm NLO} &=& \frac{1}{1 + 3
  g^2_A}\,\frac{E_0}{m_N}\, \Big[- 2 g_A \big(g_A + (\kappa + 1)\big)
  + \big(8 g^2_A + 2 (\kappa + 1) g_A + 2\big)\, \frac{m_e}{E_0}
  \nonumber\\  
\hspace{-0.3in}&+& 4 g_A \big(3 g_A + (\kappa + 1)\big)\,
\frac{E_e}{E_0}\Big] + a_0 \zeta(E_e)_{\rm NLO}, \nonumber\\
\hspace{-0.3in} S(E_e)_{\rm NLO} &=& \frac{1}{1 + 3
  g^2_A}\,\frac{m_e}{m_N}\,\Big(- 5 g^2_A - (\kappa - 4)\,g_A + (\kappa
+ 1)\Big), \nonumber\\
\hspace{-0.3in} T(E_e)_{\rm NLO} &=& \frac{1}{1 + 3
  g^2_A}\,\frac{E_0}{m_N}\,\Big[2 g_A \Big(g_A + (\kappa + 1)\Big) -
\Big(7g^2_A + (3 \kappa + 8) g_A + (\kappa + 1)\Big)\,
\frac{E_e}{E_0}\Big] + B_0\zeta(E_e)_{\rm NLO}, \nonumber\\
\hspace{-0.3in} U(E_e)_{\rm NLO} &=& 0.
\end{eqnarray}
For the definition of the $O(E^2_e/m^2_N)$ corrections it is
convenient to adduce the following expressions, calculated in
\cite{Ivanov2013, Ivanov2017, Ivanov2019a}:
\begin{eqnarray}\label{eq:B.3}
\hspace{-0.3in}\bar{a}(E_e)_{\rm NLO} &=& \frac{1}{1 + 3
  g^2_A}\, \frac{E_0}{m_N}\,\Big[2 g_A\,\big(g_A + (\kappa +
  1)\big) - 4 g_A \big(3 g_A + (\kappa +
  1)\big)\,\frac{E_e}{E_0}\Big],\nonumber\\
\hspace{-0.3in} \bar{A}(E_e)_{\rm NLO} &=&\frac{1}{1 + 3 g^2_A}\,
\frac{E_0}{m_N}\,\Big[\big(g^2_A + \kappa\, g_A - (\kappa + 1)\big) -
  \big(5 g^2_A + (3\kappa - 4)\,g_A - (\kappa +
  1)\big)\,\frac{E_e}{E_0}\Big],\nonumber\\
\hspace{-0.3in}\bar{B}(E_e)_{\rm NLO} &=& \frac{1}{1
  + 3 g^2_A}\, \frac{E_0}{m_N}\,\Big[- 2\,g_A\big(g_A + (\kappa +
  1)\big)  + \big(7 g^2_A + (3 \kappa + 8)\, g_A + (\kappa +
  1)\big)\,\frac{E_e}{E_0} \nonumber\\
\hspace{-0.3in}&-& \big(g^2_A + (\kappa + 2)\, g_A + (\kappa +
1)\big)\,\frac{m^2_e}{E^2_0}\,\frac{E_0}{E_e}\Big],\nonumber\\
\hspace{-0.3in} \bar{K}_n(E_e)_{\rm NLO} &=& \frac{1}{1 + 3
  g^2_A}\,\frac{E_0}{m_N}\,\Big(5 g^2_A + (\kappa - 4)\, g_A - (\kappa +
1)\Big)\,\frac{E_e}{E_0},\nonumber\\
\hspace{-0.3in} \bar{Q}_n(E_e)_{\rm NLO} &=& \frac{1}{1 + 3
  g^2_A}\,\frac{E_0}{m_N}\,\Big[ \Big(g^2_A + (\kappa + 2)\, g_A +
  (\kappa + 1)\Big) - \Big(7 g^2_A + (\kappa + 8)\,g_A + (\kappa +
  1)\Big)\, \frac{E_e}{E_0}\Big],\nonumber\\
\hspace{-0.3in}\bar{G}(E_e)_{\rm NLO} &=& \frac{1}{1 + 3
  g^2_A}\,\frac{E_0}{m_N}\,\Big[\big(2 g^2_A + 2 (\kappa + 1)\,
  g_A\big) - \big(10 g^2_A + 4(\kappa + 1)\,g_A +
  2\big)\, \frac{E_e}{E_0} \Big]\nonumber\\
\hspace{-0.3in} \bar{H}(E_e)_{\rm NLO} &=& \frac{1}{1 + 3
  g^2_A}\,\frac{E_0}{m_N}\,\frac{m_e}{E_e}\,\Big[ - 2\, g_A \big(g_A +
  (\kappa + 1)\big) + \big(4 g^2_A + 2 (\kappa + 1)\, g_A -
  2\Big)\,\frac{E_e}{E_0}\Big],\nonumber\\
\hspace{-0.3in} \bar{N}(E_e)_{\rm NLO} &=& \frac{1}{1 + 3
  g^2_A}\,\frac{E_0}{m_N}\,\frac{m_e}{E_e}\,\Big[-
  \Big(\frac{4}{3}\,g^2_A + \Big(\frac{4}{3} \kappa -
  \frac{1}{3}\Big)\,g_A - \frac{2}{3} (\kappa + 1)\Big)
  \nonumber\\
\hspace{-0.3in}&+& \Big(\frac{16}{3}\, g^2_A + \Big(\frac{4}{3} \kappa
- \frac{16}{3}\Big)\,g_A - \frac{2}{3} (\kappa + 1)\Big)\,
\frac{E_e}{E_0}\Big],\nonumber\\
\hspace{-0.3in}\bar{Q}_e(E_e)_{\rm NLO} &=&\frac{1}{1 + 3
  g^2_A}\,\frac{E_0}{m_N}\,\Big[- \Big(\frac{4}{3}\,g^2_A +
  \Big(\frac{4}{3} \kappa - \frac{1}{3}\Big)\,g_A - \frac{2}{3}
  (\kappa + 1)\Big) + \Big(2\,g^2_A + (2 \kappa +
  1)\,g_A\Big)\,\frac{m_e}{E_0} \nonumber\\
\hspace{-0.3in}&+& \Big(\frac{22}{3}\, g^2_A +
  \Big(\frac{10}{3}\kappa - \frac{10}{3}\Big)\,g_A - \frac{2}{3}
  (\kappa + 1)\Big)\, \frac{E_e}{E_0}\Big], \nonumber\\
\hspace{-0.3in} \bar{K}_e(E_e)_{\rm NLO} &=& \frac{1}{1 + 3
  g^2_A}\,\frac{E_0}{m_N}\, \Big[- 2 g_A \Big(g_A + (\kappa + 1)\Big)
  + \Big(8 g^2_A + 2 (\kappa + 1) g_A + 2\Big)\, \frac{m_e}{E_0}
  \nonumber\\
\hspace{-0.3in}&+& 4 g_A \big(3 g_A + (\kappa + 1)\big)\,
\frac{E_0}{E_e}\Big],
\end{eqnarray}
where $\bar{K}_n(E_e)_{\rm NLO} = K_n(E_e)_{\rm NLO}$ and
$\bar{Q}_n(E_e)_{\rm NLO} = Q_n(E_e)_{\rm NLO}$.  Using the results
obtained in \cite{Ivanov2020b}, we get the following analytical
expressions for the N$^2$LO corrections $\zeta(E_e)_{\rm N^2LO}$ and
$X(E_e)_{\rm N^2LO}$ for $X(E_e) = a(E_e), A(E_e),\ldots,U(E_e)$. They
are given by
\begin{eqnarray*}
\hspace{-0.3in}&&\zeta(E_e)_{\rm N^2LO} = 6\,
\frac{E^2_e}{m^2_N}\Big\{ \Big(1 - \frac{1}{4}\,\frac{E_0}{E_e}\Big) +
\frac{1}{3} \Big[1- 2\, a_0\, \Big( 1 -
  \frac{1}{8}\,\frac{E_0}{E_e}\Big)\Big] \Big(1 -
\frac{m^2_e}{E^2_e}\Big) \Big\} + 3\, \frac{E_e}{m_N}\,
\zeta(E_e)_{\rm NLO} - \frac{E_e}{m_N}\, \bar{a}(E_e)_{\rm
  NLO}\nonumber\\
\hspace{-0.3in}&&\times\, \Big(1 - \frac{m^2_e}{E^2_e}\Big)
 - \frac{8}{1 + 3 g^2_A}\,\Big(\frac{E^2_0}{M^2_V} +
3 g^2_A \frac{E^2_0}{M^2_A}\Big)\, \frac{E_e}{E_0}\Big(1 -
\frac{E_e}{E_0}\Big),\nonumber\\
\hspace{-0.3in}&&a(E_e)_{\rm N^2LO} = 12\, \frac{E^2_e}{m^2_N}\Big\{ -
\Big(1 - \frac{1}{8}\, \frac{E_0}{E_e}\Big) + \frac{1}{3}\, a_0\,
\Big(1 - \frac{1}{4}\, \frac{E_0}{E_e}\Big)\Big\} + 3\,
\frac{E_e}{m_N}\, \bar{a}(E_e)_{\rm NLO} - \frac{E_e}{m_N}\,
\zeta(E_e)_{\rm NLO} \nonumber\\
\hspace{-0.3in}&& - a(E_e)_{\rm NLO}\, \zeta(E_e)_{\rm NLO} +
\frac{8}{1 + 3 g^2_A}\,\Big(\frac{E^2_0}{M^2_V} + 3 g^2_A
\frac{E^2_0}{M^2_A}\Big)\, \frac{E_e}{E_0}\Big(1 -
\frac{E_e}{E_0}\Big),\nonumber\\
\hspace{-0.3in}&&A(E_e)_{\rm N^2LO} = 3\, \frac{E_e}{m_N}\,
\bar{A}(E_e)_{\rm NLO} - \frac{E_e}{m_N}\,\bar{K}_n(E_e)_{\rm NLO}
\Big(1 - \frac{m^2_e}{E^2_e}\Big) - A(E_e)_{\rm NLO}\zeta(E_e)_{\rm
  NLO},\nonumber\\
\hspace{-0.3in}&&B(E_e)_{\rm N^2LO} = 6 \, \frac{E^2_e}{m^2_N}\, B_0\,
\Big(1 - \frac{1}{4}\, \frac{E_0}{E_e}\Big) + 3\, \frac{E_e}{m_N}\,
\bar{B}(E_e)_{\rm NLO} - B(E_e)_{\rm NLO}\,\zeta(E_e)_{\rm NLO} -
B_0\, \zeta(E_e)_{\rm N^2LO}\nonumber\\
\hspace{-0.15in}&& - \frac{8 g_A}{1 + 3
  g^2_A}\,\Big(\frac{E^2_0}{M^2_V} + (1 + 2
g_A)\,\frac{E^2_0}{M^2_A}\Big)\, \frac{E_e}{E_0}\Big(1 -
\frac{E_e}{E_0}\Big),\nonumber\\
\end{eqnarray*}
\begin{eqnarray}\label{eq:B.4}
\hspace{-0.3in}&& K_n(E_e)_{\rm N^2LO} =  3\,
\frac{E_e}{m_N}\, \bar{K}_n(E_e)_{\rm NLO} - 3\,\frac{E_e}{m_N}\,
\bar{A}(E_e)_{\rm NLO},\nonumber\\
\hspace{-0.3in}&& Q_n(E_e)_{\rm N^2LO} = - 12\,
\frac{E^2_e}{m^2_N}\,B_0 \, \Big(1 - \frac{1}{8}\,
\frac{E_0}{E_e}\Big) + 3\, \frac{E_e}{m_N}\, \bar{Q}_n(E_e)_{\rm NLO}
- 3\, \frac{E_e}{m_N}\, \bar{B}(E_e)_{\rm NLO} - Q_n(E_e)_{\rm
  NLO}\,\zeta(E_e)_{\rm NLO},\nonumber\\
\hspace{-0.3in}&& G(E_e)_{\rm N^2LO} = 6\, \frac{E^2_e}{m^2_N}\,
\Big\{- \Big(1 - \frac{1}{4}\,\frac{E_0}{E_e}\Big) - \frac{1}{3}\,
\Big(1 - \frac{m^2_e}{E^2_e}\Big) + \frac{2}{3}\,a_0\, \Big(1 -
\frac{1}{8}\, \frac{E_0}{E_e}\Big) \Big\} +  3\,
\frac{E_e}{m_N}\, \bar{G}(E_e)_{\rm NLO} - \frac{E_e}{m_N}\,
\bar{H}(E_e)_{\rm NLO} \nonumber\\
\hspace{-0.3in}&& - \frac{E_e}{m_N}\, \bar{K}_e(E_e)_{\rm NLO}\Big(1 -
\frac{m_e}{E_e}\Big) - G(E_e)_{\rm NLO}\,\zeta(E_e)_{\rm NLO} +
\zeta(E_e)_{\rm N^2LO} + \frac{8}{1 + 3
  g^2_A}\,\Big(\frac{E^2_0}{M^2_V} + 3 g^2_A
\frac{E^2_0}{M^2_A}\Big)\, \frac{E_e}{E_0}\Big(1 -
\frac{E_e}{E_0}\Big),\nonumber\\
\hspace{-0.3in}&& H(E_e)_{\rm N^2LO} = - \bar{H}(E_e)_{\rm NLO}\,
\zeta(E_e)_{\rm NLO}, \nonumber\\
\hspace{-0.3in}&& N(E_e)_{\rm N^2LO} = - \bar{N}(E_e)_{\rm NLO}\,
\zeta(E_e)_{\rm NLO}, \nonumber\\
\hspace{-0.3in}&& Q_e(E_e)_{\rm N^2LO} = 3\, \frac{E_e}{m_N} \,
\bar{Q}_e(E_e)_{\rm NLO} - Q_e(E_e)_{\rm NLO}\,\zeta(E_e)_{\rm
  NLO},\nonumber\\
\hspace{-0.3in}&& K_e(E_e)_{\rm N^2LO} = 12\,
\frac{E^2_e}{m^2_N}\,\Big\{ \Big(1 - \frac{1}{8}\,
\frac{E_0}{E_e}\Big)\Big(1 + \frac{m_e}{E_e}\Big) - \frac{1}{2}\,
a_0\, \Big(1 - \frac{1}{4}\, \frac{E_0}{E_e}\Big)\Big\} + 3\,
\frac{E_e}{m_N}\, \bar{K}_e(E_e)_{\rm NLO} - 3\, \frac{E_e}{m_N}\,
\bar{G}(E_e)_{\rm NLO},\nonumber\\
\hspace{-0.3in}&& \times \Big(1 + \frac{m_e}{E_e}\Big) - K_e(E_e)_{\rm
  NLO}\, \zeta(E_e)_{\rm NLO} - \frac{8}{1 + 3
  g^2_A}\,\Big(\frac{E^2_0}{M^2_V} + 3 g^2_A \frac{E^2_0}{M^2_A}\Big)
\frac{E_e}{E_0}\Big(1 - \frac{E_e}{E_0}\Big),\nonumber\\
\hspace{-0.3in}&&S(E_e)_{\rm N^2LO} = \frac{1}{1 + 3g^2_A}
\frac{E^2_0}{2 m^2_N}\,\frac{m_e}{E_0}\Big\{\Big(7 g^2_A +(\kappa -
6)\, g_A + (\kappa + 1)\Big) + \Big(- 50 g^2_A + 50\, g_A - 2 (\kappa
+ 1)\Big)\,\frac{E_e}{E_0}\Big\}\nonumber\\
\hspace{-0.3in}&& - 3\,\frac{E_e}{m_N}\,\bar{N}(E_e)_{\rm NLO} -
S(E_e)_{\rm NLO}\, \zeta(E_e)_{\rm NLO},\nonumber\\
\hspace{-0.1in}&&T(E_e)_{\rm N^2LO} = \frac{1}{1 + 3g^2_A}
\frac{E^2_0}{2 m^2_N} \Big\{\Big(\big(- g^2_A - 4 (\kappa + 1)\, g_A -
(\kappa + 1)^2\big) + \big(g^2_A + \kappa +
1\big)\,\frac{m^2_e}{E^2_0}\Big)\nonumber\\
 \hspace{-0.1in}&&+\Big(5 g^2_A + (3 \kappa + 11)\, g_A + 3\, (\kappa +
 1)^2\Big)\, \frac{E_e}{E_0} + \Big(- 22 g^2_A - 26 g_A - 2 (\kappa +
 1)(\kappa + 2)\Big)\, \frac{E^2_e}{E^2_0}\Big\}+ B_0 \zeta(E_e)_{\rm
   N^2LO} \nonumber\\
\hspace{-0.1in}&&- T(E_e)_{\rm NLO}\,\zeta(E_e)_{\rm NLO} + \frac{8
  g_A}{1 + 3 g^2_A}\,\Big(\frac{E^2_0}{M^2_V} + (1 + 2
g_A)\,\frac{E^2_0}{M^2_A}\Big)\, \frac{E_e}{E_0}\Big(1 -
\frac{E_e}{E_0}\Big),\nonumber\\
\hspace{-0.3in}&&U(E_e)_{\rm N^2LO} = \frac{1}{1 + 3g^2_A}
\frac{E^2_0}{2 m^2_N}\,\frac{m_e}{E_0}\,\Big\{\Big((2\kappa + 1)\,g_A
+ \kappa (\kappa + 1)\Big) - \kappa \big(g_A + \kappa + 1\big)\,
\frac{E_e}{E_0}\Big\}.
\end{eqnarray}
The terms, proportional to $E^2_0/M^2_V$ and $E^2_0/M^2_A$ are induced
by the vector and axial-vector form factors of the neutron beta decay
\cite{Bernard1995} (see also \cite{Ivanov2020b}).  The
slope-parameters $M_V$ and $M_A$ are related to the charge radius of
the proton $r_p = 0.841\,{\rm fm}$ \cite{Antognini2013} and the axial
radius $r_A = 0.635\,{\rm fm}$ of the nucleon \cite{Liesenfeld1999}
(see also \cite{Ivanov2019c}), respectively. This gives $M_V =
\sqrt{12}/r_p = 813\,{\rm MeV}$ and $M_A = \sqrt{12}/r_A = 1077\, {\rm
  MeV}$.The analytical expressions for the corrections
$\zeta(E_e)_{\rm NLO}$, $X(E_e)_{\rm NLO}$ and $\bar{X}(E_e)_{\rm
  NLO}$ for $X = a, A,\ldots, K_e, S, T, U$ are given in
Eqs.(\ref{eq:B.2}) and (\ref{eq:B.3}) respectively.  We define
corrections $O(E^2_e/m^2_N)$, caused by weak magnetism and proton
recoil, at the level of $10^{-5}$ with the theoretical accuracy of
about $10^{-6}$ \cite{MathW}. The numerical analysis of relative
contributions has been carried out for the axial coupling constant
$g_A = 1.2764$ \cite{Abele2018}, the value of which agrees well with
the recommended value of the axial coupling constant obtained by means
of the global analysis of the experimental data on the axial coupling
constant by Czarnecki {\it et al.}  \cite{Sirlin2018}.

\section*{Appendix C: Wilkinson's corrections of order $10^{-5}$ to
  the neutron beta decay}
\renewcommand{\theequation}{C-\arabic{equation}}
\setcounter{equation}{0}

According to Wilkinson \cite{Wilkinson1982}, the corrections,
additional to those calculated in Appendices A and B, should be caused
by i) the proton recoil in the electron--proton final--state Coulomb
interaction, ii) the finite proton radius, iii) the proton--lepton
convolution and iv) the higher--order outer radiative
corrections. These corrections to the neutron lifetime and the
correlation coefficients $a(E_e), A(E_e), \ldots, K_e(E_e)$ have been
calculated in \cite{Ivanov2013, Ivanov2017, Ivanov2019a}. The
contributions of the proton recoil, caused by the phase-volume of the
neutron beta decay proportional to $1/m_N$ and $1/m^2_N$, which are
defined in \cite{Wilkinson1982} (see also \cite{Shekhter1959}) by the
function $S(E_e, E_0, m_N)$, we have taken into account for the
calculation of the NLO and N$^2$LO corrections in the large nucleon
mass $m_N$ expansion induced by weak magnetism and proton recoil (see
Appendix B).

\subsection*{\bf Wilkinson's corrections, induced by  proton
  recoil in the Coulomb electron--proton final--state interaction}

For the calculation of the contribution of the proton recoil in the
Fermi function we replace the electron velocity $\vec{\beta}$ by a
velocity of a relative motion of the electron-proton pair
\cite{Ivanov2013}. To NLO in the large nucleon mass $m_N$ expansion a
velocity of a relative motion of the electron-proton pair is defined
by
\begin{eqnarray}\label{eq:C.1}
\hspace{-0.1in} \vec{\beta} \to \vec{v}_{\rm rel.} = \vec{\beta} -
\frac{\vec{k}_p}{m_N}.
\end{eqnarray}
To LO in the large nucleon mass $m_N$ expansion the second term in
r.h.s. of Eq.(\ref{eq:C.1}) vanishes, and a velocity of a relative
motion of the electron-proton pair reduces to an electron velocity. As
has been shown in \cite{Ivanov2013} the Fermi function $F(E_e, Z = 1)$
with a replacement $\beta \to v_{\rm rel.}$, caused by a relative
motion of the electron-proton pair in the final state of the neutron
beta decay, undergoes the following change (see Appendix H of
Ref.\cite{Ivanov2013})
\begin{eqnarray}\label{eq:C.2}
\hspace{-0.3in}F(E_e, Z = 1) \stackrel{\vec{\beta} \to \vec{v}_{\rm
    rel.}}\longrightarrow F(E_e, Z = 1)\,\Big(1 - \frac{\pi
  \alpha}{\beta}\,\frac{E_e}{m_N} - \frac{\pi
  \alpha}{\beta^3}\,\frac{E_0 - E_e}{m_N}\,\frac{\vec{k}_e\cdot
  \vec{k}_{\nu}}{E_e E_{\bar{\bar{\nu}}}}\Big),
\end{eqnarray}
where we have taken into account the NLO terms in the large nucleon
mass $m_N$ expansion. The corrections, caused by the proton recoil in
the electron-proton final-state interactions to the neutron lifetime
$\zeta(E_e)_{\rm WF}$ and the correlation coefficients $X(E_e)_{\rm
  WF}$ for $X = a, A, B, \ldots,T$ and $U$ are equal to
\begin{eqnarray}\label{eq:C.3}
\hspace{-0.3in}\zeta(E_e)_{\rm WF} &=& - \frac{\pi
  \alpha}{\beta}\,\frac{E_e}{m_N} - \frac{1}{3}\,a_0\,\frac{\pi
  \alpha}{\beta}\,\frac{E_0 - E_e}{m_N} = - 3.16 \times
10^{-5}\frac{E_e}{\beta E_0} + 1.12 \times 10^{-6}\,\frac{E_0 -
  E_e}{\beta E_0}, \nonumber\\
\hspace{-0.3in}a(E_e)_{\rm WF} &=& \frac{1}{3}\,a^2_0\,\frac{\pi
  \alpha}{\beta}\,\frac{E_0 - E_e}{m_N} - \frac{\pi
  \alpha}{\beta^3}\,\frac{E_0 - E_e}{m_N} = 1.20 \times
10^{-7}\,\frac{E_0 - E_e}{\beta E_0} - 3.16 \times 10^{-5}\,\frac{E_0
  - E_e}{\beta^3 E_0}, \nonumber\\
\hspace{-0.3in}A(E_e)_{\rm WF} &=&\frac{1}{3}\,a_0 A_0\,\frac{\pi
  \alpha}{\beta}\,\frac{E_0 - E_e}{m_N} = 1.35 \times
10^{-7}\,\frac{E_0 - E_e}{\beta E_0},\nonumber\\
\hspace{-0.3in}B(E_e)_{\rm WF} &=&\frac{1}{3}\,a_0 B_0\,\frac{\pi
  \alpha}{\beta}\,\frac{E_0 - E_e}{m_N} = - 1.11 \times
10^{-6}\,\frac{E_0 - E_e}{\beta E_0},\nonumber\\
\hspace{-0.3in}K_n(E_e)_{\rm WF} &=& - A_0\,\frac{\pi
  \alpha}{\beta^3}\,\frac{E_0 - E_e}{m_N} = 3.78 \times 10^{-6}\,
\frac{E_0 - E_e}{\beta^3 E_0},\nonumber\\
\hspace{-0.3in}Q_n(E_e)_{\rm WF} &=& - B_0\,\frac{\pi
  \alpha}{\beta^3}\,\frac{E_0 - E_e}{m_N} = - 3.12 \times 10^{-5}\,
\frac{E_0 - E_e}{\beta^3 E_0},\nonumber\\
\hspace{-0.3in}G(E_e)_{\rm WF} &=& - \frac{1}{3}\,a_0\,\big(1 -
\beta^2\big)\, \frac{\pi \alpha}{\beta^3}\, \frac{E_0 - E_e}{m_N} =
1.12 \times 10^{-6}\,(1 - \beta^2)\, \frac{E_0 - E_e}{\beta^3
  E_0},\nonumber\\
\hspace{-0.3in}H(E_e)_{\rm WF} &=&- \frac{1}{3}\,a^2_0\,\sqrt{1 -
  \beta^2}\,\frac{\pi \alpha}{\beta}\,\frac{E_0 - E_e}{m_N} = - 1.20
\times 10^{-7}\,\sqrt{1 - \beta^2}\,\frac{E_0 - E_e}{\beta E_0},
\nonumber\\
\hspace{-0.3in}N(E_e)_{\rm WF} &=&- \frac{1}{3}\,a_0 A_0\,\sqrt{1 -
  \beta^2}\,\frac{\pi \alpha}{\beta}\,\frac{E_0 - E_e}{m_N} = 1.35
\times 10^{-7}\,\sqrt{1 - \beta^2}\,\frac{E_0 - E_e}{\beta
  E_0},\nonumber\\
\hspace{-0.3in}Q_e(E_e)_{\rm WF} &=&- \frac{1}{3}\,a_0 A_0\,\frac{\pi
  \alpha}{\beta}\,\frac{E_0 - E_e}{m_N} + \frac{1}{3}\,B_0\, \big(1 +
\sqrt{1 - \beta^2}\,\big) \,\frac{\pi \alpha}{\beta^3}\,\frac{E_0 -
  E_e}{m_N} = \nonumber\\
\hspace{-0.3in}&=& - 1.35 \times 10^{-7}\, \frac{E_0 - E_e}{\beta E_0} + 1.04 \times 10^{-4}\, \big(1
+ \sqrt{1 - \beta^2}\,\big)\, \frac{E_0 - E_e}{\beta^3 E_0},
\nonumber\\
\hspace{-0.3in}K_e(E_e)_{\rm WF} &=&- \frac{1}{3}\,a^2_0\, \frac{\pi
  \alpha}{\beta}\,\frac{E_0 - E_e}{m_N} + \big(1 + \sqrt{1 -
  \beta^2}\,\big) \,\frac{\pi \alpha}{\beta^3}\,\frac{E_0 - E_e}{m_N}
=\nonumber\\
\hspace{-0.3in}&=& - 1.20 \times 10^{-7}\,\frac{E_0 - E_e}{\beta E_0}
+ 3.16 \times 10^{-5}\, \big(1 + \sqrt{1 - \beta^2}\,\big) \,\frac{E_0
  - E_e}{\beta^3 E_0},\nonumber\\
\hspace{-0.3in}S(E_e)_{\rm WF} &=& \sqrt{1 - \beta^2}\,A_0\,\frac{\pi
  \alpha}{\beta^3}\,\frac{E_0 - E_e}{m_N} = 3.78 \times
10^{-6}\,\sqrt{1 - \beta^2}\,\frac{E_0 - E_e}{\beta^3 E_0},\nonumber\\
\hspace{-0.3in}T(E_e)_{\rm WF} &=& - \frac{1}{3}\,a_0 B_0\,\frac{\pi
  \alpha}{\beta}\,\frac{E_0 - E_e}{m_N} = 1.11 \times 10^{-6}\,
\frac{E_0 - E_e}{\beta E_0},\nonumber\\
\hspace{-0.3in}U(E_e)_{\rm WF} &=& 0.
\end{eqnarray}
In comparison with the result, obtained in \cite{Ivanov2017}, an
additional term in $Q_e(E_e)_{\rm WF}$ appears because of the
contribution off the correlation coefficient $T(E_e)$.  The
contributions of Wilkinson's corrections, caused by proton recoil in
the electron-proton final-state Coulomb interaction, to the
electron-energy and angular distribution of the neutron beta decay
with correlation structures beyond the standard correlation structures
by Jackson {\it et al.}  \cite{Jackson1957a} and Ebel and Feldman
\cite{Ebel1957} (see Eq.(\ref{eq:1})) are given in Appendix F.  We
would like to emphasize that Wilkinson's corrections, induced by
proton recoil in the final-state Coulomb electron-proton interaction
are well defined in the experimental electron-energy region
$0.811\,{\rm MeV} \le E_e \le 1.211\,{\rm MeV}$ \cite{Abele2018}.

\subsection*{\bf Wilkinson's corrections, induced by  i) the finite
proton--radius $r_p$, ii) the lepton--nucleon convolution and iii) the
higher--order outer radiative corrections}

The corrections under consideration, caused by i) the finite
proton--radius $r_p$, ii) the lepton--nucleon convolution and iii) the
higher--order outer radiative corrections, are defined by the
functions $L(E_e, Z = 1)$, $C(E_e, Z = 1)$ and $J(Z = 1)$,
respectively \cite{Wilkinson1982} (see also
\cite{Ivanov2017}). According to Wilkinson \cite{Wilkinson1982}, the
contribution of $J(Z=1)$ is equal to $J(Z=1) = 1 + 3.92\times 10^{-4}$
(see also \cite{Ivanov2017}). The corrections $L(E_e, Z=1)$ and
$C(E_e,Z=1)$, adapted for the neutron beta decay, are determined by
\cite{Ivanov2017}:
\begin{eqnarray}\label{eq:C.4}
\hspace{-0.3in}&&L(E_e, Z = 1) = 1 + \frac{13}{60}\,\alpha^2 -
\alpha\,r_p\,E_e\,\Big(1 - \frac{1}{2}\,\frac{m^2_e}{E^2_e}\Big) = 1 +
1.15\times 10^{-5} - 4.02\times 10^{-5}\,\frac{E_e}{E_0},\nonumber\\
\hspace{-0.3in}&&C(E_e, Z = 1) = 1 + \Big[\Big(-
  \frac{9}{20}\,\alpha^2 + \frac{1}{5}\,m^2_e r^2_p -
  \frac{1}{5}\,E^2_0 r^2_p\Big) + \Big( -
  \frac{1}{5}\,\alpha\,r_p\,E_0 - \frac{2}{15}\,E^2_0
  r^2_p\Big)\,\frac{1 - g^2_A}{1 + 3 g^2_A}\Big]\nonumber\\
\hspace{-0.3in}&&+\Big[\Big(-
  \frac{3}{5}\,\alpha\,r_p\, E_0 + \frac{2}{5}\,E^2_0\,r^2_p\Big) +
  \Big(\frac{1}{5}\,\alpha\,r_p\, E_0 -
  \frac{2}{15}\,E^2_0\,r^2_p\Big)\, \frac{1 - g^2_A}{1 + 3
    g^2_A}\Big]\,\frac{E_e}{E_0} +
\frac{2}{15}\,m_e\,E_0\,r^2_p\,\frac{1 - g^2_A}{1 + 3
  g^2_A}\,\frac{m_e}{E_e}\nonumber\\
\hspace{-0.3in}&&+ \frac{2}{5}\,\Big(- 1 + \frac{1}{3}\,\frac{1 -
  g^2_A}{1 + 3 g^2_A}\Big)\,E^2_0 r^2_p\,\frac{E^2_e}{E^2_0} = 1 -
2.78 \times 10^{-5} - 1.24\times 10^{-5}\,\frac{E_e}{E_0} - 1.26\times
10^{-5}\,\frac{E^2_e}{E^2_0}.
\end{eqnarray}
For the calculation of the numerical values of the constant terms and
coefficients in front of the powers of $E_e/E_0$ we use $r_p =
0.841\,{\rm fm}$ \cite{Antognini2013}.  Following Wilkinson
\cite{Wilkinson1982} we define Wilkinson's correction, caused by i)
the finite proton--radius $r_p$, ii) the lepton--nucleon convolution
and iii) the higher--order outer radiative corrections, as
follows
\begin{eqnarray}\label{eq:C.5}
\hspace{-0.3in}L(E_e, Z = 1)C(E_e, Z = 1)J(Z = 1) = 1 + \delta
L(E_e,Z=1) + \delta C(E_e,Z=1) + \delta J(Z=1) = 1 + \zeta(E_e)_{\rm
  WR},
\end{eqnarray}
where $\zeta(E_e)_{\rm WR}$ is equal to
\begin{eqnarray}\label{eq:C.6}
\hspace{-0.3in}&&\zeta(E_e)_{\rm WR} = 3.76 \times 10^{-4} - 5.26 \times
10^{-5}\,\frac{E_e}{E_0} - 1.26 \times 10^{-5}\,\frac{E^2_e}{E^2_0}.
\end{eqnarray}
The contributions of Wilkinson's corrections, caused by i) the finite
proton--radius $r_p$, ii) the lepton--nucleon convolution and iii) the
higher--order outer radiative corrections, to the correlation
coefficients are equal to
\begin{eqnarray*}
\hspace{-0.3in}&&a(E_e)_{\rm WR} = - a_0\zeta(E_e)_{\rm WR} = -
a_0 \Big( 3.76 \times 10^{-4} - 5.26 \times 10^{-5}\,\frac{E_e}{E_0} -
1.26 \times 10^{-5}\,\frac{E^2_e}{E^2_0}\Big),\nonumber\\
\hspace{-0.3in}&&A(E_e)_{\rm WR} = - A_0\zeta(E_e)_{\rm WR} = -A_0
\Big( 3.76 \times 10^{-4} - 5.26 \times 10^{-5}\,\frac{E_e}{E_0} -
1.26 \times 10^{-5}\,\frac{E^2_e}{E^2_0}\Big), \nonumber\\
\hspace{-0.3in}&&B(E_e)_{\rm WR} = - B_0\zeta(E_e)_{\rm WR} = - B_0
\Big( 3.76 \times 10^{-4} - 5.26 \times 10^{-5}\,\frac{E_e}{E_0} -
1.26 \times 10^{-5}\,\frac{E^2_e}{E^2_0}\Big), \nonumber\\
\hspace{-0.3in}&&K_n(E_e)_{\rm WR} = Q_n(E_e)_{\rm WR} = 0,
\nonumber\\ 
\hspace{-0.3in}&&G(E_e)_{\rm WR} = \zeta(E_e)_{\rm WR} = 3.76 \times
10^{-4} - 5.26 \times 10^{-5}\,\frac{E_e}{E_0} - 1.26 \times
10^{-5}\,\frac{E^2_e}{E^2_0}, \nonumber\\
\hspace{-0.3in}&&H(E_e)_{\rm WR} =
\frac{m_e}{E_e}\,a_0\,\zeta(E_e)_{\rm WR} = \frac{m_e}{E_0}\,a_0\,
\Big( 3.76 \times 10^{-4}\frac{E_0}{E_e} - 5.26 \times
10^{-5} - 1.26 \times
10^{-5}\,\frac{E_e}{E_0}\Big),\nonumber\\
\hspace{-0.3in}&&N(E_e)_{\rm WR} = \frac{m_e}{E_e}\,A_0\zeta(E_e)_{\rm
  WR} = \frac{m_e}{E_0}\,A_0\, \Big( 3.76 \times
10^{-4}\frac{E_0}{E_e} - 5.26 \times 10^{-5} - 1.26 \times
10^{-5}\,\frac{E_e}{E_0}\Big),
\nonumber\\ \hspace{-0.3in}&&Q_e(E_e)_{\rm WR} = A_0\zeta(E_e)_{\rm
  WR} = A_0 \Big( 3.76 \times 10^{-4} - 5.26 \times
10^{-5}\,\frac{E_e}{E_0} - 1.26 \times
10^{-5}\,\frac{E^2_e}{E^2_0}\Big),\nonumber\\
\hspace{-0.3in}&&K_e(E_e)_{\rm WR} = a_0\, \zeta(E_e)_{\rm WR} = a_0
\Big( 3.76 \times 10^{-4} - 5.26 \times 10^{-5}\,\frac{E_e}{E_0} -
1.26 \times 10^{-5}\,\frac{E^2_e}{E^2_0}\Big), \nonumber\\
\end{eqnarray*}
\begin{eqnarray}\label{eq:C.7}
\hspace{-0.3in}&&S(E_e)_{\rm WR} = U(E_e)_{\rm WR} = 0, \nonumber\\
\hspace{-0.3in}&&T(E_e)_{\rm WR} = B_0 \zeta(E_e)_{\rm WR} = B_0 \Big(
3.76 \times 10^{-4} - 5.26 \times 10^{-5}\,\frac{E_e}{E_0} - 1.26
\times 10^{-5}\,\frac{E^2_e}{E^2_0}\Big).
\end{eqnarray}
Now we may obtain the total contributions of Wilkinson's corrections
to the correlation function $\zeta(E_e)$ and the correlation
coefficients. We would like to emphasize that the correction
$\zeta(E_e)_{\rm WR}$ does not depend practically on the value of the
axial coupling constant $g_A$.

\vspace{-0.1in}
\subsection*{\bf Wilkinson's corrections to the neutron
beta decay}
\vspace{-0.05in} Summing the contributions in Eqs.(\ref{eq:C.3}),
(\ref{eq:C.6}) and (\ref{eq:C.7}) we define total Wilkinson's
corrections to the neutron lifetime and the correlation coefficients
of the neutron beta decay
\begin{eqnarray}\label{eq:C.8}
\hspace{-0.3in}\zeta(E_e)_{\rm WC} &=& 3.76 \times 10^{-4} - 5.26
\times 10^{-5}\,\frac{E_e}{E_0} - 1.26 \times
10^{-5}\,\frac{E^2_e}{E^2_0} - \frac{\pi
  \alpha}{\beta}\,\frac{E_e}{m_N}, \nonumber\\
\hspace{-0.3in}a(E_e)_{\rm WC} &=& - 3.76\times 10^{-4}\,a_0 -
\frac{\pi \alpha}{\beta^3}\,\frac{E_0 - E_e}{m_N},\nonumber\\
\hspace{-0.3in}A(E_e)_{\rm WC} &=& - - 3.76\times 10^{-4}\,A_0,
\nonumber\\
\hspace{-0.3in}B(E_e)_{\rm WC} &=& - B_0\Big( 3.76 \times 10^{-4} -
5.26 \times 10^{-5}\,\frac{E_e}{E_0} - 1.26 \times
10^{-5}\,\frac{E^2_e}{E^2_0}\Big), \nonumber\\
\hspace{-0.3in}K_n(E_e)_{\rm WC} &=& 0,
\nonumber\\
\hspace{-0.3in}Q_n(E_e)_{\rm WC} &=& - B_0\,\frac{\pi
  \alpha}{\beta^3}\, \frac{E_0 - E_e}{\beta^3 E_0}, \nonumber\\
\hspace{-0.3in}G(E_e)_{\rm WC} &=&  3.76 \times
10^{-4} - 5.26 \times 10^{-5}\,\frac{E_e}{E_0} - 1.26 \times
10^{-5}\,\frac{E^2_e}{E^2_0}, \nonumber\\
\hspace{-0.3in}H(E_e)_{\rm WC} &=& N(E_e)_{\rm WC} = 0,
\nonumber\\ \hspace{-0.3in}Q_e(E_e)_{\rm WC} &=& 3.76\times
10^{-4}\,A_0 + \frac{1}{3}\,B_0\, \big(1 + \sqrt{1 -
  \beta^2}\,\big)\,\frac{\pi \alpha}{\beta^3}\,\frac{E_0 - E_e}{m_N},
\nonumber\\
\hspace{-0.3in}K_e(E_e)_{\rm WC} &=& 3.76\times 10^{-4}\, a_0 +
\big(1 + \sqrt{1 - \beta^2}\,\big)\,\frac{\pi \alpha}{\beta^3}\,\frac{E_0 - E_e}{m_N}, \nonumber\\
\hspace{-0.3in}S(E_e)_{\rm WC} &=& U(E_e)_{\rm W} = 0, \nonumber\\
\hspace{-0.3in}T(E_e)_{\rm WC} &=& B_0\Big( 3.76 \times
10^{-4} - 5.26 \times 10^{-5}\,\frac{E_e}{E_0} - 1.26 \times
10^{-5}\,\frac{E^2_e}{E^2_0}\Big).
\end{eqnarray}
Wilkinson's corrections in Eq.(\ref{eq:C.8}) are calculated at the
account for the contributions, which are not smaller than a few parts
of $10^{-5}$ with a theoretical accuracy of about a few parts of
$10^{-6}$ in the experimental electron-energy region $0.811\,{\rm MeV}
\le E_e \le 1.211\,{\rm MeV}$ \cite{Abele2018}.

\vspace{-0.1in}
\section*{Appendix D: Radiative corrections of order
  $O(\alpha E_e/m_N)$, induced by Sirlin's outer radiative corrections
  of order $O(\alpha/\pi)$ and the phase-volume of the neutron beta
  decay} \renewcommand{\theequation}{D-\arabic{equation}}
\setcounter{equation}{0}

In this Appendix we analyze the contributions of the radiative
corrections $O(\alpha E_e/m_N)$, induced by Sirlin's outer radiative
corrections of order $O(\alpha/\pi)$ and the phase-volume of the
neutron beta decay taken to NLO in the large nucleon mass $m_N$
expansion. For this aim we rewrite Eq.(\ref{eq:A.13}) keeping the
contributions of the $O(\alpha/\pi)$ corrections only. We get
\begin{eqnarray}\label{eq:D.1}
\hspace{-0.3in}&&\sum_{\rm pol.}\frac{|M(n \to p e^-
  \bar{\nu}_e)|^2}{(1 + 3 g^2_A)|G_V|^2 64 m^2_n E_e E_{\bar{\nu}}} =
\Big[\Big(1 + \frac{\alpha}{\pi}\,f_{\beta}(E_e,\mu)\Big) +
  B_0\Big(1 + \frac{\alpha}{\pi}\,f_{\beta}(E_e, \mu)\Big)\,
  \frac{\vec{\xi}_n \cdot \vec{k}_{\bar{\nu}}}{E_{\bar{\nu}}}\Big]\,
\Big(1 - \frac{m_e}{E_e}\,\zeta^0_e\Big) \nonumber\\
\hspace{-0.3in}&& + \Big[a_0\Big(1 +
  \frac{\alpha}{\pi}\,f_{\beta}(E_e, \mu)\Big)
  \frac{\vec{k}_{\bar{\nu}}}{E_{\bar{\nu}}} + A_0\Big(1 +
  \frac{\alpha}{\pi}\,f_{\beta}(E_e, \mu)\Big) \vec{\xi}_n\Big]\cdot
\Big(\frac{\vec{k}_e}{E_e} - \frac{m_e}{E_e}\,\vec{\zeta}_e\Big) +
\Big[- \frac{\alpha}{\pi}\, \frac{\sqrt{1 - \beta^2}}{2\beta}{\ell
    n}\Big(\frac{1 + \beta}{1 - \beta}\Big) \nonumber\\
\hspace{-0.3in}&& - B_0\, \frac{\alpha}{\pi}\, \frac{\sqrt{1 -
    \beta^2}}{2\beta} {\ell n}\Big(\frac{1 + \beta}{1 - \beta}\Big) \,
\frac{\vec{\xi}_n \cdot \vec{k}_{\bar{\nu}}}{E_{\bar{\nu}}}\Big]\,
\frac{m_e}{E_e} + \Big[A_0\, \frac{\alpha}{\pi}\, \frac{\sqrt{1 -
      \beta^2}}{2\beta}{\ell n}\Big(\frac{1 + \beta}{1 - \beta}\Big)
  \vec{\xi}_n + a_0 \, \frac{\alpha}{\pi}\, \frac{\sqrt{1 -
      \beta^2}}{2\beta}{\ell n}\Big(\frac{1 + \beta}{1 - \beta}\Big)\nonumber\\
\hspace{-0.3in}&& \times
\frac{\vec{k}_{\bar{\nu}}}{E_{\bar{\nu}}}\Big] \cdot
\Big(\vec{\zeta}_e - \frac{\vec{k}_e}{E_e}\,\zeta^0_e\Big) + \ldots,
\end{eqnarray}
where the ellipsis denotes the contributions, which are not important
for the aim of this Appendix. In terms of irreducible correlation
structures and using Eq.(\ref{eq:A.7}) we transcribe the r.h.s. of
Eq.(\ref{eq:D.1}) into the form
\begin{eqnarray}\label{eq:D.2}
\hspace{-0.3in}&&\sum_{\rm pol.}\frac{|M(n \to p e^-
  \bar{\nu}_e)|^2}{(1 + 3 g^2_A)|G_V|^2 64 m^2_n E_e E_{\bar{\nu}}} =
\Big[1 + \frac{\alpha}{\pi}\Big(\bar{g}_n(E_e) -
  g^{(1)}_{\beta\gamma}(E_e, \mu)\Big)\Big] + a_0\,\Big[1 +
  \frac{\alpha}{\pi}\Big(\bar{g}_n(E_e) + \frac{\sqrt{1 -
      \beta^2}}{2\beta}{\ell n}\Big(\frac{1 + \beta}{1 -
    \beta}\Big)\nonumber\\
\hspace{-0.3in}&& - g^{(1)}_{\beta\gamma}(E_e, \mu)\Big)\Big]\,
\frac{\vec{k}_e\cdot \vec{k}_{\bar{\nu}}}{E_e E_{\bar{\nu}}} + A_0
\,\Big[1 + \frac{\alpha}{\pi}\,\Big(\bar{g}_n(E_e) + \frac{1 -
    \beta^2}{2\beta}\,{\ell n}\Big(\frac{1 + \beta}{1 - \beta}\Big) -
  g^{(1)}_{\beta\gamma}(E_e,\mu)\Big)\Big] \,\frac{\vec{\xi}_n\cdot
  \vec{k}_e}{E_e} + B_0 \,\Big[1 +
  \frac{\alpha}{\pi}\,\Big(\bar{g}_n(E_e)\nonumber\\
\hspace{-0.3in}&& - g^{(1)}_{\beta\gamma}(E_e,\mu)\Big)\Big]\,
\frac{\vec{\xi}_n\cdot \vec{k}_{\bar{\nu}}}{E_{\bar{\nu}}} - \Big[1 +
  \frac{\alpha}{\pi}\,\Big(\bar{g}_n(E_e) + \frac{1 -
    \beta^2}{2\beta}\, {\ell n}\Big(\frac{1 + \beta}{1 - \beta}\Big) -
  g^{(1)}_{\beta\gamma}(E_e,\mu)\Big)\Big] \frac{\vec{\xi}_e \cdot
  \vec{k}_e}{E_e} - \frac{m_e}{E_e}\,a_0 \Big[1 +
\frac{\alpha}{\pi}\,\Big(\bar{g}_n(E_e) \nonumber\\
\hspace{-0.3in}&& - \frac{\beta}{2}\,{\ell n}\Big(\frac{1 + \beta}{1 -
  \beta}\Big) - g^{(1)}_{\beta\gamma}(E_e,\mu)\Big)\Big]\,
\frac{\vec{\xi}_e \cdot \vec{k}_{\bar{\nu}}}{E_{\bar{\nu}}} -
\frac{m_e}{E_e}\,A_0 \Big[1 + \frac{\alpha}{\pi}\,\Big(\bar{g}_n(E_e)
  - \frac{\beta}{2}\,{\ell n}\Big(\frac{1 + \beta}{1 - \beta}\Big) -
  g^{(1)}_{\beta\gamma}(E_e,\mu)\Big)\Big]\, \vec{\xi}_n\cdot
\vec{\xi}_e \nonumber\\
\hspace{-0.3in}&& - A_0 \,\Big[1 +
  \frac{\alpha}{\pi}\,\Big(\bar{g}_n(E_e) + \big(1 + \sqrt{1 -
    \beta^2}\,\big) \frac{\sqrt{1 - \beta^2}}{2\beta}\,{\ell
    n}\Big(\frac{1 + \beta}{1 - \beta}\Big) -
  g^{(1)}_{\beta\gamma}(E_e,\mu)\Big)\Big]\, \frac{(\vec{\xi}_n\cdot
  \vec{k}_e)( \vec{k}_e\cdot \vec{\xi}_e)}{(E_e + m_e) E_e} - a_0
\Big[1 + \frac{\alpha}{\pi}\Big(\bar{g}_n(E_e)\nonumber\\
\hspace{-0.3in}&& + (1 + \sqrt{1 - \beta^2}\,)\frac{\sqrt{1 -
    \beta^2}}{2 \beta}\,{\ell n}\Big(\frac{1 + \beta}{1 - \beta}\Big)
- g^{(1)}_{\beta\gamma}(E_e,\mu)\Big)\Big]\, \frac{(\vec{\xi}_e\cdot
  \vec{k}_e)( \vec{k}_e\cdot \vec{k}_{\bar{\nu}})}{(E_e + m_e)E_e
  E_{\bar{\nu}}} - B_0\Big(1 + \frac{\alpha}{\pi}\,\Big(\bar{g}_n(E_e)
+ \frac{1 - \beta^2}{2\beta}\,{\ell n}\Big(\frac{1 + \beta}{1 -
  \beta}\Big) \nonumber\\
\hspace{-0.3in}&& - g^{(1)}_{\beta\gamma}(E_e,\mu)\Big)\Big)
\frac{(\vec{\xi}_n \cdot \vec{k}_{\bar{\nu}})(\vec{\xi}_e \cdot
  \vec{k}_e)}{E_e E_{\bar{\nu}}}.
\end{eqnarray}
Taking into account the contribution of the phase-volume of the
neutron beta decay Eq.(\ref{eq:A.2}) we obtain
\begin{eqnarray}\label{eq:D.3}
\hspace{-0.3in}&&\Phi_n(\vec{k}_e, \vec{k}_{\bar{\nu}_e})\sum_{\rm
  pol.}\frac{|M(n \to p e^- \bar{\nu}_e)|^2}{(1 + 3 g^2_A)|G_V|^2 64
  m^2_n E_e E_{\bar{\nu}}} = \Big[1 + 3 \frac{E_e}{m_N}\Big(1 -
  \frac{\vec{k}_e\cdot \vec{k}_{\bar{\nu}}}{E_e
    E_{\bar{\nu}}}\Big)\Big]\bigg\{ \Big[1 +
  \frac{\alpha}{\pi}\Big(\bar{g}_n(E_e) -
  g^{(1)}_{\beta\gamma}(E_e, \mu)\Big)\Big] \nonumber\\
\hspace{-0.3in}&& + a_0\,\Big[1 +
  \frac{\alpha}{\pi}\Big(\bar{g}_n(E_e) + \frac{\sqrt{1 -
      \beta^2}}{2\beta}{\ell n}\Big(\frac{1 + \beta}{1 - \beta}\Big)-
  g^{(1)}_{\beta\gamma}(E_e, \mu)\Big)\Big]\, \frac{\vec{k}_e\cdot
  \vec{k}_{\bar{\nu}}}{E_e E_{\bar{\nu}}} + A_0 \,\Big[1 +
  \frac{\alpha}{\pi}\,\Big(\bar{g}_n(E_e) + \frac{1 -
    \beta^2}{2\beta}\,{\ell n}\Big(\frac{1 + \beta}{1 - \beta}\Big)
  \nonumber\\
\hspace{-0.3in}&& - g^{(1)}_{\beta\gamma}(E_e,\mu)\Big)\Big]
\,\frac{\vec{\xi}_n\cdot \vec{k}_e}{E_e} + B_0 \,\Big[1 +
  \frac{\alpha}{\pi}\,\Big(\bar{g}_n(E_e) -
  g^{(1)}_{\beta\gamma}(E_e,\mu)\Big)\Big]\, \frac{\vec{\xi}_n\cdot
  \vec{k}_{\bar{\nu}}}{E_{\bar{\nu}}} - \Big[1 +
  \frac{\alpha}{\pi}\,\Big(\bar{g}_n(E_e) + \frac{1 -
    \beta^2}{2\beta}\, {\ell n}\Big(\frac{1 + \beta}{1 - \beta}\Big)
  \nonumber\\
\hspace{-0.3in}&& - g^{(1)}_{\beta\gamma}(E_e,\mu)\Big)\Big]
\frac{\vec{\xi}_e \cdot \vec{k}_e}{E_e} - \frac{m_e}{E_e}\,a_0 \Big[1
+ \frac{\alpha}{\pi}\,\Big(\bar{g}_n(E_e) - \frac{\beta}{2}\,{\ell
  n}\Big(\frac{1 + \beta}{1 - \beta}\Big) -
g^{(1)}_{\beta\gamma}(E_e,\mu)\Big)\Big]\, \frac{\vec{\xi}_e \cdot
  \vec{k}_{\bar{\nu}}}{E_{\bar{\nu}}} - \frac{m_e}{E_e}\,A_0 \Big[1 +
  \frac{\alpha}{\pi}\,\Big(\bar{g}_n(E_e)\nonumber\\  
\hspace{-0.3in}&& - \frac{\beta}{2}\,{\ell n}\Big(\frac{1 + \beta}{1 -
  \beta}\Big) - g^{(1)}_{\beta\gamma}(E_e,\mu)\Big)\Big]\,
\vec{\xi}_n\cdot \vec{\xi}_e - A_0 \,\Big[1 +
  \frac{\alpha}{\pi}\,\Big(\bar{g}_n(E_e) + \big(1 + \sqrt{1 -
    \beta^2}\,\big) \frac{\sqrt{1 - \beta^2}}{2\beta}\,{\ell
    n}\Big(\frac{1 + \beta}{1 - \beta}\Big) -
  g^{(1)}_{\beta\gamma}(E_e,\mu)\Big)\Big] \nonumber\\
\hspace{-0.3in}&& \times \, \frac{(\vec{\xi}_n\cdot \vec{k}_e)(
  \vec{k}_e\cdot \vec{\xi}_e)}{(E_e + m_e) E_e} - a_0 \Big[1 +
  \frac{\alpha}{\pi}\Big(\bar{g}_n(E_e) + (1 + \sqrt{1 -
    \beta^2}\,)\frac{\sqrt{1 - \beta^2}}{2 \beta}\,{\ell
    n}\Big(\frac{1 + \beta}{1 - \beta}\Big) -
  g^{(1)}_{\beta\gamma}(E_e,\mu)\Big)\Big]\, \frac{(\vec{\xi}_e\cdot
  \vec{k}_e)( \vec{k}_e\cdot \vec{k}_{\bar{\nu}})}{(E_e + m_e)E_e
  E_{\bar{\nu}}} \nonumber\\
\hspace{-0.3in}&& - B_0\Big[1 +
  \frac{\alpha}{\pi}\,\Big(\bar{g}_n(E_e) + \frac{1 -
    \beta^2}{2\beta}\,{\ell n}\Big(\frac{1 + \beta}{1 - \beta}\Big) -
  g^{(1)}_{\beta\gamma}(E_e,\mu)\Big)\Big]\, \frac{(\vec{\xi}_n \cdot
  \vec{k}_{\bar{\nu}})(\vec{\xi}_e \cdot \vec{k}_e)}{E_e
  E_{\bar{\nu}}}\bigg\}.
\end{eqnarray}
The electron-energy and angular distribution of the neutron beta
decay, taking into account the radiative corrections of order
$O(\alpha/\pi)$ and the NLO corrections in the large nucleon mass
$m_N$ expansion, induced by the phase-volume, is given by
\begin{eqnarray*}
\hspace{-0.3in}&&\frac{d^5 \lambda_{\beta}(E_e, \vec{k}_e,
  \vec{k}_{\bar{\nu}}, \vec{\xi}_n, \vec{\xi}_e)}{dE_e d\Omega_e
  d\Omega_{\bar{\nu}}} = (1 + 3 g^2_A)\,\frac{|G_V|^2}{16 \pi^5}\,(E_0
- E_e)^2 \,\sqrt{E^2_e - m^2_e}\, E_e\,F(E_e, Z =
1)\, \Big[1 + 3 \frac{E_e}{m_N}\Big(1 -
  \frac{\vec{k}_e\cdot \vec{k}_{\bar{\nu}}}{E_e
    E_{\bar{\nu}}}\Big)\Big]\nonumber\\
\hspace{-0.3in}&& \times \bigg\{ \Big[1 +
  \frac{\alpha}{\pi}\Big(\bar{g}_n(E_e) - g^{(1)}_{\beta\gamma}(E_e,
  \mu)\Big)\Big] + a_0\,\Big[1 + \frac{\alpha}{\pi}\Big(\bar{g}_n(E_e)
  + \frac{\sqrt{1 - \beta^2}}{2\beta}{\ell n}\Big(\frac{1 + \beta}{1 -
    \beta}\Big)- g^{(1)}_{\beta\gamma}(E_e, \mu)\Big)\Big]\,
\frac{\vec{k}_e\cdot \vec{k}_{\bar{\nu}}}{E_e E_{\bar{\nu}}} \nonumber\\
\hspace{-0.3in}&& + A_0 \,\Big[1 +
  \frac{\alpha}{\pi}\,\Big(\bar{g}_n(E_e) + \frac{1 -
    \beta^2}{2\beta}\,{\ell n}\Big(\frac{1 + \beta}{1 - \beta}\Big) -
  g^{(1)}_{\beta\gamma}(E_e,\mu)\Big)\Big] \,\frac{\vec{\xi}_n\cdot
  \vec{k}_e}{E_e} + B_0 \,\Big[1 +
  \frac{\alpha}{\pi}\,\Big(\bar{g}_n(E_e) -
  g^{(1)}_{\beta\gamma}(E_e,\mu)\Big)\Big]\, \frac{\vec{\xi}_n\cdot
  \vec{k}_{\bar{\nu}}}{E_{\bar{\nu}}} \nonumber\\
\hspace{-0.3in}&& - \Big[1 + \frac{\alpha}{\pi} \Big(\bar{g}_n(E_e) +
  \frac{1 - \beta^2}{2\beta} {\ell n}\Big(\frac{1 + \beta}{1 -
    \beta}\Big) - g^{(1)}_{\beta\gamma}(E_e,\mu)\Big)\Big]
\frac{\vec{\xi}_e \cdot \vec{k}_e}{E_e} - \frac{m_e}{E_e} a_0 \Big[1 +
  \frac{\alpha}{\pi} \Big(\bar{g}_n(E_e) - \frac{\beta}{2}\,{\ell
    n}\Big(\frac{1 + \beta}{1 - \beta}\Big) -
  g^{(1)}_{\beta\gamma}(E_e,\mu)\Big)\Big] \nonumber\\ 
\hspace{-0.3in}&& \times \, \frac{\vec{\xi}_e \cdot
  \vec{k}_{\bar{\nu}}}{E_{\bar{\nu}}} - \frac{m_e}{E_e}\,A_0 \Big[1 +
  \frac{\alpha}{\pi}\,\Big(\bar{g}_n(E_e) - \frac{\beta}{2}\,{\ell
    n}\Big(\frac{1 + \beta}{1 - \beta}\Big) -
  g^{(1)}_{\beta\gamma}(E_e,\mu)\Big)\Big]\, \vec{\xi}_n\cdot
\vec{\xi}_e - A_0 \,\Big[1 + \frac{\alpha}{\pi}\,\Big(\bar{g}_n(E_e) +
  \big(1 + \sqrt{1 - \beta^2}\,\big)\nonumber\\
\end{eqnarray*}
\begin{eqnarray}\label{eq:D.4}  
\hspace{-0.3in}&& \times \frac{\sqrt{1 - \beta^2}}{2\beta}\,{\ell
  n}\Big(\frac{1 + \beta}{1 - \beta}\Big) -
g^{(1)}_{\beta\gamma}(E_e,\mu)\Big)\Big] \, \frac{(\vec{\xi}_n\cdot
  \vec{k}_e)( \vec{k}_e\cdot \vec{\xi}_e)}{(E_e + m_e) E_e} - a_0
\Big[1 + \frac{\alpha}{\pi}\Big(\bar{g}_n(E_e) + (1 + \sqrt{1 -
    \beta^2}\,)\frac{\sqrt{1 - \beta^2}}{2 \beta}\,{\ell
    n}\Big(\frac{1 + \beta}{1 - \beta}\Big) \nonumber\\    
\hspace{-0.3in}&& - g^{(1)}_{\beta\gamma}(E_e,\mu)\Big)\Big]\,
\frac{(\vec{\xi}_e\cdot \vec{k}_e)( \vec{k}_e\cdot
  \vec{k}_{\bar{\nu}})}{(E_e + m_e)E_e E_{\bar{\nu}}} - B_0\Big[1 +
  \frac{\alpha}{\pi}\,\Big(\bar{g}_n(E_e) + \frac{1 -
    \beta^2}{2\beta}\,{\ell n}\Big(\frac{1 + \beta}{1 - \beta}\Big) -
  g^{(1)}_{\beta\gamma}(E_e,\mu)\Big)\Big]\, \frac{(\vec{\xi}_n \cdot
  \vec{k}_{\bar{\nu}})(\vec{\xi}_e \cdot \vec{k}_e)}{E_e
  E_{\bar{\nu}}}\bigg\}.
\end{eqnarray}
In order to remove the dependence of the electron-energy and angular
distribution of the neutron beta decay on the infrared cut-off $\mu$
we have to take into account the contribution of the neutron radiative
beta decay $n \to p + e^- + \bar{\nu}_e + \gamma$, where $\gamma$ is a
real photon. It is well-known \cite{Berman1958, Kinoshita1959,
  Berman1962, Kaellen1967, Abers1968} (see also \cite{Sirlin1967,
  Shann1971} and \cite{Ivanov2013, Ivanov2017, Ivanov2019a}) that the
contribution of the neutron radiative beta decay is extremely needed
for cancellation of the infrared divergences in the radiative
corrections of order $O(\alpha/\pi)$, caused by one-virtual photon
exchanges.

For the removal of the infrared dependence we use the following
electron-photon-energy and angular distribution, calculated in
\cite{Ivanov2021b} (see Eq.B-14) in Ref.\cite{Ivanov2021b}):
\begin{eqnarray}\label{eq:D.5}
\hspace{-0.15in}&&\frac{d^6\lambda_{\beta \gamma}(E_e,\omega,\vec{k}_e,
  \vec{k}_{\bar{\nu}}, \vec{\xi}_n, \vec{\xi}_e)}{d\omega d E_e
  d\Omega_ed\Omega_{\bar{\nu}}} = (1 + 3 g^2_A) \,
\frac{\alpha}{\pi}\,\frac{|G_V|^2}{16 \pi^5}\,\sqrt{E^2_e -
  m^2_e}\,E_e\,F(E_e, Z = 1)\,(E_0 - E_e - \omega)^2
\Phi_{n\gamma}(\vec{k}_e, \vec{k}_{\bar{\nu}_e}, \omega)
\nonumber\\
\hspace{-0.3in}&& \times \, \bigg\{\frac{1}{\omega}\,
\Big[\Big(1 + \frac{\omega}{E_e} +
  \frac{1}{2}\,\frac{\omega^2}{E^2_e}\Big)\,\Big[\frac{1}{\beta}\,{\ell
      n}\Big(\frac{1 + \beta}{1 - \beta}\Big) - 2\Big] +
  \frac{\omega^2}{E^2_e}\Big] + a_0\,\frac{\vec{k}_e\cdot
  \vec{k}_{\bar{\nu}}}{E_e E_{\bar{\nu}}}\, \frac{1}{\omega}\, \Big(1
+ \frac{1}{\beta^2}\,\frac{\omega}{E_e} + \frac{1}{2
  \beta^2}\,\frac{\omega^2}{E^2_e}\Big) \,\Big[\frac{1}{\beta}\, {\ell n}\Big(\frac{1
    + \beta}{1 - \beta}\Big) - 2\Big] \nonumber\\
\hspace{-0.3in}&& + A_0\, \frac{\vec{\xi}_n \cdot
  \vec{k}_e}{E_e}\,\frac{1}{\omega} \, \Big(1 +
\frac{1}{\beta^2}\,\frac{\omega}{E_e}+ \frac{1}{2
  \beta^2}\,\frac{\omega^2}{E^2_e}\Big) \,\Big[\frac{1}{\beta}\, {\ell
    n}\Big(\frac{1 + \beta}{1 - \beta}\Big) - 2\Big] +
B_0\,\frac{\vec{\xi}_n \cdot
  \vec{k}_{\bar{\nu}}}{E_{\bar{\nu}}}\,\frac{1}{\omega} \, \Big[\Big(1
  + \frac{\omega}{E_e} + \frac{1}{2}\,\frac{\omega^2}{E^2_e}\Big)
  \,\Big[\frac{1}{\beta}\,{\ell n}\Big(\frac{1 + \beta}{1 -
      \beta}\Big) - 2\Big] \nonumber\\ 
\hspace{-0.3in}&& + \frac{\omega^2}{E^2_e}\Big] + (- 1)\,
\frac{\vec{\xi}_e \cdot \vec{k}_e}{E_e}\,\frac{1}{\omega} \,\Big(1 +
\frac{1}{\beta^2}\,\frac{\omega}{E_e} + \frac{1}{2
  \beta^2}\,\frac{\omega^2}{E^2_e}\Big)\Big]
  \,\Big[\frac{1}{\beta}\,{\ell n}\Big(\frac{1 + \beta}{1 -
      \beta}\Big) - 2\Big] + (- 1)\,\frac{m_e}{E_e}\,a_0\,
  \frac{\vec{\xi}_e \cdot \vec{k}_{\bar{\nu}}}{E_{\bar{\nu}}}
  \,\frac{1}{\omega} \,\Big(1 -
  \frac{1}{2\beta^2}\,\frac{\omega^2}{E^2_e}\Big)\nonumber\\ 
\hspace{-0.3in}&& \times \, \Big[\frac{1}{\beta}\,{\ell n}\Big(\frac{1
    + \beta}{1 - \beta}\Big) - 2\Big] + (- 1)\, \frac{m_e}{E_e}\, A_0
\,\vec{\xi}_n \cdot \vec{\xi}_e\,\frac{1}{\omega} \,\Big(1 -
\frac{1}{2\beta^2}\,\frac{\omega^2}{E^2_e}\Big)\,
\Big[\frac{1}{\beta}\,{\ell n}\Big(\frac{1 + \beta}{1 - \beta}\Big) -
  2\Big] + (- 1) \,A_0\, \frac{\vec{\xi}_n \cdot
  \vec{k}_e)(\vec{\xi}_e\cdot \vec{k}_e)}{(E_e + m_e) E_e}\nonumber\\
\hspace{-0.3in}&& \times \, \Big\{\frac{1}{\omega}\,\Big(1 -
\frac{1}{2\beta^2}\,\frac{\omega^2}{E^2_e}\Big)
\,\Big[\frac{1}{\beta}\,{\ell n}\Big(\frac{1 + \beta}{1 - \beta}\Big)
  - 2\Big] + (1 + \sqrt{1 - \beta^2}\,)\Big[\frac{1}{\beta^2}\,
  \frac{\omega}{E_e} \,\Big[\frac{1}{\beta}\, {\ell n}\Big(\frac{1 +
      \beta}{1 - \beta}\Big) - 2\Big] + \frac{1}{2 \beta^2}\,
  \frac{\omega^2}{E^2_e}\, \,\Big(\frac{3 -
    \beta^2}{\beta^2}\nonumber\\
\hspace{-0.3in}&& \times \Big[\frac{1}{\beta}\,{\ell n}\Big(\frac{1 +
    \beta}{1 - \beta}\Big) - 2\Big] - 2\Big)\Big]\Big\} +
(-1)\,a_0\,\frac{(\vec{\xi}_e \cdot \vec{k}_e)(\vec{k}_e \cdot
  \vec{k}_{\bar{\nu}})}{(E_e + m_e) E_e
  E_{\bar{\nu}}}\,\Big\{\frac{1}{\omega}\,\Big(1 -
\frac{1}{2\beta^2}\,\frac{\omega^2}{E^2_e}\Big)\Big[\frac{1}{\beta}\,{\ell
    n}\Big(\frac{1 + \beta}{1 - \beta}\Big) - 2\Big] + (1 + \sqrt{1 -
  \beta^2}\,) \nonumber\\
\hspace{-0.3in}&& \times \Big[\frac{1}{\beta^2}\, \frac{\omega}{E_e}
  \,\Big[\frac{1}{\beta}\, {\ell n}\Big(\frac{1 + \beta}{1 -
      \beta}\Big) - 2\Big] + \frac{1}{2 \beta^2}\,
  \frac{\omega^2}{E^2_e}\, \,\Big(\frac{3 -
    \beta^2}{\beta^2}\Big[\frac{1}{\beta}\,{\ell n}\Big(\frac{1 +
      \beta}{1 - \beta}\Big) - 2\Big] - 2\Big) \Big] \Big\} - B_0
\frac{(\vec{\xi}_n\cdot \vec{k}_{\bar{\nu}})(\vec{\xi}_e\cdot
  \vec{k}_e)}{E_e E_{\bar{\nu}}}\,\frac{1}{\omega} \, \Big(1 +
\frac{1}{\beta^2}\,\frac{\omega}{E_e} \nonumber\\
\hspace{-0.3in}&&+ \frac{1}{2 \beta^2}\,\frac{\omega^2}{E^2_e}\Big)
\,\Big[\frac{1}{\beta}\, {\ell n}\Big(\frac{1 + \beta}{1 - \beta}\Big)
  - 2\Big] \bigg\},
\end{eqnarray}
where $\Phi_{n\gamma}(\vec{k}_e, \vec{k}_{\bar{\nu}_e}, \omega)$ is
the contribution of the phase-volume of the neutron radiative beta
decay \cite{Ivanov2017a}
\begin{eqnarray}\label{eq:D.6}
\hspace{-0.21in}\Phi_{n\gamma}(\vec{k}_e, \vec{k}_{\bar{\nu}_e},
\omega ) = 1 + 3\,\frac{E_e}{m_N}\Big(1 -
\frac{\vec{k}_e\cdot \vec{k}_{\bar{\nu}_e}}{E_e E_{\bar{\nu}_e}}\Big)
+ O\Big(\frac{\omega}{m_N}\Big)
\end{eqnarray}
The contribution of the phase-volume Eq.(\ref{eq:D.6}) is the rest of
the expression, calculated to NLO in the large nucleon mass $m_N$
expansion and the integration over the directions of the photon
3-momentum \cite{Ivanov2017a}. As has been shown in
\cite{Ivanov2017a}, the contributions of the terms $O(\omega/m_N)$ are
of order of $10^{-6}$ and even smaller. So, the $O(\alpha E_e/m_N)$
corrections can be induced only by the second term in
Eq.(\ref{eq:D.6}).

\begin{figure}
 \includegraphics[height=0.18\textheight]{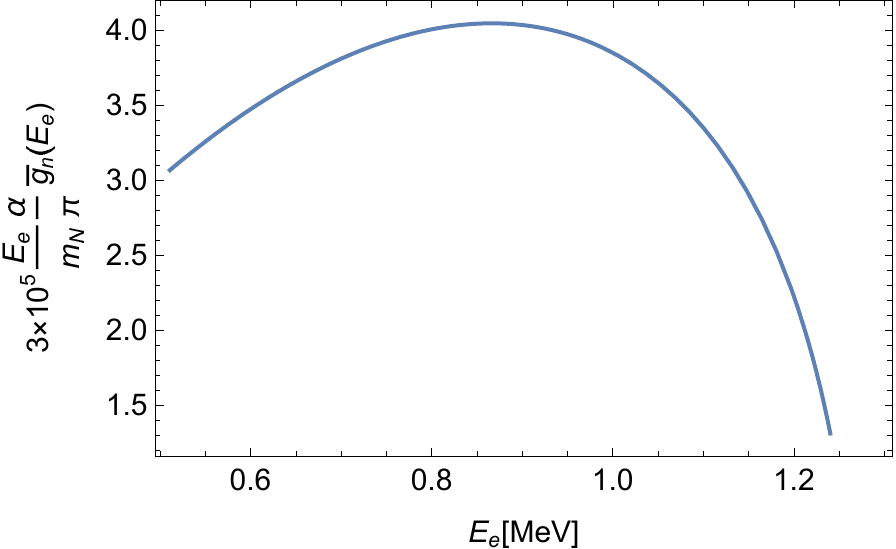} 
 \includegraphics[height=0.18\textheight]{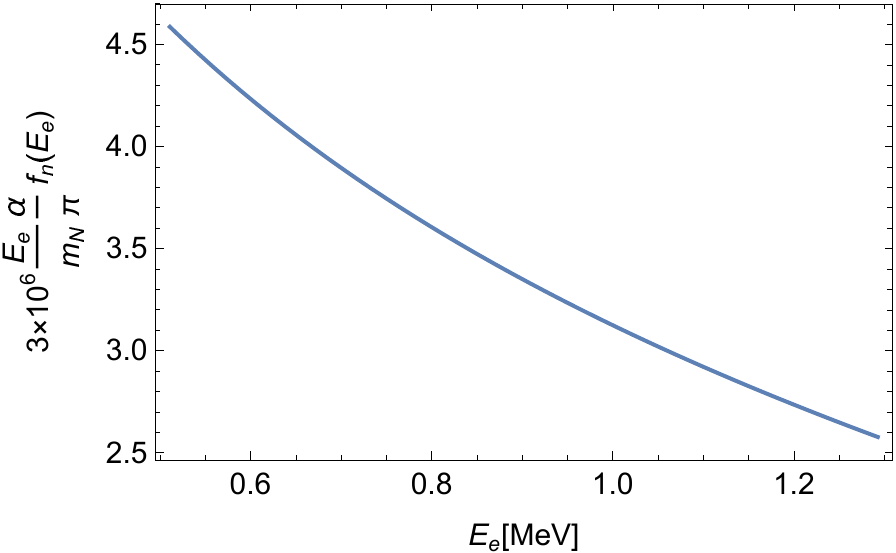}
 \includegraphics[height=0.18\textheight]{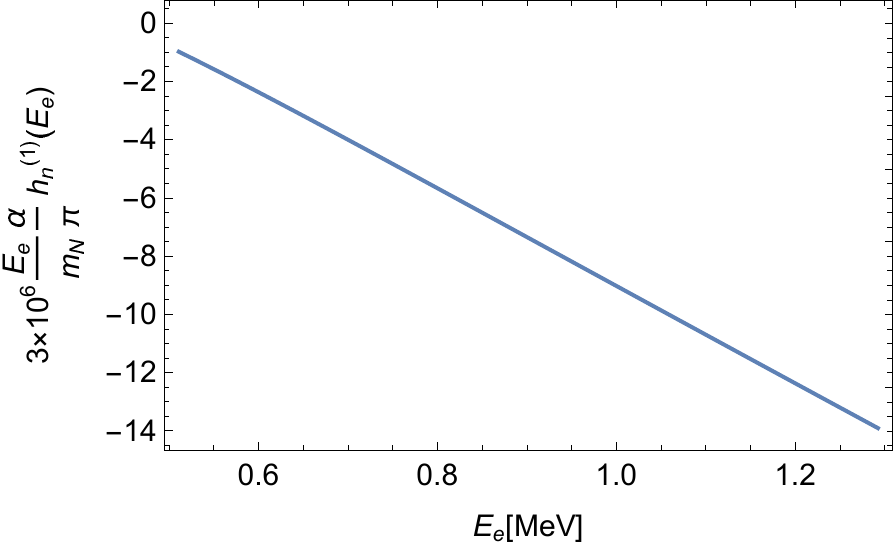}
  \includegraphics[height=0.18\textheight]{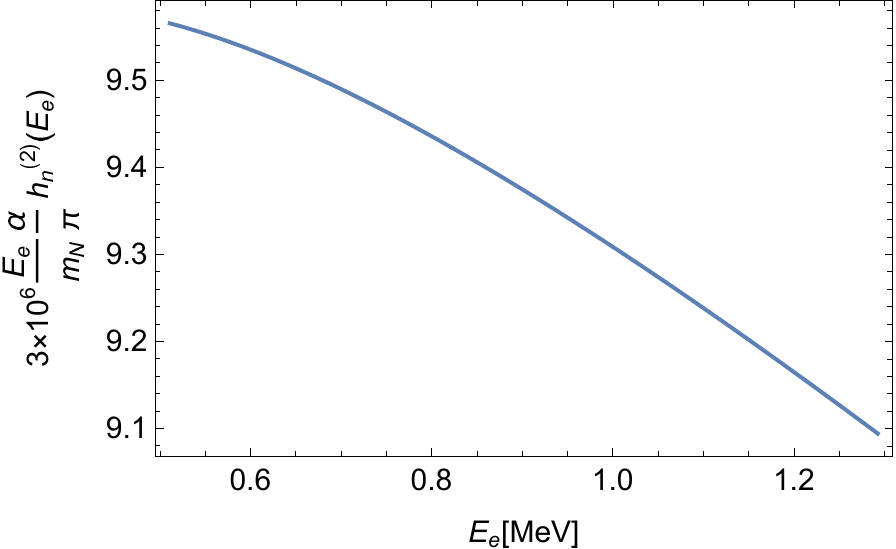}
  \caption{The outer radiative corrections of order $O(\alpha
    E_e/m_N)$ in the electron-energy region $m_e \le E_e < E_0$,
    induced by the outer radiative corrections of order
    $O(\alpha/\pi)$ and the phase-volume of the neutron beta decay,
    calculated to NLO in the large nucleon mass $m_N$ expansion. }
\label{fig:fig1}
\end{figure}

Having integrated over $\omega$ we arrive at the following expression
\begin{eqnarray}\label{eq:D.7}
\hspace{-0.15in}&& \frac{d^6\lambda_{\beta \gamma}(E_e,\vec{k}_e,
  \vec{k}_{\bar{\nu}}, \vec{\xi}_n, \vec{\xi}_e)}{d E_e
  d\Omega_ed\Omega_{\bar{\nu}}} = (1 + 3 g^2_A) \,
\frac{\alpha}{\pi}\,\frac{|G_V|^2}{16 \pi^5}\,\sqrt{E^2_e -
  m^2_e}\,E_e\,F(E_e, Z = 1)\,(E_0 - E_e)^2\Big[1 +
  3\,\frac{E_e}{m_N}\Big(1 - \frac{\vec{k}_e\cdot
    \vec{k}_{\bar{\nu}_e}}{E_e E_{\bar{\nu}_e}}\Big)\Big] \nonumber\\
\hspace{-0.3in}&& \times \, \bigg\{g^{(1)}_{\beta \gamma}(E_e, \mu) +
a_0\,\frac{\vec{k}_e\cdot \vec{k}_{\bar{\nu}}}{E_e E_{\bar{\nu}}}\,
g^{(2)}_{\beta \gamma}(E_e, \mu) + A_0\, g^{(2)}_{\beta \gamma}(E_e,
\mu) + B_0\,\frac{\vec{\xi}_n \cdot
  \vec{k}_{\bar{\nu}}}{E_{\bar{\nu}}}\,g^{(1)}_{\beta \gamma}(E_e,
\mu) + (- 1)\, \frac{\vec{\xi}_e \cdot \vec{k}_e}{E_e}\,g^{(2)}_{\beta
  \gamma}(E_e, \mu) \nonumber\\
\hspace{-0.3in}&& + (- 1)\,\frac{m_e}{E_e}\,a_0\, \frac{\vec{\xi}_e \cdot
  \vec{k}_{\bar{\nu}}}{E_{\bar{\nu}}} \, g^{(3)}_{\beta\gamma}(E_e,
\mu) + (- 1)\, \frac{m_e}{E_e}\, A_0 \,\vec{\xi}_n \cdot
\vec{\xi}_e\,g^{(3)}_{\beta \gamma}(E_e, \mu) + (- 1) \,A_0\,
\frac{\vec{\xi}_n \cdot \vec{k}_e)(\vec{\xi}_e\cdot \vec{k}_e)}{(E_e +
  m_e) E_e} \,g^{(4)}_{\beta \gamma}(E_e, \mu) \nonumber\\
\hspace{-0.3in}&& + (-1)\,a_0\,\frac{(\vec{\xi}_e \cdot
  \vec{k}_e)(\vec{k}_e \cdot \vec{k}_{\bar{\nu}})}{(E_e + m_e) E_e
  E_{\bar{\nu}}}\, g^{(4)}_{\beta\gamma}(E_e, \mu) - B_0
\frac{(\vec{\xi}_n\cdot \vec{k}_{\bar{\nu}})(\vec{\xi}_e\cdot
  \vec{k}_e)}{E_e E_{\bar{\nu}}}\,g^{(2)}_{\beta \gamma}(E_e, \mu)
\bigg\}.
\end{eqnarray}
The functions $g^{(j)}_{\beta \gamma}(E_e, \mu)$ for $j = 1,2,3,4$
have been calculated in \cite{Ivanov2013, Ivanov2017, Ivanov2019a,
  Ivanov2021a}. As has been shown in \cite{Ivanov2013, Ivanov2017,
  Ivanov2019a, Ivanov2021a}, the difference between functions
$g^{(1)}_{\beta\gamma}(E_e,\mu) - g^{(j)}_{\beta\gamma}(E_e, \mu)$ for
$j = 2,3,4$ does not depend on the regularization, i.e.
\begin{eqnarray}\label{eq:D.8}
\hspace{-0.15in} \lim_{\mu \to 0}\Big(g^{(1)}_{\beta\gamma}(E_e,\mu) -
g^{(j)}_{\beta\gamma}(E_e, \mu)\Big) = \lim_{\omega_{\rm min} \to
  0}\Big(g^{(1)}_{\beta\gamma}(E_e,\omega_{\rm min}) -
g^{(j)}_{\beta\gamma}(E_e,\omega_{\rm min})\Big),
\end{eqnarray}
where $\omega_{\rm min}$ is a non-covariant infrared cut-off, which
can be also treated as the photon--energy threshold of the detector
\cite{Ivanov2013, Ivanov2017, Ivanov2019a, Ivanov2021a}. Summing up
Eqs.(\ref{eq:D.4}) and (\ref{eq:D.7}) we obtain the total
electron-energy and angular distribution of the neutron beta decay
\begin{eqnarray}\label{eq:D.9}
\hspace{-0.3in}&&\frac{d^5 \lambda_n(E_e, \vec{k}_e,
  \vec{k}_{\bar{\nu}}, \vec{\xi}_n, \vec{\xi}_e)}{dE_e d\Omega_e
  d\Omega_{\bar{\nu}}} = (1 + 3 g^2_A)\,\frac{|G_V|^2}{16 \pi^5}\,(E_0
- E_e)^2 \,\sqrt{E^2_e - m^2_e}\, E_e\,F(E_e, Z = 1)\, \Big[1 + 3
  \frac{E_e}{m_N}\Big(1 - \frac{\vec{k}_e\cdot
    \vec{k}_{\bar{\nu}}}{E_e E_{\bar{\nu}}}\Big)\Big] \nonumber\\
\hspace{-0.3in}&&\bigg\{ \Big(1 +
\frac{\alpha}{\pi}\,\bar{g}_n(E_e)\Big) + a_0\,\Big[1 +
  \frac{\alpha}{\pi}\Big(\bar{g}_n(E_e) + f_n(E_e)\Big)\Big]\,
\frac{\vec{k}_e\cdot \vec{k}_{\bar{\nu}}}{E_e E_{\bar{\nu}}} + A_0
\,\Big[1 + \frac{\alpha}{\pi}\,\Big(\bar{g}_n(E_e) +
  f_n(E_e)\Big)\Big] \,\frac{\vec{\xi}_n\cdot \vec{k}_e}{E_e}\nonumber\\
\hspace{-0.3in}&& + B_0 \,\Big(1 +
\frac{\alpha}{\pi}\,\bar{g}_n(E_e)\Big)\, \frac{\vec{\xi}_n\cdot
  \vec{k}_{\bar{\nu}}}{E_{\bar{\nu}}} - \Big[1 +
  \frac{\alpha}{\pi}\,\Big(\bar{g}_n(E_e) + f_n(E_e)\Big)\Big]
\frac{\vec{\xi}_e \cdot \vec{k}_e}{E_e} - \frac{m_e}{E_e}\,a_0 \Big[1
  + \frac{\alpha}{\pi}\,\Big(\bar{g}_n(E_e) +
  h^{(1)}_n(E_e)\Big)\Big]\, \frac{\vec{\xi}_e \cdot
  \vec{k}_{\bar{\nu}}}{E_{\bar{\nu}}} \nonumber\\
\hspace{-0.3in}&&- \frac{m_e}{E_e}\,A_0 \Big[1 +
  \frac{\alpha}{\pi}\,\Big(\bar{g}_n(E_e) + h^{(1)}_n(E_e)\Big]\,
\vec{\xi}_n\cdot \vec{\xi}_e - A_0 \,\Big[1 +
  \frac{\alpha}{\pi}\,\Big(\bar{g}_n(E_e) + h^{(2)}_n(E_e)\Big]\,
\frac{(\vec{\xi}_n\cdot \vec{k}_e)( \vec{k}_e\cdot \vec{\xi}_e)}{(E_e
  + m_e) E_e} 
\nonumber\\
\hspace{-0.3in}&& - a_0 \Big[1 + \frac{\alpha}{\pi}\Big(\bar{g}_n(E_e)
  + h^{(2)}_n(E_e)\Big)\Big]\, \frac{(\vec{\xi}_e\cdot \vec{k}_e)(
  \vec{k}_e\cdot \vec{k}_{\bar{\nu}})}{(E_e + m_e)E_e E_{\bar{\nu}}} -
B_0\Big[1 + \frac{\alpha}{\pi}\,\Big(\bar{g}_n(E_e) +
  f_n(E_e)\Big)\Big]\, \frac{(\vec{\xi}_n \cdot
  \vec{k}_{\bar{\nu}})(\vec{\xi}_e \cdot \vec{k}_e)}{E_e
  E_{\bar{\nu}}}\bigg\}.
\end{eqnarray}
The detailed calculation of the functions $(\alpha/\pi)\bar{g}_n(E_e)$
\cite{Sirlin1967}, $(\alpha/\pi)f_n(E_e)$ \cite{Shann1971},
$(\alpha/\pi)h^{(1)}_n(E_e)$ and $(\alpha/\pi)h^{(2)}_n(E_e)$ one can
find in \cite{Ivanov2013, Ivanov2017, Ivanov2019a, Ivanov2021a,
  Ivanov2021b}. The behaviour of the functions
$(\alpha/\pi)\bar{g}_n(E_e)$, $(\alpha/\pi)f_n(E_e)$,
$(\alpha/\pi)h^{(1)}_n(E_e)$ and $(\alpha/\pi)h^{(2)}_n(E_e)$
multiplied by the factor $3 (E_e/m_N)$, caused by the phase-volume of
the neutron beta decay, is shown in Fig.\,\ref{fig:fig1}.

One can see that in the electron-energy region $m_e \le E_e < E_0$ the
functions $3(\alpha/\pi)(E_e/m_N) \bar{g}_n(E_e)$,
$3(\alpha/\pi)(E_e/m_N)h^{(1)}_n(E_e)$ and $3
(\alpha/\pi)(E_e/m_N)h^{(2)}_n(E_e)$ are of order of a few parts of
$10^{-5}$, whereas the function $3(\alpha/\pi)(E_e/m_N) f_n(E_e)$ is
of order of a few parts of $10^{-6}$. However, the functions
$3(\alpha/\pi)(E_e/m_N)h^{(j)}_n(E_e)$ for $j = 1,2$ are multiplied by
either $(m_e/E_e)\,X_0$ or $X_0$, where $X_0 = a_0, A_0 \sim -
0.1$. As a result, the values of the functions $3(\alpha/\pi)(m_e/m_N)
X_0 h^{(j)}_n(E_e)$ and $3(\alpha/\pi)(E_e/m_N) X_0 h^{(j)}_n(E_e)$
become of order of a few parts of $10^{-6}$. Hence, the outer
radiative corrections of order $O(\alpha E_e/m_N)$ are defined by the
function $3(\alpha/\pi)(E_e/m_N) \bar{g}_n(E_e)$ only.

As result, the correlation function $\zeta(E_e)$ and the correlation
coefficients $a(E_e)$, $Q_n(E_e)$, $Q_e(E_e)$ and $K_e(E_e)$ acquire
the following outer or model-independent $O(\alpha E_e/m_N)$ radiative
corrections :
\begin{eqnarray}\label{eq:D.10}
\zeta(E_e)_{\rm RC-PhV} &=&
3\,\frac{\alpha}{\pi}\,\frac{E_e}{m_N}\,\bar{g}_n(E_e) \quad,\quad
a(E_e)_{\rm RC-PhV} = -
3\,\frac{\alpha}{\pi}\,\frac{E_e}{m_N}\,\bar{g}_n(E_e),
\nonumber\\ Q_n(E_e)_{\rm RC-PhV} &=& - 3\,
B_0\,\frac{\alpha}{\pi}\,\frac{E_e}{m_N}\,\bar{g}_n(E_e)\quad,\quad
Q_e(E_e)_{\rm RC-PhV} =
B_0\,\frac{\alpha}{\pi}\,\frac{E_e}{m_N}\,\bar{g}_n(E_e)\,\Big(1 +
\frac{m_e}{E_e}\Big), \nonumber\\ K_e(E_e)_{\rm RC-PhV} &=&
3\,\frac{\alpha}{\pi}\,\frac{E_e}{m_N}\,\bar{g}_n(E_e)\,
\Big(1+\frac{m_e}{E_e}\Big).
\end{eqnarray}
These radiative corrections are of order of a few parts of $10^{-5}$
in the experimental electron-energy region $0.811\,{\rm MeV} \le E_e
\le 1.211\,{\rm MeV}$ \cite{Abele2018}. They are plotted in
\cite{MathW}. The analytical expression of the function
$\bar{g}_n(E_e)$ is given in Eq.(\ref{eq:A.19}).  We have to notice
that there is the contribution, proportional to
$3(\alpha/\pi)(E_e/m_N) \bar{g}_n(E_e)$, to the electron-energy and
angular distribution of the neutron beta decay with the correlation
structure beyond the standard correlation structures in
Eq.(\ref{eq:1}) (see the last term in Eq.(\ref{eq:F.2})).

\section*{Appendix E: Analytical expressions for the correlation
  function $\zeta(E_e)$ and correlation coefficients $a(E_e)$,
  $A(E_e)$, $B(E_e)$, $K_n(E_e)$, $Q_n(E_e)$ and $A^{(\beta)}(E_e)$}
\renewcommand{\theequation}{E-\arabic{equation}}
\setcounter{equation}{0}

In this Appendix we give the analytical expressions for the
correlation function $\zeta(E_e)$ and the correlation coefficients
$X(E_e)$ for $X = a, A, B, \ldots, U$ and also for the correlation
coefficient $A^{(\beta)}(E_e) = A(E_e) + \frac{1}{3}\,Q_n(E_e)$ as
functions of the electron energy $E_e$ and the axial coupling constant
$g_A$.  The correlation function $\zeta(E_e)$ and correlation
coefficients are calculated with a theoretical accuracy of about
$10^{-6}$.

For the correlation function $\zeta(E_e)$ we obtain the following
expression
\begin{eqnarray}\label{eq:E.1}
\hspace{-0.15in}\zeta(E_e) &=& 1 + \frac{\alpha}{\pi}\,\bar{g}_n(E_e) +
\frac{1}{1 + 3 g^2_A}\,\frac{E_0}{m_N}\, \Big[ - 2\,g_A\big(g_A +
  (\kappa + 1)\big) + \big(10 g^2_A + 4(\kappa + 1)\, g_A +
  2\big)\,\frac{E_e}{E_0}\nonumber\\
\hspace{-0.3in}&-& 2 g_A\,\big(g_A + (\kappa +
1)\big)\,\frac{m^2_e}{E^2_0}\,\frac{E_0}{E_e}\Big] + \zeta(E_e)_{\rm
  RC-NLO} + \zeta(E_e)_{\rm RC-PhV} + \zeta(E_e)_{\rm N^2L0} +
\zeta(E_e)_{\rm WC}.
\end{eqnarray}
Here $\bar{g}_n(E_e)$ is Sirlin's function \cite{Sirlin1967} (see also
Eq.(\ref{eq:A.19}. Then, the term proportional to $E_0/m_N$ defines
the well-known corrections $O(E_e/m_N)$, caused by weak magnetism and
proton recoil (see, for example, \cite{Gudkov2006, Ivanov2013}). The
corrections $\zeta(E_e)_{\rm RC-NLO}$, $\zeta(E_e)_{\rm RC-PhV}$, $
\zeta(E_e)_{\rm N^2L0}$ and $ \zeta(E_e)_{\rm WC}$ are defined by the
$O(\alpha E_e/m_N)$ inner and outer radiative corrections, the
$O(E^2_e/m^2_N)$ corrections, caused by weak magnetism and proton
recoil, and Wilkinson's corrections, respectively. They are equal to
\begin{eqnarray}\label{eq:E.2}
\hspace{-0.15in} && \zeta(E_e)_{\rm RC-NLO} =\frac{1}{1 + 3 g^2_A}\,
\frac{\alpha}{\pi}\, \frac{E_e}{m_N} \Big( f_V(E_e) + \sqrt{1 -
  \beta^2}\,f_S(E_e) + g_A g_V(E_e)\, \beta^2 + (1 + 2 g_A)
\,h_A(E_e)\, \beta^2 \nonumber\\
\hspace{-0.3in}&&+ g_A \frac{E_0 - E_e}{E_e}\,\sqrt{1 - \beta^2}\,
h_S(E_e) + 3 g^2_A f_A(E_e) + 3 g^2_A \sqrt{1 - \beta^2}\,
f_T(E_e)\Big) + \frac{3 g^2_A}{1 + 3 g^2_A} \, \frac{\alpha}{\pi}\,
\frac{5}{2}\, \frac{m^2_N}{M^2_W}\,{\ell n}\frac{M^2_W}{m^2_N}
\nonumber\\
\hspace{-0.3in}&&+ \frac{1}{1 + 3 g^2_A}\,\frac{\alpha}{\pi}\,
\Big(\bar{g}_{\rm st}(E_e) + 3 g_A \bar{f}_{\rm st}(E_e)\Big),\nonumber\\
\hspace{-0.15in} && \zeta(E_e)_{\rm RC-PhV} =
3\,\frac{\alpha}{\pi}\,\frac{E_e}{m_N}\,\bar{g}_n(E_e),\nonumber\\
\hspace{-0.15in} && \zeta(E_e)_{\rm N^2L0} =  6\,
\frac{E^2_e}{m^2_N}\Big\{ \Big(1 - \frac{1}{4}\,\frac{E_0}{E_e}\Big) +
\frac{1}{3} \Big[1- 2\, a_0\, \Big( 1 -
  \frac{1}{8}\,\frac{E_0}{E_e}\Big)\Big] \Big(1 -
\frac{m^2_e}{E^2_e}\Big) \Big\} + 3\, \frac{E_e}{m_N}\,
\zeta(E_e)_{\rm NLO} - \frac{E_e}{m_N}\, \bar{a}(E_e)_{\rm
  NLO}\nonumber\\
\hspace{-0.3in}&&\times\, \Big(1 - \frac{m^2_e}{E^2_e}\Big)
 - \frac{8}{1 + 3 g^2_A}\,\Big(\frac{E^2_0}{M^2_V} +
3 g^2_A \frac{E^2_0}{M^2_A}\Big)\, \frac{E_e}{E_0}\Big(1 -
\frac{E_e}{E_0}\Big),\nonumber\\
\hspace{-0.15in}&& \zeta(E_e)_{\rm WC} = 3.76 \times 10^{-4} -
5.26 \times 10^{-5}\,\frac{E_e}{E_0} - 1.26 \times
10^{-5}\,\frac{E^2_e}{E^2_0} - \frac{\pi
  \alpha}{\beta}\,\frac{E_e}{m_N}.
\end{eqnarray}
The terms, proportional to $E^2_0/M^2_V$ and $E^2_0/M^2_A$ , appear
from the contributions of the vector and axial-vector form-factors of
the neutron beta decay. For numerical analysis we use $M_V = 813\,{\rm
  MeV}$ and $M_A = 1077\,{\rm MeV}$ \cite{Ivanov2020b}. The first
three terms in $ \zeta(E_e)_{\rm WC}$ do not depend practically on the
axial coupling constant $g_A$.

Of course, the correlation function $\zeta(E_e)$ in Eq.(\ref{eq:E.1})
should be supplemented by the inner radiative corrections $\Delta^V_R$
and $\Delta^A_R$ of order $O(\alpha/\pi)$, caused by the Feynman
$\gamma W^-$-box diagrams and calculated in \cite{Sirlin1986,
  Sirlin2004, Sirlin2006, Seng2018, Seng2018a, Sirlin2019, Hayen2020,
  Hayen2021, Gorchtein2021}).

For the correlation coefficient $a(E_e)$ we obtain the following
expression
\begin{eqnarray}\label{eq:E.3}
\hspace{-0.3in}a(E_e) &=& a_0\Big\{1 + \frac{\alpha}{\pi}\,f_n(E_e) +
\frac{1}{(1 - g^2_A)(1 + 3 g^2_A)}\,\frac{E_0}{m_N}\,\Big(a_1 + a_2
\frac{E_e}{E_0} + a_3\frac{E_0}{E_e}\Big)\Big\}\nonumber\\
\hspace{-0.3in}&+&  a(E_e)_{\rm RC-NLO} +  a(E_e)_{\rm RC-PhV}
+ a(E_e)_{\rm N^2LO} + a(E_e)_{\rm WC},
\end{eqnarray}
where the function $f_n(E_e)$ has been calculated by Shann
\cite{Shann1971} (see also Eq.(\ref{eq:A.19}) and \cite{Ivanov2013,
  Ivanov2017}) and the coefficients $a_1$, $a_2$ and $a_3$ are equal
to \cite{Ivanov2013}
\begin{eqnarray}\label{eq:E.4}
\hspace{-0.3in}a_1 &=& 4 g_A (g^2_A + 1)\Big(g_A + (\kappa +
1)\Big),\nonumber\\
\hspace{-0.3in}a_2 &=& - 26 g^4_A - 8(\kappa + 1)\,g^3_A - 20
g^2_A - 8(\kappa + 1)\,g_A - 2,\nonumber\\
\hspace{-0.3in}a_3 &=& - 2 g_A (g^2_A - 1)\Big(g_A + (\kappa +
1)\Big)\,\frac{m^2_e}{E^2_0}.
\end{eqnarray}
The corrections $a(E_e)_{\rm RC-NLO}$, $ a(E_e)_{\rm RC-PhV}$, $
a(E_e)_{\rm N^2L0}$ and $ a(E_e)_{\rm WC}$, defined by the $O(\alpha
E_e/m_N)$ inner and outer radiative corrections, the $O(E^2_e/m^2_N)$
corrections, caused by weak magnetism and proton recoil, and
Wilkinson's corrections, respectively, are given by
\begin{eqnarray}\label{eq:E.5}
\hspace{-0.15in}&&a(E_e)_{\rm RC-NLO} = \frac{1}{1 + 3
  g^2_A}\,\frac{\alpha}{\pi}\,\frac{E_e}{m_N} \Big(f_V(E_e) + g_A
\sqrt{1 - \beta^2}\,g_S(E_e) + g_A\, g_V(E_e)+ (1 - 2 g_A)\, h_A(E_e)
- g^2_A f_A(E_e)\Big) \nonumber\\
\hspace{-0.15in}&&- \frac{g^2_A}{1 + 3
  g^2_A}\,\frac{\alpha}{\pi}\,\frac{5}{2}\, \frac{m^2_N}{M^2_W}\,{\ell
  n}\frac{M^2_W}{m^2_N} + \frac{1}{1 + 3 g^2_A}\,\frac{\alpha}{\pi}\,
\Big(\bar{g}_{\rm st}(E_e) - g_A \bar{f}_{\rm st}(E_e)\Big) - a_0\,
 \zeta(E_e)_{\rm RC-NLO},\nonumber\\
\hspace{-0.15in}&&a(E_e)_{\rm RC-PhV} = - 3\,
\frac{\alpha}{\pi}\,\frac{E_e}{m_N}\,\bar{g}_n(E_e),\nonumber\\
\hspace{-0.15in}&&a(E_e)_{\rm N^2L0} = 12\, \frac{E^2_e}{m^2_N}\Big\{ -
\Big(1 - \frac{1}{8}\, \frac{E_0}{E_e}\Big) + \frac{1}{3}\, a_0\,
\Big(1 - \frac{1}{4}\, \frac{E_0}{E_e}\Big)\Big\} + 3\,
\frac{E_e}{m_N}\, \bar{a}(E_e)_{\rm NLO} - \frac{E_e}{m_N}\,
\zeta(E_e)_{\rm NLO} \nonumber\\
\hspace{-0.3in}&& - a(E_e)_{\rm NLO}\, \zeta(E_e)_{\rm NLO} +
\frac{8}{1 + 3 g^2_A}\,\Big(\frac{E^2_0}{M^2_V} + 3 g^2_A
\frac{E^2_0}{M^2_A}\Big)\, \frac{E_e}{E_0}\Big(1 -
\frac{E_e}{E_0}\Big),\nonumber\\
\hspace{-0.15in}&& a(E_e)_{\rm WC} = - 3.76 \times 10^{-4}\,a_0 -
\frac{\alpha \pi}{\beta^3}\,\frac{E_0 - E_e}{m_N}.
\end{eqnarray}
For the correlation coefficient $A(E_e)$ we obtain the following
expression
\begin{eqnarray}\label{eq:E.6}
\hspace{-0.15in}A(E_e) &=& A_0\,\Big\{1 +
\frac{\alpha}{\pi}\,f_n(E_e) + \frac{1}{2 g_A(1 - g_A)(1
  + 3 g^2_A)}\,\frac{E_0}{m_N} \Big(A _1  + A_2 \frac{E_e}{E_0} +
A_3\frac{E_0}{E_e}\Big)\Big\}\nonumber\\
\hspace{-0.3in}&+&  A(E_e)_{\rm RC-NLO} +  A(E_e)_{\rm N^2LO}
+ A(E_e)_{\rm WC},
\end{eqnarray}
where the function $f_n(E_e)$ has been calculated by Shann
\cite{Shann1971} (see also Eq.(\ref{eq:A.19})) and the coefficients
$A_1$, $A_2$ and $A_3$ are given by \cite{Ivanov2013}
\begin{eqnarray}\label{eq:E.7}
\hspace{-0.3in}A_1 &=& - g^4_A - \kappa \,g^3_A + (\kappa +
2)\,g^2_A + \kappa\, g_A - (\kappa + 1),\nonumber\\
\hspace{-0.3in}A_2 &=& 5 g^4_A - \kappa\, g^3_A - (5 \kappa +
6)\,g^2_A - 3 \kappa\,g_A + (\kappa + 1),\nonumber\\
\hspace{-0.3in}A_3 &=& -\,4 g^2_A(g_A -  1)\,\Big(g_A +  (\kappa +
1)\Big)\,\frac{m^2_e}{E^2_0}.
\end{eqnarray}
The corrections $A(E_e)_{\rm RC-NLO}$, $A(E_e)_{\rm N^2L0}$ and
$A(E_e)_{\rm WC}$, defined by the $O(\alpha E_e/m_N)$ inner radiative
corrections, the $O(E^2_e/m^2_N)$ corrections, caused by weak
magnetism and proton recoil, and Wilkinson's corrections,
respectively, are equal to
\begin{eqnarray*}
\hspace{-0.15in}&&A(E_e)_{\rm RC-NLO} = \frac{1}{1 + 3 g^2_A}\,
\frac{\alpha}{\pi}\, \frac{E_e}{m_N}\, \Big(g_A f_V(E_e) + \sqrt{1 -
  \beta^2}\, g_S(E_e) + g_V(E_e) - g_A h_A(E_e) + g_A(1 - 2 g_A) \,
f_A(E_e)\Big) \nonumber\\
\hspace{-0.15in}&&+ \frac{1}{1 + 3 g^2_A}\, g_A(1 - 2
g_A)\,\frac{\alpha}{\pi}\, \frac{5}{2}\, \frac{m^2_N}{M^2_W}\,{\ell
  n}\frac{M^2_W}{m^2_N} + \frac{1}{1 + 3 g^2_A}\,\frac{\alpha}{\pi}\,
\Big(g_A \bar{g}_{\rm st}(E_e) + (1 - 2 g_A)\, \bar{f}_{\rm
  st}(E_e)\Big) - A_0 \,\zeta(E_e)_{\rm RC-NLO}, \nonumber\\
\end{eqnarray*}
\begin{eqnarray}\label{eq:E.8} 
\hspace{-0.15in}&& A(E_e)_{\rm N^2L0} = 3\, \frac{E_e}{m_N}\,
\bar{A}(E_e)_{\rm NLO} - \frac{E_e}{m_N}\,\bar{K}_n(E_e)_{\rm NLO}
\Big(1 - \frac{m^2_e}{E^2_e}\Big) - A(E_e)_{\rm NLO}\zeta(E_e)_{\rm
  NLO}\nonumber\\
\hspace{-0.15in} &&A(E_e)_{\rm WC} = - 3.76\times 10^{-4}\,A_0.
\end{eqnarray}
For the correlation coefficient $B(E_e)$ we obtain the following
expression
\begin{eqnarray}\label{eq:E.9}
\hspace{-0.15in}B(E_e) &=&B_0\,\Big\{1 + \frac{1}{2 g_A(1 + g_A)(1 + 3
  g^2_A)}\,\frac{E_0}{m_N}\Big(B_1 + B_2 \frac{E_e}{E_0} +
B_3\frac{E_0}{E_e}\Big)\Big\}\nonumber\\
\hspace{-0.3in}&+& B(E_e)_{\rm RC-NLO} +  B(E_e)_{\rm N^2LO}
+  B(E_e)_{\rm WC},
\end{eqnarray}
where the coefficients $B_1$, $B_2$ and $B_3$ are given by
\cite{Ivanov2013}
\begin{eqnarray}\label{eq:E.10}
\hspace{-0.3in}B_1 &=& - 2\,g_A (1 - g_A)^2 \Big(g_A + (\kappa
+ 1)\Big),\nonumber\\
\hspace{-0.3in}B_2 &=& g^4_A + (\kappa - 4)\,g^3_A - (5 \kappa
+ 2)\,g^2_A + (3 \kappa + 4)\,g_A + (\kappa + 1),\nonumber\\
\hspace{-0.3in}B_3 &=& (g^2_A - 1)(1 + g_A)\Big(g_A + (\kappa +
1)\Big)\,\frac{m^2_e}{E^2_0}.
\end{eqnarray} 
The corrections $B(E_e)_{\rm RC-NLO}$, $ B(E_e)_{\rm N^2L0}$ and $
B(E_e)_{\rm WC}$, defined by the $O(\alpha E_e/m_N)$ inner radiative
corrections, the $O(E^2_e/m^2_N)$ corrections, caused by weak
magnetism and proton recoil, and Wilkinson's corrections,
respectively, are equal to
\begin{eqnarray}\label{eq:E.11} 
\hspace{-0.15in}&&B(E_e)_{\rm RC-NLO} = \frac{1}{1 + 3 g^2_A}\,
\frac{\alpha}{\pi}\, \frac{E_e}{m_N}\,\Big(g_A f_V(E_e) + g_A \sqrt{1
  - \beta^2}\, f_S(E_e) + \frac{E_0 - E_e}{E_e}\, \sqrt{1 - \beta^2}\,
h_S(E_e) + g_A g_V(E_e)\, \beta^2 \nonumber\\
\hspace{-0.15in}&&- (1 - 2 g_A)\, h_A(E_e)\, \beta^2 + g_A(1 + 2 g_A)
\, f_A(E_e) + g_A(1 + 2 g_A) \sqrt{1 - \beta^2}\, f_T(E_e) \Big)
+ \frac{1}{1 + 3 g^2_A}\,g_A(1 + 2
g_A)\nonumber\\
\hspace{-0.15in}&&\times \,\frac{\alpha}{\pi}\, \frac{5}{2}\,
\frac{m^2_N}{M^2_W}\,{\ell n}\frac{M^2_W}{m^2_N} + \frac{1}{1 + 3
  g^2_A}\,\frac{\alpha}{\pi}\, \Big(g_A \bar{g}_{\rm st}(E_e) + (1 + 2
g_A)\, \bar{f}_{\rm st}(E_e)\Big) - B_0 \,\zeta(E_e)_{\rm
  RC-NLO},\nonumber\\
\hspace{-0.15in}&& B(E_e)_{\rm N^2L0} = 6 \, \frac{E^2_e}{m^2_N}\, B_0\,
\Big(1 - \frac{1}{4}\, \frac{E_0}{E_e}\Big) + 3\, \frac{E_e}{m_N}\,
\bar{B}(E_e)_{\rm NLO} - B(E_e)_{\rm NLO}\,\zeta(E_e)_{\rm NLO} -
B_0\, \zeta(E_e)_{\rm N^2LO}\nonumber\\
\hspace{-0.15in}&& - \frac{8 g_A}{1 + 3
  g^2_A}\,\Big(\frac{E^2_0}{M^2_V} + (1 + 2
g_A)\,\frac{E^2_0}{M^2_A}\Big)\, \frac{E_e}{E_0}\Big(1 -
\frac{E_e}{E_0}\Big),\nonumber\\
\hspace{-0.15in} && B(E_e)_{\rm WC} = B_0\Big( 3.76 \times
10^{-4} - 5.26 \times 10^{-5}\,\frac{E_e}{E_0} - 1.26 \times
10^{-5}\,\frac{E^2_e}{E^2_0}\Big) + \frac{1}{3}\,a_0 B_0 \frac{\pi
  \alpha}{\beta}\,\frac{E_0 - E_e}{m_N}.
\end{eqnarray}
The coefficients of Wilkinson's term in the parentheses do not
practically depend on the axial coupling constant $g_A$.

For the correlation coefficient $K_n(E_e)$ we obtain the following
expressions
\begin{eqnarray}\label{eq:E.12}
\hspace{-0.15in}K_n(E_e) &=& \frac{1}{1 + 3 g^2_A}\,\frac{E_e}{m_N}
\Big(5 g^2_A + (\kappa - 4)\, g_A - (\kappa + 1)\Big) + K_n(E_e)_{\rm
  RC} + K_n(E_e)_{\rm N^2LO} + K_n(E_e)_{\rm WC}.
\end{eqnarray}
The corrections $K_n(E_e)_{\rm RC-NLO}$, $ K_n(E_e)_{\rm N^2L0}$ and
$K_n(E_e)_{\rm WC}$, defined by the $O(\alpha E_e/m_N)$ inner
radiative corrections, the $O(E^2_e/m^2_N)$ corrections, caused by
weak magnetism and proton recoil, and Wilkinson's corrections,
respectively, are equal to
\begin{eqnarray}\label{eq:E.13}
\hspace{-0.15in}&&K_n(E_e)_{\rm RC-NLO} = \frac{1}{1 + 3 g^2_A}\,
\frac{\alpha}{\pi}\, \frac{E_e}{m_N}\,\Big((1 - g_A)\,g_V(E_e) + (1 +
2 g_A)\, h_A(E_e)\Big),\nonumber\\
\hspace{-0.15in}&& K_n(E_e)_{\rm N^2L0} = 3\, \frac{E_e}{m_N}\,
\bar{K}_n(E_e)_{\rm NLO} - 3\,\frac{E_e}{m_N}\, \bar{A}(E_e)_{\rm
  NLO},\nonumber\\
\hspace{-0.15in} && K_n(E_e)_{\rm WC} = - A_0\,\frac{\pi
  \alpha}{\beta^3}\,\frac{E_0 - E_e}{m_N}.
\end{eqnarray}
For the correlation coefficient $Q_n(E_e)$ we obtain the following
expression
\begin{eqnarray}\label{eq:E.14}
\hspace{-0.15in}Q_n(E_e) &=& \frac{1}{1 + 3 g^2_A}\,\frac{E_0}{m_N}
\Big[ \Big(g^2_A + (\kappa + 2)g_A + (\kappa + 1)\Big) -
  \Big(7 g^2_A + (\kappa + 8)\,g_A + (\kappa + 1)\Big)
  \frac{E_e}{E_0}\Big]\nonumber\\
\hspace{-0.15in}&+& Q_n(E_e)_{\rm RC-NLO} +  Q_n(E_e)_{\rm
  RC-PhV} + Q_n(E_e)_{\rm N^2LO} + Q_n(E_e)_{\rm WC},
\end{eqnarray}
where the corrections $Q_n(E_e)_{\rm RC-NLO}$, $Q_n(E_e)_{\rm
  RC-PhV}$, $ Q_n(E_e)_{\rm N^2L0}$ and $Q_n(E_e)_{\rm WC}$, defined
by the $O(\alpha E_e/m_N)$ inner and outer radiative corrections, the
$O(E^2_e/m^2_N)$ corrections, caused by weak magnetism and proton
recoil, and Wilkinson's corrections, respectively, are equal to
\begin{eqnarray}\label{eq:E.15}
  \hspace{-0.15in}&&Q_n(E_e)_{\rm RC-NLO} = 0,\nonumber\\
   \hspace{-0.15in}&& Q_n(E_e)_{\rm RC-PhV} = -
   3\,B_0\,\frac{\alpha}{\pi}\,\frac{E_e}{m_N}\,\bar{g}_n(E_e),\nonumber\\
\hspace{-0.15in}&& Q_n(E_e)_{\rm N^2L0} = - 12\,
\frac{E^2_e}{m^2_N}\,B_0 \, \Big(1 - \frac{1}{8}\,
\frac{E_0}{E_e}\Big) + 3\, \frac{E_e}{m_N}\, \bar{Q}_n(E_e)_{\rm NLO}
- 3\, \frac{E_e}{m_N}\, \bar{B}(E_e)_{\rm NLO} - Q_n(E_e)_{\rm
  NLO}\,\zeta(E_e)_{\rm NLO},\nonumber\\
\hspace{-0.15in}&& Q_n(E_e)_{\rm WC} = - B_0\,\frac{\pi
  \alpha}{\beta^3}\,\frac{E_0 - E_e}{m_N}.
\end{eqnarray}
Now we are able to give the analytical expression for the correlation
coefficient $A^{(\beta)}(E_e)$ defined by \cite{Wilkinson1982}
\begin{eqnarray}\label{eq:E.16}
\hspace{-0.15in}&& A^{(\beta)}(E_e) = A(E_e) + \frac{1}{3}\,Q_n(E_e).
\end{eqnarray}
This correlation coefficient is responsible for the electron (beta)
asymmetry in the neutron beta decay \cite{Abele2018} (see also
\cite{Ivanov2013}). The correlation coefficient $A^{(\beta)}(E_e)$ we
give in the following form \cite{Wilkinson1982} (see also
\cite{Ivanov2013})
\begin{eqnarray}\label{eq:E.17}
\hspace{-0.15in}A^{(\beta)}(E_e) &=& A_0\Big\{1 +
\frac{\alpha}{\pi}\,f_n(E_e) + \frac{g_A + \kappa + 1}{g_A(1 - g_A)(1
  + 3 g^2_A)}\,\frac{E_0}{m_N}\Big(A^{(\beta)}_1 +
A^{(\beta)}_2\,\frac{E_e}{E_0} +
A^{(\beta)}_3\,\frac{E_0}{E_e}\Big)\Big\}\nonumber\\ &+& 
A^{(\beta)}(E_e)_{\rm RC-NLO} + A^{(\beta)}(E_e)_{\rm RC-PhV} + 
A^{(\beta)}(E_e)_{\rm N^2LO} +  A^{(\beta)}(E_e)_{\rm WC},
\end{eqnarray}
 The function $f_n(E_e)$ is given in Eq.(\ref{eq:A.19}).  The
 coefficients $A^{(\beta)}_1$, $A^{(\beta)}_2$ and $A^{(\beta)}_3$ are
 equal to \cite{Wilkinson1982} (see also \cite{Ivanov2013})
\begin{eqnarray}\label{eq:E.18}
\hspace{-0.15in}A^{(\beta)}_1 &=& g^2_A + \frac{2}{3}\,g_A -
\frac{1}{3},\nonumber\\ A^{(\beta)}_2 &=& - g^3_A - 3 g^2_A -
\frac{5}{3}\, g_A + \frac{1}{3},\nonumber\\ A^{(\beta)}_3 &=&  2
g^2_A(1 - g_A)\,\frac{m^2_e}{E^2_0}
\end{eqnarray}
and the terms $ A^{(\beta)}(E_e)_{\rm RC-NLO}$, $
A^{(\beta)}(E_e)_{\rm RC-PhV}$, $A^{(\beta)}(E_e)_{\rm N^2L0}$
and $ A^{(\beta)}(E_e)_{\rm WC}$, defined by the radiative
corrections $O(\alpha E_e/m_N)$, the corrections $O(E^2_e/m^2_N)$,
caused by weak magnetism and proton recoil, and Wilkinson's
corrections, respectively, are equal to
\begin{eqnarray}\label{eq:E.19}
\hspace{-0.15in}&&A^{(\beta)}(E_e)_{\rm RC-NLO} = \frac{1}{1 + 3 g^2_A}\,
\frac{\alpha}{\pi}\, \frac{E_e}{m_N}\, \Big(g_A f_V(E_e) + \sqrt{1 -
  \beta^2}\, g_S(E_e) + g_V(E_e) - g_A h_A(E_e) + g_A(1 - 2 g_A) \,
f_A(E_e)\Big) \nonumber\\
\hspace{-0.15in}&&+ \frac{1}{1 + 3 g^2_A}\, g_A(1 - 2
g_A)\,\frac{\alpha}{\pi}\, \frac{5}{2}\, \frac{m^2_N}{M^2_W}\,{\ell
  n}\frac{M^2_W}{m^2_N} + \frac{1}{1 + 3 g^2_A}\,\frac{\alpha}{\pi}\,
\Big(g_A \bar{g}_{\rm st}(E_e) + (1 - 2 g_A)\, \bar{f}_{\rm
  st}(E_e)\Big) - A_0\, \zeta(E_e)_{\rm RC-NLO},\nonumber\\
\hspace{-0.15in}&&A^{(\beta)}(E_e)_{\rm RC-PhV} =
\frac{1}{3}\,Q_n(E_e)_{\rm RC-PhV} = - B_0\,\frac{\alpha}{\pi}\,
  \frac{E_e}{m_N}\, \bar{g}_n(E_e),\nonumber\\
\hspace{-0.15in}&&A^{(\beta)}(E_e)_{\rm N^2LO} = A(E_e)_{\rm N^2LO} +
\frac{1}{3}\,Q_n(E_e)_{\rm N^2LO}, \nonumber\\
\hspace{-0.15in}&&A^{(\beta)}(E_e)_{\rm WC} = 3.76 \times 10^{-4}\, A_0 -
\frac{1}{3}\,B_0\,\frac{\pi \alpha}{\beta^3}\,\frac{E_0 - E_e}{m_N}.
\end{eqnarray}
The functions $\zeta(E_e)_{\rm NLO}$, $a(E_e)_{\rm NLO}$, $A(E_e)_{\rm
  NLO}$, $B(E_e)_{\rm NLO}$ and $Q_n(E_e)_{\rm NLO}$ are given in
Eq.(\ref{eq:B.2}).

For the experimental analysis of the antineutrino asymmetry in the
neutron beta decay one has to use Eqs.(27) and (28) in
Ref.\cite{Ivanov2013} and the correlation coefficients $a(E_e)$,
$A(E_e)$, $B(E_e)$, $K_n(E_e)$ and $Q_n(E_e)$ given in this Appendix.
For the account for the contribution of the Fierz interference term
$b$ in the antineutrino asymmetry one may use Eqs.(19) and (20) in
Ref.\cite{Ivanov2019y}, where the correlation coefficients $X(E_e)$
are replaced by $X(E_e)/(1 + b m_e/E_e)$ for $X(E_e) = a(E_e), A(E_e),
B(E_e), K_n(E_e)$ and $Q_n(E_e)$, respectively.

For the correlation coefficients $G(E_e)$, $H(E_e)$, $N(E_e)$,
$Q_e(E_e)$, $K_e(E_e)$, $S(E_e)$, $T(E_e)$ and $U(E_e)$ as functions
of the electron energy $E_e$ and the axial coupling constant $g_A$ we
give the analytical expressions in the following form
\begin{eqnarray}\label{eq:E.20}
\hspace{-0.30in}G(E_e) &=& - \Big(1 +
\frac{\alpha}{\pi}\,f_n(E_e)\Big) + G(E_e)_{\rm NLO} + G(E_e)_{\rm
  RC-NLO} + G(E_e)_{\rm N^2LO} + G(E_e)_{\rm WC},\nonumber\\
\hspace{-0.30in}H(E_e) &=& - \frac{m_e}{E_e}\,a_0 \Big(1 +
\frac{\alpha}{\pi}\,h^{(1)}_n(E_e)\Big) + H(E_e)_{\rm NLO} +
H(E_e)_{\rm RC-NLO} + H(E_e)_{\rm N^2LO} + H(E_e)_{\rm WC},\nonumber\\
\hspace{-0.30in}N(E_e) &=& - \frac{m_e}{E_e}\,A_0 \Big( 1 +
\frac{\alpha}{\pi}\,h^{(1)}_n(E_e)\Big) + N(E_e)_{\rm NLO} +
N(E_e)_{\rm RC-NLO} + N(E_e)_{\rm N^2LO} + N(E_e)_{\rm WC},\nonumber\\
\hspace{-0.30in}Q_e(E_e) &=& - A_0 \,\Big(1 +
\frac{\alpha}{\pi}\,h^{(2)}_n(E_e)\Big) + Q_e(E_e)_{\rm NLO} +
Q_e(E_e)_{\rm RC-NLO} + Q_e(E_e)_{\rm RC-PhV}+ Q_e(E_e)_{\rm N^2LO} +
Q_e(E_e)_{\rm WC},\nonumber\\
\hspace{-0.30in}K_e(E_e) &=& - a_0 \,\Big(1 +
\frac{\alpha}{\pi}\,h^{(2)}_n(E_e)\Big) + K_e(E_e)_{\rm NLO} +
K_e(E_e)_{\rm RC-NLO} + K_e(E_e)_{\rm RC-PhV} + K_e(E_e)_{\rm N^2LO} +
K_e(E_e)_{\rm WC},\nonumber\\
\hspace{-0.30in}S(E_e) &=&S(E_e)_{\rm NLO}  + S(E_e)_{\rm RC-NLO} +
S(E_e)_{\rm N^2LO},\nonumber\\
\hspace{-0.3in}T(E_e) &=& - B_0\Big(1 +
\frac{\alpha}{\pi}\,f_n(E_e)\Big) + T(E_e)_{\rm NLO} + T(E_e)_{\rm
  RC-NLO}+ T(E_e)_{\rm N^2LO} + T(E_e)_{\rm WC},\nonumber\\
\hspace{-0.30in}U(E_e) &=& U(E_e)_{\rm RC-NLO}  +
U(E_e)_{\rm N^2LO}.
\end{eqnarray}
The analytical expressions of the functions $f_n(E_e)$,
$h^{(1)}_n(E_e)$ and $h^{(2)}_n(E_e)$ are given in
Eq.(\ref{eq:A.19}). The corrections $X(E_e)_{\rm NLO}$, $X(E_e)_{\rm
  RC-NLO}$, $X(E_e)_{\rm RC-PhV}$, $X(E_e)_{\rm N^2LO}$ and
$X(E_e)_{\rm WC}$ for $X = G, H, N, Q_e, K_e, S, T, U$ are adduced in
Eq.(\ref{eq:B.2}), Eq.(\ref{eq:A.20}), Eq.(\ref{eq:D.10}),
Eq.(\ref{eq:B.4}) and Eq.(\ref{eq:C.8}), respectively.

For the practical applications and numerical analysis the analytical
expressions of the correlation function $\zeta(E_e)$ and correlation
coefficients $X(E_e)$ for $X = a, A, B; \ldots, U$ and
$A^{(\beta)}(E_e)$ are programmed in \cite{MathW}.

\section*{Appendix F: Contributions to the  electron-energy and angular
  distribution of the neutron beta decay with correlation structures
  beyond Eq.(\ref{eq:1})}
\renewcommand{\theequation}{F-\arabic{equation}}
\setcounter{equation}{0}

The electron-energy and angular distribution of the neutron beta decay
Eq.(\ref{eq:1}), supplemented by the contributions with correlation
structures beyond the standard ones, takes the form
\begin{eqnarray}\label{eq:F.1}
\hspace{-0.15in}&&\frac{d^5 \lambda_n(E_e, \vec{k}_e,
  \vec{k}_{\bar{\nu}}, \vec{\xi}_n, \vec{\xi}_e)}{dE_e d\Omega_e
  d\Omega_{\bar{\nu}}} = (1 + 3 g^2_A)\,\frac{|G_V|^2}{16 \pi^5}\,(E_0
- E_e)^2 \,\sqrt{E^2_e - m^2_e}\, E_e\,F(E_e, Z =
1)\,\zeta(E_e)\,\Big\{1 + b(E_e)\,\frac{m_e}{E_e}\nonumber\\
\hspace{-0.15in}&& + a(E_e)\,\frac{\vec{k}_e\cdot
  \vec{k}_{\bar{\nu}}}{E_e E_{\bar{\nu}}} +
A(E_e)\,\frac{\vec{\xi}_n\cdot \vec{k}_e}{E_e} + B(E_e)\,
\frac{\vec{\xi}_n\cdot \vec{k}_{\bar{\nu}}}{E_{\bar{\nu}}} +
K_n(E_e)\,\frac{(\vec{\xi}_n\cdot \vec{k}_e)(\vec{k}_e\cdot
  \vec{k}_{\bar{\nu}})}{E^2_e E_{\bar{\nu}}}+
Q_n(E_e)\,\frac{(\vec{\xi}_n\cdot \vec{k}_{\bar{\nu}})(\vec{k}_e\cdot
  \vec{k}_{\bar{\nu}})}{ E_e E^2_{\bar{\nu}}}\nonumber\\
\hspace{-0.15in}&& + D(E_e)\,\frac{\vec{\xi}_n\cdot (\vec{k}_e\times
  \vec{k}_{\bar{\nu}})}{E_e E_{\bar{\nu}}} + G(E_e)\,\frac{\vec{\xi}_e
  \cdot \vec{k}_e}{E_e} + H(E_e)\,\frac{\vec{\xi}_e \cdot
  \vec{k}_{\bar{\nu}}}{E_{\bar{\nu}}} + N(E_e)\,\vec{\xi}_n\cdot
\vec{\xi}_e + Q_e(E_e)\,\frac{(\vec{\xi}_n\cdot \vec{k}_e)(
  \vec{k}_e\cdot \vec{\xi}_e)}{(E_e + m_e) E_e}\nonumber\\
\hspace{-0.15in}&& + K_e(E_e)\,\frac{(\vec{\xi}_e\cdot \vec{k}_e)(
  \vec{k}_e\cdot \vec{k}_{\bar{\nu}})}{(E_e + m_e)E_e E_{\bar{\nu}}} +
R(E_e)\,\frac{\vec{\xi}_n\cdot(\vec{k}_e \times \vec{\xi}_e)}{E_e} +
L(E_e)\,\frac{\vec{\xi}_e\cdot(\vec{k}_e \times
  \vec{k}_{\bar{\nu}})}{E_eE_{\bar{\nu}}} +
S(E_e)\,\frac{(\vec{\xi}_n\cdot \vec{\xi}_e)(\vec{k}_e \cdot
  \vec{k}_{\bar{\nu}})}{E_e E_{\bar{\nu}}} \nonumber\\
\hspace{-0.15in}&& + T(E_e)\,\frac{(\vec{\xi}_n \cdot
  \vec{k}_{\bar{\nu}})(\vec{\xi}_e \cdot \vec{k}_e)}{E_e
  E_{\bar{\nu}}} + U(E_e)\, \frac{(\vec{\xi}_n\cdot
  \vec{k}_e)(\vec{\xi}_e \cdot \vec{k}_{\bar{\nu}})}{E_e
  E_{\bar{\nu}}} + V(E_e)\, \frac{\vec{\xi}_n\cdot (\vec{\xi}_e \times
  \vec{k}_{\bar{\nu}})}{E_{\bar{\nu}}} + W(E_e)\,
\frac{\vec{\xi}_n\cdot (\vec{k}_e \times
  \vec{k}_{\bar{\nu}})(\vec{\xi}_e \cdot \vec{k}_e)}{(E_e + m_e) E_e
  E_{\bar{\nu}}} \Big\}\nonumber\\
\hspace{-0.15in}&&+ \frac{d^5 \lambda_n(E_e, \vec{k}_e,
  \vec{k}_{\bar{\nu}}, \vec{\xi}_n, \vec{\xi}_e)}{dE_e d\Omega_e
  d\Omega_{\bar{\nu}}}\Big|_{\rm RC-NLO} + \frac{d^5 \lambda_n(E_e,
  \vec{k}_e, \vec{k}_{\bar{\nu}}, \vec{\xi}_n, \vec{\xi}_e)}{dE_e
  d\Omega_e d\Omega_{\bar{\nu}}}\Big|_{\rm NLO} + \sum^4_{m =
  1}\frac{d^5 \lambda^{(m)}_n(E_e, \vec{k}_e, \vec{k}_{\bar{\nu}},
  \vec{\xi}_n, \vec{\xi}_e)}{dE_e d\Omega_e
  d\Omega_{\bar{\nu}}}\Big|_{\rm N^2LO}\nonumber\\
\hspace{-0.15in}&& + \frac{d^5 \lambda_n(E_e, \vec{k}_e,
  \vec{k}_{\bar{\nu}}, \vec{\xi}_n, \vec{\xi}_e)}{dE_e
  d\Omega_e d\Omega_{\bar{\nu}}}\Big|_{\rm W}\Big\},
\end{eqnarray}
where the last seven terms are defined by the following expressions
(see also \cite{Ivanov2020b}):
\begin{eqnarray}\label{eq:F.2}
\hspace{-0.15in}&&\frac{d^5 \lambda_n(E_e, \vec{k}_e,
  \vec{k}_{\bar{\nu}}, \vec{\xi}_n, \vec{\xi}_e)}{dE_e d\Omega_e
  d\Omega_{\bar{\nu}}}\Big|_{\rm RC-NLO} = (1 + 3
g^2_A)\,\frac{|G_V|^2}{16\pi^5}\,(E_0 - E_e)^2 \,\sqrt{E^2_e -
  m^2_e}\, E_e\,F(E_e, Z = 1)\,\zeta(E_e) \nonumber\\
\hspace{-0.15in}&& \times \Big\{\frac{2 E_e}{1 + 3 g^2_A}\,\Big(2g_A
U_5 - (1 - g_A)\, U_7 - (1 + g_A)\, U_8\Big)\, \frac{(\vec{\xi}_n
  \cdot \vec{k}_e)(\vec{\xi}_e \cdot \vec{k}_e)(\vec{k}_e \cdot
  \vec{k}_{\bar{\nu}})}{(E_e + m_e) E^2_e E_{\bar{\nu}}} - \frac{2
  E_e}{1 + 3 g^2_A}\, (1 + g_A)\, U_6 \nonumber\\
\hspace{-0.3in}&& \times \Big[\Big(\frac{(\vec{\xi}_n \cdot
    \vec{k}_{\bar{\nu}})(\vec{\xi}_e \cdot
    \vec{k}_{\bar{\nu}})}{E^2_{\bar{\nu}}} - \frac{1}{3}\,\vec{\xi}_e
  \cdot \vec{\xi}_e\Big) + \Big(- \frac{(\vec{\xi}_n \cdot
    \vec{k}_{\bar{\nu}})(\vec{\xi}_e \cdot \vec{k}_e)(\vec{k}_e \cdot
    \vec{k}_{\bar{\nu}})}{(E_e + m_e) E_e E^2_{\bar{\nu}}} +
  \frac{1}{3}\,\frac{(\vec{\xi}_n \cdot \vec{k}_e)(\vec{\xi}_e \cdot
    \vec{k}_e)}{(E_e + m_e) E_e}\Big)\Big] \nonumber\\
\hspace{-0.3in}&& + 3\, B_0\,
\frac{\alpha}{\pi}\,\frac{E_e}{m_N}\,\bar{g}_n(E_e)\Big(\frac{(\vec{\xi}_n
  \cdot \vec{k}_{\bar{\nu}})(\vec{\xi}_e \cdot \vec{k}_e)(\vec{k}_e
  \cdot \vec{k}_{\bar{\nu}})}{E^2_e E^2_{\bar{\nu}}} -
\frac{1}{3}\,\frac{(\vec{\xi}_n \cdot \vec{k}_e)(\vec{\xi}_e \cdot
  \vec{k}_e)}{E^2_e}\Big)\Big\},
\end{eqnarray}
where the functions $U_5, U_6, U_7$ and $U_8$ are given in
Eq.(\ref{eq:A.4}), and
\begin{eqnarray}\label{eq:F.3}
\hspace{-0.15in}&&\frac{d^5 \lambda_n(E_e, \vec{k}_e,
  \vec{k}_{\bar{\nu}}, \vec{\xi}_n, \vec{\xi}_e)}{dE_e d\Omega_e
  d\Omega_{\bar{\nu}}}\Big|_{\rm NLO} = (1 + 3
g^2_A)\,\frac{|G_V|^2}{16\pi^5}\,(E_0 - E_e)^2 \,\sqrt{E^2_e -
  m^2_e}\, E_e\,F(E_e, Z = 1)\,\zeta(E_e) \nonumber\\
\hspace{-0.15in}&& \times \,\frac{E_e}{m_N}\,\Big\{ -
3\, \frac{1 - g^2_A}{1 + 3 g^2_A}
\,\Big(\frac{(\vec{k}_e\cdot \vec{k}_{\bar{\nu}})^2}{E^2_e
  E^2_{\bar{\nu}}} - \frac{1}{3}\,\frac{k^2_e}{E^2_e}\Big) +
3\,\frac{1 - g^2_A}{1 + 3 g^2_A}\, \Big(\frac{(\vec{\xi}_e\cdot
  \vec{k}_{\bar{\nu}})(\vec{k}_e\cdot \vec{k}_{\nu})}{E_e
  E^2_{\bar{\nu}}} - \frac{1}{3}\,\frac{\vec{\xi}_e\cdot
  \vec{k}_e}{E_e}\,\Big)\,\frac{m_e}{E_e} \nonumber\\
\hspace{-0.15in}&& + 3\,\frac{1 - g^2_A}{1 + 3 g^2_A}\,
\Big(\frac{(\vec{\xi}_e\cdot \vec{k}_e)(\vec{k}_e\cdot
  \vec{k}_{\bar{\nu}})^2}{(E_e + m_e)E^2_e E^2_{\bar{\nu}}} -
\frac{1}{3}\,\Big(1 - \frac{m_e}{E_e}\Big)\,\frac{\vec{\xi}_e\cdot
  \vec{k}_e}{E_e}\,\Big)\Big\}
\end{eqnarray}
and 
\begin{eqnarray}\label{eq:F.4}
\hspace{-0.15in}&&\frac{d^5 \lambda^{(1)}_n(E_e, \vec{k}_e,
  \vec{k}_{\bar{\nu}}, \vec{\xi}_n, \vec{\xi}_e)}{dE_e d\Omega_e
  d\Omega_{\bar{\nu}}}\Big|_{\rm N^2LO} = (1 + 3
g^2_A)\,\frac{|G_V|^2}{16\pi^5}\,(E_0 - E_e)^2 \,\sqrt{E^2_e -
  m^2_e}\, E_e\,F(E_e, Z = 1)\,\zeta(E_e) \nonumber\\
\hspace{-0.15in}&& \times \,\bigg\{\frac{E^2_0}{2 m^2_N}\, \Big\{ -
\frac{1}{1 + 3 g^2_A}\, (g_A + 2\kappa + 1)\, \frac{E_e
  E_{\bar{\nu}}}{E^2_0}\Big(\frac{(\vec{k}_e\cdot
  \vec{k}_{\bar{\nu}})^2}{E^2_e E^2_{\bar{\nu}}} - \frac{1}{3}\,
\frac{k^2_e}{E^2_e}\Big) - \frac{1}{1 + 3 g^2_A}\, (g^2_A + 2 \kappa +
1) \, \frac{E_{\bar{\nu}}}{E_e + m_e}\nonumber\\
\hspace{-0.3in}&&\times \,\Big(
\frac{(\vec{k}_e\cdot \vec{k}_{\bar{\nu}})^2}{E^2_e E^2_{\bar{\nu}}} -
\frac{1}{3}\, \frac{k^2_e}{E^2_e}\Big)\, \frac{\vec{\xi}_e \cdot
  \vec{k}_e}{E_e} - \frac{1}{1 + 3 g^2_A}\, (g^2_A + 2\kappa + 1) \frac{m_e
  E_{\bar{\nu}}}{E^2_0}\Big(\frac{(\vec{\xi}_e \cdot
  \vec{k}_{\bar{\nu}})(\vec{k}_e \cdot \vec{k}_{\bar{\nu}})}{E_e
  E^2_{\bar{\nu}}} - \frac{1}{3}\,\frac{\vec{\xi}_e \cdot
  \vec{k}_e}{E_e} \Big) \nonumber\\
\hspace{-0.15in}&&+ \frac{1}{1 + 3 g^2_A} \, 2 \kappa g_A\, \frac{E_e
  E_{\bar{\nu}}}{E^2_0} \Big( \frac{(\vec{\xi}_n \cdot
  \vec{k}_{\bar{\nu}})(\vec{k}_e \cdot \vec{\xi}_e)(\vec{k}_e \cdot
  \vec{k}_{\bar{\nu}})}{E^2_e E^2_{\bar{\nu}}} - \frac{1}{3}\,
\frac{(\vec{\xi}_n \cdot \vec{k}_e)(\vec{k}_e \cdot \vec{\xi}_e)}{
  E^2_e } \Big) + \frac{1}{1 + 3 g^2_A} \, \Big(2 \kappa g_A \,
\frac{E_e E_{\bar{\nu}}}{E^2_0}\Big)\nonumber\\
\hspace{-0.15in}&& \times \, \frac{(\vec{\xi}_n \cdot
  \vec{k}_e)(\vec{k}_e \cdot \vec{\xi}_e)(\vec{k}_e \cdot
  \vec{k}_{\bar{\nu}})}{(E_e + m_e) E^2_e E_{\bar{\nu}}}\Big\} +
\Big\{ \frac{8 g_A}{1 + 3 g^2_A}\,\Big(\frac{E^2_0}{M^2_V} + (1 + 2
g_A)\, \frac{E^2_0}{M^2_A}\Big) \Big(\frac{(\vec{\xi}_n \cdot
  \vec{k}_{\bar{\nu}})(\vec{k}_e \cdot \vec{k}_{\bar{\nu}})}{E_e
  E^2_{\bar{\nu}}} - \frac{1}{3} \frac{\vec{\xi}_n \cdot
  \vec{k}_e}{E_e}\Big) \nonumber\\
\hspace{-0.15in}&&  - \frac{8 g_A}{1 + 3 g^2_A}
  \Big(\frac{E^2_0}{M^2_V} + (1 + 2 g_A) \frac{E^2_0}{M^2_A}\Big)
  \Big( \frac{(\vec{\xi}_n \cdot \vec{k}_{\bar{\nu}})(\vec{\xi}_e
    \cdot \vec{k}_e)(\vec{k}_e \cdot \vec{k}_{\bar{\nu}})}{E^2_e
    E^2_{\bar{\nu}}} - \frac{1}{3} \frac{(\vec{\xi}_n \cdot
    \vec{k}_e)(\vec{\xi}_e \cdot
    \vec{k}_e)}{E^2_e}\Big)\Big\}\,\frac{E_e}{E_0}\,\Big(1 -
  \frac{E_e}{E_0}\Big)\bigg\}
\end{eqnarray}
and
\begin{eqnarray*}
\hspace{-0.15in}&&\frac{d^5 \lambda^{(2)}_n(E_e, \vec{k}_e,
  \vec{k}_{\bar{\nu}}, \vec{\xi}_n, \vec{\xi}_e)}{dE_e d\Omega_e
  d\Omega_{\bar{\nu}}}\Big|_{\rm N^2LO} = (1 + 3
g^2_A)\,\frac{|G_V|^2}{16\pi^5}\,(E_0 - E_e)^2 \,\sqrt{E^2_e -
  m^2_e}\, E_e\,F(E_e, Z = 1)\,\zeta(E_e) \nonumber\\
\hspace{-0.15in}&& \times \, 6\, \frac{E^2_e}{m^2_N} \Big\{ \Big[1 -
  2\, \frac{1 - g^2_A}{1 + 3 g^2_A}\, \Big(1 - \frac{1}{8}\,
  \frac{E_0}{E_e}\Big)\Big] \,\Big(\frac{(\vec{k}_e\cdot
  \vec{k}_{\bar{\nu}})^2}{E^2_e E^2_{\bar{\nu}}} - \frac{1}{3}\,
\frac{k^2_e}{E^2_e}\Big) + 2 \, \frac{g_A(1 - g_A)}{1 + 3 g^2_A}
\Big(\frac{(\vec{k}_e\cdot \vec{k}_{\bar{\nu}})^2}{E^2_e
  E^2_{\bar{\nu}}} - \frac{1}{3}\, \frac{k^2_e}{E^2_e}\Big)
\frac{\vec{\xi}_n \cdot \vec{k}_e}{E_e} \nonumber\\
\hspace{-0.15in}&& + \Big[ - 1 + 2 \, \frac{1 - g^2_A}{1 + 3
    g^2_A}\,\Big(1 - \frac{1}{8}\, \frac{E_0}{E_e}\Big)\,
  \frac{E_e}{E_e + m_e}\Big] \, \Big(\frac{(\vec{k}_e\cdot
  \vec{k}_{\bar{\nu}})^2}{E^2_e E^2_{\bar{\nu}}} - \frac{1}{3}\,
\frac{k^2_e}{E^2_e}\Big)\, \frac{\vec{\xi}_e \cdot \vec{k}_e}{E_e} +
2\, \frac{1 - g^2_A}{1 + 3 g^2_A}\,\frac{m_e}{E_e}\,\Big(1 -
\frac{1}{8}\, \frac{E_0}{E_e}\Big) \nonumber\\
\hspace{-0.15in}&& \times \Big(\frac{(\vec{\xi}_e \cdot
  \vec{k}_{\bar{\nu}})(\vec{k}_e \cdot \vec{k}_{\bar{\nu}})}{E_e
  E^2_{\bar{\nu}}}- \frac{1}{3}\, \frac{\vec{\xi}_e \cdot
  \vec{k}_e}{E_e}\Big) - 2 \, \frac{g_A(1 - g_A)}{1 + 3 g^2_A}\,
\frac{m_e}{E_e}\,\Big(\frac{(\vec{k}_e\cdot
  \vec{k}_{\bar{\nu}})^2}{E^2_e E^2_{\bar{\nu}}} - \frac{1}{3}\,
\frac{k^2_e}{E^2_e}\Big) \, (\vec{\xi}_n \cdot \vec{\xi}_e) + 4\,
\frac{g_A(1 + g_A)}{1 + 3 g^2_A}\, \Big(1 - \frac{1}{8}\,
\frac{E_0}{E_e}\Big)\nonumber\\
\end{eqnarray*}
\begin{eqnarray}\label{eq:F.5}
\hspace{-0.15in}&& \times \Big(\frac{(\vec{\xi}_n \cdot
  \vec{k}_{\bar{\nu}})(\vec{k}_e \cdot \vec{\xi}_e)(\vec{k}_e \cdot
  \vec{k}_{\bar{\nu}})}{E^2_e E^2_{\bar{\nu}}} - \frac{1}{3}\,
\frac{(\vec{\xi}_n \cdot \vec{k}_e)(\vec{k}_e \cdot
  \vec{\xi}_e)}{E^2_e}\Big) + 2 \, \frac{g_A(1 - g_A)}{1 + 3 g^2_A}\,
\frac{m_e}{E_e}\,\Big(\frac{(\vec{k}_e\cdot
  \vec{k}_{\bar{\nu}})^2}{E^2_e E^2_{\bar{\nu}}} - \frac{1}{3}\,
\frac{k^2_e}{E^2_e}\Big) \,\frac{(\vec{\xi}_n \cdot
  \vec{k}_e)(\vec{k}_e \cdot \vec{\xi}_e)}{(E_e + m_e) E_e}
\nonumber\\
\hspace{-0.15in}&& + 4\, \frac{g_A (1 - g_A)}{1 + 3 g^2_A}\, \Big(1 -
\frac{1}{8}\, \frac{E_0}{E_e}\Big)\, \frac{(\vec{\xi}_n \cdot
  \vec{k}_e) (\vec{k}_e \cdot \vec{\xi}_e) (\vec{k}_e \cdot
  \vec{k}_{\bar{\nu}})}{(E_e + m_e) E^2_e E_{\bar{\nu}}} + 2 \,
\frac{g_A(1 + g_A)}{1 + 3 g^2_A}\frac{(\vec{\xi}_n \cdot
  \vec{k}_{\bar{\nu}})(\vec{k}_e \cdot \vec{k}_{\bar{\nu}})^2}{E^2_e
  E^3_{\bar{\nu}}}- 2 \, \frac{g_A(1 + g_A)}{1 + 3 g^2_A}\nonumber\\
\hspace{-0.15in}&& \times \frac{(\vec{\xi}_n \cdot
  \vec{k}_{\bar{\nu}})(\vec{k}_e \cdot \vec{\xi}_e)(\vec{k}_e \cdot
  \vec{k}_{\bar{\nu}})^2}{E^3_e E^3_{\bar{\nu}}} - \frac{1 - g^2_A}{1
  + 3 g^2_A}\, \frac{m_e}{E_e}\, \frac{(\vec{\xi}_e \cdot
  \vec{k}_{\bar{\nu}})(\vec{k}_e \cdot \vec{k}_{\bar{\nu}})^2}{E^2_e
  E^3_{\bar{\nu}}} - \frac{1 - g^2_A}{1 + 3 g^2_A}\,
\frac{(\vec{\xi}_e \cdot \vec{k}_e)(\vec{k}_e \cdot
  \vec{k}_{\bar{\nu}})^3}{(E_e + m_e) E^3_e E^3_{\bar{\nu}}}- \frac{1
  - g^2_A}{1 + 3 g^2_A}\, \frac{(\vec{k}_e \cdot
  \vec{k}_{\bar{\nu}})^3}{ E^3_e E^3_{\bar{\nu}}}\Big\}\nonumber\\
\hspace{-0.15in}&& 
\end{eqnarray}
and
\begin{eqnarray*}
\hspace{-0.30in}&&\frac{d^5 \lambda^{(3)}_n(E_e, \vec{k}_e,
  \vec{k}_{\bar{\nu}}, \vec{\xi}_n, \vec{\xi}_e)}{dE_e d\Omega_e
  d\Omega_{\bar{\nu}}}\Big|_{\rm N^2LO} = (1 + 3
g^2_A)\,\frac{|G_V|^2}{16\pi^5}\,(E_0 - E_e)^2 \,\sqrt{E^2_e -
  m^2_e}\, E_e\,F(E_e, Z = 1)\,\zeta(E_e) \nonumber\\
\hspace{-0.30in}&& \times \,\frac{E^2_0}{2 m^2_N}\,\Big\{ - \frac{1}{1
  + 3 g^2_A}\, \Big[(g^2_A + (\kappa + 1)^2)\,
  \frac{E_{\bar{\nu}}}{E_0}\Big] \Big(\frac{(\vec{k}_e \cdot
  \vec{k}_{\bar{\nu}})^2}{E^2_e E^2_{\bar{\nu}}} -
\frac{1}{3}\,\frac{k^2_e}{E^2_e}\Big) + \frac{1}{1 + 3 g^2_A}\,
 \Big[(g^2_A + (\kappa + 1)^2)\,
   \frac{m_e}{E_0}\, \frac{E_{\bar{\nu}}}{E_0}\Big]\nonumber\\
\end{eqnarray*}
\begin{eqnarray}\label{eq:F.6} 
\hspace{-0.30in}&& \times \Big(\frac{(\vec{\xi}_e \cdot
  \vec{k}_{\bar{\nu}})(\vec{k}_e \cdot \vec{k}_{\bar{\nu}})}{E_e
  E^2_{\bar{\nu}}} - \frac{1}{3}\, \frac{\vec{\xi}_e \cdot
  \vec{k}_e}{E_e}\Big) + \frac{1}{1 + 3 g^2_A}\,\Big[(g^2_A + (\kappa
  + 1)^2)\, \frac{E_e E_{\bar{\nu}}}{E^2_0}\Big] \Big(\frac{(\vec{k}_e
  \cdot \vec{k}_{\bar{\nu}})^2}{E^2_e E^2_{\bar{\nu}}} -
\frac{1}{3}\,\frac{k^2_e}{E^2_e}\Big)\, \frac{\vec{\xi}_e \cdot
  \vec{k}_e}{E_e + m_e} \nonumber\\
\hspace{-0.30in}&& + \frac{1}{1 + 3 g^2_A}\, \,
\frac{m_e}{E_e}\, \Big[(\kappa + 1)\big(g_A + (\kappa + 1)\big)\,
\frac{E^2_{\bar{\nu}}}{E^2_0} + \big(g_A + (\kappa + 1)\big)\,
\frac{E_e E_{\bar{\nu}}}{E^2_0} + g_A \big(g_A + (\kappa +
1)\big)\,\frac{E_{\bar{\nu}}}{E_0}\Big] \nonumber\\
\hspace{-0.30in}&& \times \Big(\frac{(\vec{\xi}_n \cdot
  \vec{k}_{\bar{\nu}})(\vec{k}_{\bar{\nu}} \cdot
  \vec{\xi}_e)}{E^2_{\bar{\nu}}} - \frac{1}{3}\, \vec{\xi}_n \cdot
\vec{\xi}_e\Big) + \frac{1}{1 + 3 g^2_A}\, \Big[g_A \big(g_A + (\kappa
  + 1)\big)\, \frac{E_{\bar{\nu}}}{E_0} + \big(g_A + (\kappa +
  1)\big)\, \frac{E_e E_{\bar{\nu}}}{E^2_0} \nonumber\\
\hspace{-0.30in}&& + (\kappa + 1)\big(g_A + (\kappa +
1)\big)\,\frac{E^2_{\bar{\nu}}}{E^2_0} - (\kappa + 2)\big(g_A +
(\kappa + 1)\big)\,\frac{E_e E_{\bar{\nu}}}{E^2_0}\Big(1 +
\frac{m_e}{E_e}\Big)\Big] \Big(\frac{(\vec{\xi}_n \cdot
  \vec{k}_{\bar{\nu}})(\vec{k}_e \cdot \vec{k}_{\bar{\nu}})}{E_e
  E^2_{\bar{\nu}}} - \frac{1}{3}\, \frac{\vec{\xi}_n \cdot
  \vec{k}_e}{E_e}\Big)\nonumber\\
\hspace{-0.30in}&& \times \frac{\vec{k}_e \cdot \vec{\xi}_e}{E_e +
  m_e} + \frac{1}{1 + 3 g^2_A}\,\Big[\kappa \big(g_A - (\kappa +
  1)\big)\, \frac{m_e}{E_0}\, \frac{E_e}{E_0} - g_A \big(g_A - (\kappa
  + 1)\big) \, \frac{E_e}{E_0} + (\kappa + 1) \big(g_A - (\kappa +
  1)\big) \, \frac{E^2_e}{E^2_0} \nonumber\\
\hspace{-0.30in}&& + (\kappa + 1)\big(3 g_A + (\kappa + 1)\big)\,
\frac{E_e E_{\bar{\nu}}}{E^2_0}\Big]\, \frac{(\vec{\xi}_n \cdot
  \vec{k}_e)(\vec{k}_e \cdot \vec{\xi}_e)(\vec{k}_e \cdot
  \vec{k}_{\bar{\nu}})}{(E_e + m_e) E^2_e E_{\bar{\nu}}}\Big\}
\end{eqnarray}
and
\begin{eqnarray}\label{eq:F.7}
\hspace{-0.15in}&&\frac{d^5 \lambda^{(4)}_n(E_e, \vec{k}_e,
  \vec{k}_{\bar{\nu}}, \vec{\xi}_n, \vec{\xi}_e)}{dE_e d\Omega_e
  d\Omega_{\bar{\nu}}}\Big|_{\rm N^2LO} = (1 + 3
g^2_A)\,\frac{|G_V|^2}{16\pi^5}\,(E_0 - E_e)^2 \,\sqrt{E^2_e -
  m^2_e}\, E_e\,F(E_e, Z = 1)\,\zeta(E_e) \nonumber\\
\hspace{-0.15in}&& \times \,3\, \frac{E_e}{m_N} \Big\{ -
\bar{K}_n(E_e)_{\rm NLO}\Big(\frac{(\vec{k}_e \cdot
  \vec{k}_{\bar{\nu}})^2}{E^2_e E^2_{\bar{\nu}}} - \frac{1}{3}\,
\frac{k^2_e}{E^2_e}\Big) - \bar{Q}_n(E_e)_{\rm NLO}
\frac{(\vec{\xi}_n \cdot \vec{k}_{\bar{\nu}})(\vec{k}_e \cdot
  \vec{k}_{\bar{\nu}})^2}{E^2_e E^3_{\bar{\nu}}} -
\bar{H}(E_e)_{\rm NLO}\nonumber\\
\hspace{-0.15in}&& \times \,\Big(\frac{(\vec{\xi}_e \cdot
  \vec{k}_{\bar{\nu}})(\vec{k}_e \cdot \vec{k}_{\bar{\nu}})}{E_e
  E^2_{\bar{\nu}}} - \frac{1}{3}\, \frac{\vec{\xi}_e \cdot
  \vec{k}_e}{E_e}\Big) - \bar{Q}_e(E_e)_{\rm NLO} \,
\frac{(\vec{\xi}_n \cdot \vec{k}_e)(\vec{k}_e \cdot
  \vec{\xi}_e)(\vec{k}_e \cdot \vec{k}_{\bar{\nu}})}{(E_e + m_e) E^2_e
  E_{\bar{\nu}}} \nonumber\\
\hspace{-0.15in}&& - \bar{K}_e(E_e)_{\rm NLO}\Big(\frac{(\vec{k}_e
  \cdot \vec{k}_{\bar{\nu}})^2}{E^2_e E^2_{\bar{\nu}}} - \frac{1}{3}\,
\frac{k^2_e}{E^2_e}\Big)\, \frac{\vec{\xi}_e \cdot \vec{k}_e}{E_e +
  m_e}\Big\} + 3\, \frac{E_e}{m_N}\Big(1
- \frac{\vec{k}_e \cdot \vec{k}_{\bar{\nu}}}{E_e
    E_{\bar{\nu}}}\Big)\frac{d^5 \lambda_n(E_e, \vec{k}_e,
    \vec{k}_{\bar{\nu}}, \vec{\xi}_n, \vec{\xi}_e)}{dE_e d\Omega_e
    d\Omega_{\bar{\nu}}}\Big|_{\rm NLO},
\end{eqnarray}
where $\bar{K}_n(E_e)_{\rm NLO}$ and $\bar{Q}_n(E_e)_{\rm NLO}$
coincide with $K_n(E_e)_{\rm NLO}$ and $Q_n(E_e)_{\rm NLO}$, which are
given in Eq.(\ref{eq:B.2}), whereas $\bar{H}(E_e)$, $\bar{Q}_e(E_e)$
and $\bar{K}_e(E_e)$ are defined by the expressions in
Eq.(\ref{eq:B.3}). The last term in Eq.(\ref{eq:F.1}) is equal to
\begin{eqnarray}\label{eq:F.8}
\hspace{-0.15in}&&\frac{d^5 \lambda_n(E_e, \vec{k}_e,
  \vec{k}_{\bar{\nu}}, \vec{\xi}_n, \vec{\xi}_e)}{dE_e d\Omega_e
  d\Omega_{\bar{\nu}}}\Big|_{\rm WC} = (1 + 3
g^2_A)\,\frac{|G_V|^2}{16 \pi^5}\,(E_0 - E_e)^2 \,\sqrt{E^2_e -
  m^2_e}\, E_e\,F(E_e, Z = 1)\nonumber\\
\hspace{-0.15in}&& \times \,\Big\{ - B_0\,
\frac{\pi\alpha}{\beta^3}\,\frac{E_0 - E_e}{m_N}
\Big(\frac{(\vec{\xi}_n \cdot \vec{k}_{\bar{\nu}})(\vec{\xi}_e \cdot
  \vec{k}_e)(\vec{k}_e \cdot
  \vec{k}_{\bar{\nu}})}{E^2_eE^2_{\bar{\nu}}} - \frac{1}{3}\,
\frac{(\vec{\xi}_n \cdot \vec{k}_e)(\vec{\xi}_e \cdot
  \vec{k}_e)}{E^2_e}\Big)\Big\}.
\end{eqnarray}
These contributions to the electron-energy and angular distribution of
the neutron beta decay vanish after the integration over the
directions of the antineutrino 3-momentum
$\vec{k}_{\bar{\nu}}$. Because of the contributions of Wilkinson's
corrections, caused by the proton recoil in the electron-proton
final-state Coulomb interaction, and the $O(\alpha E_e/m_N)$ outer
radiative corrections (see Eq.(\ref{eq:D.10})) the electron-energy and
angular distributions in Eqs.(\ref{eq:F.2}) - (\ref{eq:F.8}) are well
defined in the experimental electron-energy region $0.811\,{\rm MeV}
\le E_e \le 1.211\,{\rm MeV}$ \cite{Abele2018}.


\newpage


\begin{thebibliography}{9}
\bibitem{Abele2008} H. Abele, {\it The neutron. Its properties and
  basic interactions}, Progr. Part. Nucl. Phys. {\bf 60}, 1 (2008);
  \\ DOI: https://doi.org/10.1016/j.ppnp.2007.05.002.

\bibitem{Nico2009} J. S. Nico, {\it Neutron beta decay}, J. Phys. G:
  Nucl. Part. Phys. {\bf 36}, 104001 (2009); \\ DOI:
  https://doi.org/10.1088/0954-3899/36/10/104001.

\bibitem{Paul2009} S. Paul, {\it The puzzle of neutron lifetime},
  Nucl. Instrum. Meth. A {\bf 611}, 157 (2009); \\ DOI:
  10.1016/j.nima.2009.07.095.

\bibitem{Abele2016} H. Abele, {\it Precision experiments with cold and
  ultra-cold neutrons}, Hyperfine Interact. {\bf 237}, 155  (2016);
  \\ DOI: https://doi.org/10.1007/s10751-016-1352-z.


\bibitem{Abele2018} B. M\"arkisch, H. Mest, H. Saul, X. Wang,
  H. Abele, D. Dubbers, M. Klopf, A. Petoukhov, C. Roick, T. Soldner,
  and D. Werder, {\it Measurement of the weak axial-vector coupling
    constant in the decay of free neutrons using a pulsed cold neutron
    beam}, Phys. Rev. Lett. {\bf 122}, 242501 (2019); \\ DOI:
  https://doi.org/10.1103/PhysRevLett.122.242501; arXiv: 1812.04666
  [nucl-ex].

\bibitem{Sirlin2018} A. Czarnecki, W. J. Marciano, and A. Sirlin, {\it
  Neutron lifetime and axial coupling connection},
  Phys. Rev. Lett. {\bf 120}, 202002 (2018); \\ DOI:
  https://doi.org/10.1103/PhysRevLett.120.202002; arXiv: 1802.01804
  [hep-ph].

\bibitem{Dubbers2021} D. Dubbers and B. M\"arkisch, {\it Precise
  measurements of the decay of free neutrons}, Annual Review of
  Nuclear and Particle Science, {\bf 71}, 139 (2021); \\ DOI:
  10.1146/annurev-nucl-102419-043156.

  
\bibitem{Bodek2019} K. Bodek, L. De Keukeleere, M. Kolodziej,
  A. Kozela, M. Kuzniak, K. Lojek, M. Perkowski, H. Przybilski,
  K. Pysz, D. Rozpedzik, N. Severijns, T. Soldner, A. R. Young, and
  J. Zejma, {\it BRAND – Search for BSM physics at TeV scale by
    exploring transverse polarization of electrons emitted in neutron
    decay}, International Workshop on Particle Physics at Neutron
  Sources 2018 (PPNS 2018), EPJ Web of Conferences {\bf 219}, 04001
  (2019); \\ DOI: https://doi.org/10.1051/epjconf/201921904001.

\bibitem{DGH2014} J. F. Gonoghue, E. Golowich, and B. R. Holstein, in
  {\it Dynamics of the Standard Model}, 2nd edition, Cambridge
  University Press, Cambridge 2014; \\ DOI:
  https://doi.org/10.1017/CBO9780511803512.

\bibitem{PDG2020} P. A. Zyla {\it et al.}, {\it Review of particle
  physics} (Particle Data Group), Prog. Theor. Exp. Phys. {\bf 2020},
  083C01 (2020); \\ DOI: https://doi.org/10.1093/ptep/ptaa104.

\bibitem{Ivanov2013} A. N. Ivanov, M. Pitschmann, and
  N. I. Troitskaya, {\it Neutron $\beta$--decay as a laboratory for
    testing the standard model}, Phys. Rev. D {\bf 88}, 073002 (2013);
  \\ DOI: https://doi.org/10.1103/PhysRevD.88.073002; arXiv:1212.0332
     [hep--ph].

 \bibitem{Ivanov2017} A. N. Ivanov, R. H\"ollwieser, N. I. Troitskaya,
   M. Wellenzohn, and Ya. A. Berdnikov, {\it Precision analysis of
     electron energy spectrum and angular distribution of neutron beta
     decay with polarized neutron and electron}, Phys. Rev. C {\bf
     95}, 055502 (2017); \\ DOI: 10.1103/PhysRevC.95.055502;
   arXiv:1705.07330 [hep-ph].

\bibitem{Ivanov2018} A. N. Ivanov, R. H\"ollwieser, N. I. Troitskaya,
  M. Wellenzohn, and Ya. A. Berdnikov, {\it Tests of the standard
    model in neutron beta decay with polarized neutron and electron
    and an unpolarized proton}, Phys. Rev. C 98, 035503 (2018);
  \\ DOI: https://doi.org/10.1103/PhysRevC.98.035503; arXiv:1805.03880 [hep-ph].

\bibitem{Ivanov2019a} A. N. Ivanov, R. H\"ollwieser, N. I. Troitskaya,
  M. Wellenzohn, and Ya. A. Berdnikov, {\it Test of the Standard Model
    in neutron beta decay with polarized electrons and unpolarized
    neutrons and protons}, Phys. Rev. D 99, 053004 (2019); \\ DOI:
  10.1103/PhysRevD.99.053004; arXiv:1811.04853 [hep-ph].

\bibitem{Wilkinson1982} D. H. Wilkinson, {\it Analysis of neutron beta
  decay}, Nucl. Phys. A {\bf 377}, 474 (1982); \\ DOI:
  https://doi.org/10.1016/0375-9474(82)90051-3.
  
\bibitem{Ivanov2019b} A. N. Ivanov, R. H\"ollwieser, N. I. Troitskaya,
  M. Wellenzohn, and Ya. A. Berdnikov, {\it Radiative corrections of
    order $O(\alpha E_e/m_N)$ to Sirlin's radiative corrections of
    order $O(\alpha/\pi)$ to the neutron lifetime}, Phys. Rev. D {\bf
    99}, 093006 (2019);\\ DOI:
  https://doi.org/10.1103/PhysRevD.99.093006; arXiv:1905.01178
  [hep-ph].

\bibitem{Ivanov2020a} A. N. Ivanov, R. H\"ollwieser, N. I. Troitskaya,
  M. Wellenzohn, and Ya. A. Berdnikov, {\it Radiative corrections of
    order $O(\alpha E_e/m_N)$ to Sirlin's radiative corrections of
    order $O(\alpha/\pi)$, induced by hadronic structure of the
    neutron}, Phys. Rev. D {\bf 103}, 113007 (2021); \\ DOI:
  https://link.aps.org/doi/10.1103/PhysRevD.103.113007; arXiv:
  2105.06952 [hep-ph].

\bibitem{Wilkinson1970} D. H. Wilkinson and B. E. F. Macfield, {\it
  The numerical evaluation of radiative corrections of order $\alpha$
  to allowed nuclear $\beta$--decay}, Nucl. Phys. A {\bf 158}, 110
  (1970); DOI: https://doi.org/10.1016/0375-9474(70)90055-2.  

  
\bibitem{Sirlin1967} A. Sirlin, {\it General properties of the
  electromagnetic corrections to the beta decay of a physical
  nucleon}, Phys. Rev. {\bf 164}, 1767
  (1967);\\ DOI:https://doi.org/10.1103/PhysRev.164.1767.

\bibitem{Sirlin1978} A. Sirlin, {\it Current algebra formulation of
  radiative corrections in gauge theories and the universality of the
  weak interactions}, Rev. Mod. Phys. {\bf 50}, 573 (1978);\\ DOI:
  https://doi.org/10.1103/RevModPhys.50.573.



\bibitem{Shann1971} R. T. Shann, {\it Electromagnetic effects in the
  decay of polarized neutrons}, Nuovo Cimento A {\bf 5}, 591
  (1971).\\ DOI: https://doi.org/10.1007/BF02734566.




\bibitem{Ando2004} S. Ando, H. W. Fearing, V. Gudkov, K. Kubodera,
  F. Myhrer, S. Nakamura, and T. Sato, {\it Neutron beta--decay in
    effective field theory}, Phys. Lett. B {\bf 595}, 250 (2004);
  \\ DOI: https://doi.org/10.1016/j.physletb.2004.06.037.
  
\bibitem{Gudkov2006} V. Gudkov, G. I. Greene, and J. R. Calarco, {\it
  General classification and analysis of neutron beta-decay
  experiments}, Phys. Rev. C {\bf 73}, 035501 (2006); \\ DOI:
  https://doi.org/10.1103/PhysRevC.73.035501.

\bibitem{Ivanov2020b} A. N. Ivanov, R. H\"ollwieser, N. I. Troitskaya,
  M. Wellenzohn, and Ya. A. Berdnikov, {\it Corrections of order
    $O(E^2_e/m^2_N)$, caused by weak magnetism and proton recoil, to
    the neutron lifetime and correlation coefficients of the neutron
    beta decay}, Results in Physics, {\bf 21}, 103806 (2021); \\ DOI:
  https://doi.org/10.1016/j.rinp.2020.103806; arXiv: 2010.14336
  [hep-ph].


\bibitem{Jackson1957a} J. D. Jackson, S. B. Treiman, and H. W. Wyld
  Jr., {\it Possible tests of time reversal invariance in beta decay},
  Phys. Rev. {\bf 106}, 517 (1957); \\ DOI:
  https://doi.org/10.1103/PhysRev.106.517.

\bibitem{Jackson1957b} J. D. Jackson, S. B. Treiman, and H. W. Wyld
  Jr., {\it Coulomb corrections in allowed beta transitions},
  Nucl. Phys. {\bf 4}, 206 (1957); \\ DOI:
  https://doi.org/10.1016/0029-5582(87)90019-8.


\bibitem{Jackson1958} J. D. Jackson, S. B. Treiman, and H. W. Wyld,
  Jr., {\it Note on relativistic coulomb wave functions},
  Z. Phys. {\bf 150}, 640 (1958); \\ DOI:
  https://doi.org/10.1007/BF01340460.

\bibitem{Ebel1957} M. E. Ebel and G. Feldman, {\it Further remarks on
  Coulomb corrections in allowed beta transitions}, Nucl. Phys. {\bf
  4}, 213 (1957); \\ DOI:
  https://doi.org/10.1016/0029-5582(87)90020-4.

\bibitem{Ivanov2021a} A. N. Ivanov, R. H\"ollwieser, N. I. Troitskaya,
  M. Wellenzohn, and Ya. A. Berdnikov, {\it On the correlation
    coefficient $T(E_e)$ of the neutron beta decay, caused by the
    correlation structure invariant under discrete P, C and T
    symmetries}; Phys. Lett. B {\bf 816}, 136263 (2021); \\ DOI:
  https://doi.org/10.1016/j.physletb.2021.136263; arXiv: 2101.01014
  [hep-ph].

\bibitem{Ivanov2021b} A. N. Ivanov, R. H\"ollwieser, N. I. Troitskaya,
  M. Wellenzohn, and Ya. A. Berdnikov, {\it On the structure of the
    correlation coefficients $S(E_e)$ and $U(E_e)$ of the neutron beta
    decay}; Phys. Rev. C {\bf 104}, 025503 (2021); \\ DOI:
  https://doi.org/10.1103/PhysRevC.104.025503; arXiv: 2102.02021
  [hep-ph].

\bibitem{Itzykson1980} C. Itzykson and J.--B. Zuber, in {\it Quantum
  field theory}, McGraw--Hill Inc., New York, 1980.

\bibitem{Blatt1952} 
J. M. Blatt and V. F. Weisskopf,  {\it Theoretical nuclear physics},
John Wily $\&$ Sons, New York 1952.

  
\bibitem{Antognini2013} A.  Antognini {\it et al.}, {\it Proton
   structure from the measurement of 2S-2P transition frequencies of
   muonic hydrogen}, Science {\bf 339} 417 (2013); \\ DOI:
   10.1126/science.1230016.


\bibitem{Fierz1937} M. Fierz, {\it Zur Fermischen Theorie des
  $\beta$-Zerfalls}, Z. Physik {\bf 104}, 553 (1937); \\ DOI:
  https://doi.org/10.1007/BF01330070.

\bibitem{Hardy2020} J. C. Hardy and I. S. Towner, {\it Superallowed
  $0^+ \to 0^+$ nuclear beta decays: 2020 critical survey, with
  implications for $V_{ud}$ and CKM unitarity}, Phys. Rev. C {\bf
  102}, 045501 (2020); \\ DOI:
  https://doi.org/10.1103/PhysRevC.102.045501.

\bibitem{Severijns2019} M. Gonz\'alez--Alonso, O. Naviliat--Cuncic,
  and N. Severijns, {\it New physics searches in nuclear and neutron
    beta decay}, Prog. Part. Nucl. Phys. {\bf 104}, 165 (2019);
  \\ DOI: https://doi.org/10.1016/j.ppnp.2018.08.002.

\bibitem{Abele2019} H. Saul, Ch. Roick, H. Abele, H. Mest, M. Klopf,
  A. Petukhov, T. Soldner, X. Wang, D. Werder, and B.  M\"arkisch,
  {\it Limit on the Fierz interference term b from a measurement of
    the beta asymmetry in neutron decay}, Phys. Rev. Lett. {\bf 125},
  112501 (2020); \\ DOI:
  https://doi.org/10.1103/PhysRevLett.125.112501.

\bibitem{Young2019} V. Cirigliano, A. Garcia, D. Gazit,
  O. Naviliat-Cuncic, G. Savard, and A. Young, {\it Precision beta
    decay as a probe of new physics}, arXiv:1907.02164 [nucl-ex].

\bibitem{Sun2020} X. Sun {\it et al.}, {\it Improved limits on Fierz
  interference using asymmetry measurements from the ultracold neutron
  asymmetry (UCNA) experiment} (UCNA Collaboration), Phys.  Rev. C
  {\bf 101}, 035503 (2020); \\ DOI:
  https://doi.org/10.1103/PhysRevC.101.035503.

\bibitem{Ivanov2019y} A. N. Ivanov, R. H\"ollwieser, N. I. Troitskaya,
  M. Wellenzohn, and Ya. A. Berdnikov, {\it Neutron dark matter decays
    and correlation coefficients of neutron beta decays},
  Nucl. Phys. B {\bf 938}, 114 (2019); \\ DOI:
  https://doi.org/10.1016/j.nuclphysb.2018.11.005.

\bibitem{Ivanov2019x} A. N. Ivanov, R. H\"ollwieser, N. I. Troitskaya,
  M. Wellenzohn, and Ya. A. Berdnikov, {\it Precision analysis of
    pseudoscalar interactions in neutron beta decays}, Nucl. Phys. B
  {\bf 951}, 114891 (2020); \\ DOI:
  https://doi.org/10.1016/j.nuclphysb.2019.114891; arXiv:1905.04147
  [hep-ph].

\bibitem{Lee1956a} T. D. Lee and C. N. Yang, {\it
    Charge conjugation, a new quantum number $G$ , and selection rules
    concerning a nucleon anti-nucleon system}, Nuovo Cimento {\bf 10},
  749 (1956); \\ DOI: https://doi.org/10.1007/BF02744530.

\bibitem{Weinberg1958} S. Weinberg, {\it Charge symmetry of weak
  interactions}, Phys. Rev. {\bf 112}, 1375 (1958); \\ DOI:
  https://doi.org/10.1103/PhysRev.112.1375.
  
\bibitem{Gardner2001} S. Gardner and C. Zhang, {\it Sharpening
  low-energy, Standard-Model tests via correlation coefficients in
  neutron beta decay}, Phys. Rev. Lett. {\bf 86}, 5666 (2001); \\ DOI:
  https://doi.org/10.1103/PhysRevLett.86.5666.

\bibitem{Gardner2013} S. Gardner and B. Plaster, {\it Framework for
  maximum likelihood analysis of neutron beta decay observables to
  resolve the limits of the V - A law}, Phys. Rev. C {\bf 87}, 065504
  (2013); The contribution to 4th International Conference on Particle
  Physics and Astrophysics (ICPPA-2018) 22–26 October 2018, Moscow,
  Russian Federation;\\ DOI:
  https://doi.org/10.1103/PhysRevC.87.065504.

\bibitem{MathW} The numerical analysis of the correlation function
  $\zeta(E_e)$ and correlation coefficients $a(E_e)$, $A(E_e)$,
  $\ldots$, $U(E_e)$ we have carried out by Wolfram Mathematica
  12. The analytical expressions for the correlation function and
  correlation coefficients are given in Appendix E programmed in the
  nb file, where one may find plotted radiative corrections $O(\alpha
  E_e/m_N)$ in Eqs.(\ref{eq:A.20}) and (\ref{eq:D.10}). This nb file
  can be sent by a request for practical applications of the results,
  obtained in this work.

\bibitem{Bilenky1959} S. M. Bilen'kii, R. M. Ryndin,
  Ya. A. Smorodinskii, and Ho Tso-Hsiu, {\it On the theory of the neutron
  beta decay}, JETP {\bf 37}, 1759 (1959) (in Russian);
  Sov. Phys. JETP, {\bf 37}(10), 1241 (1960).

\bibitem{Sirlin1986} W. J. Marciano and A. Sirlin, {\it Radiative
  corrections to $\beta$ decay and the possibility of a fourth
  generation}, Phys. Rev. Lett.  {\bf 56}, 22 (1986); \\ DOI:
  https://doi.org/10.1103/PhysRevLett.56.22.

\bibitem{Sirlin2004} A. Czarnecki, W. J. Marciano, and A. Sirlin, {\it
  Precision measurements and CKM unitarity}, Phys. Rev. D {\bf 70},
  093006 (2004);\\ DOI: 10.1103/PhysRevD.70.093006.

\bibitem{Sirlin2006} W. J. Marciano and A. Sirlin, {\it Improved
  calculation of electroweak radiative corrections and the value of
  $V(ud)$}, Phys. Rev. Lett.  {\bf 96}, 032002 (2006); \\ DOI:
  10.1103/PhysRevLett.96.032002.

\bibitem{Seng2018} Ch.-Y. Seng, M. Gorchtein, H. H. Patel, and
  M. J. Ramsey-Musolf, {\it Reduced hadronic uncertainty in the
    determination of $V_{ ud}$}, Phys. Rev. Lett. {\bf 121}, 241804
  (2018); \\ DOI: 10.1103/PhysRevLett.121.241804; arXiv:1807.10197
  [hep-ph].

\bibitem{Seng2018a} Ch.-Y. Seng, M. Gorchtein, and
  M. J. Ramsey-Musolf, {\it Dispersive evaluation of the inner
    radiative correction in neutron and nuclear beta decay}, Phys. Rev. D {\bf 100}, 013001(2019); \\ DOI:  10.1103/PhysRevD.100.013001;
  arXiv:1812.03352 [nucl-th].
  
\bibitem{Sirlin2019} A. Czarnecki, W. J. Marciano, and A. Sirlin, {\it
  Radiative corrections to neutron and nuclear beta decays revisited},
  Phys. Rev. D {\bf 100}, 073008 (2019);\\ DOI:
  10.1103/PhysRevD.100.073008.

\bibitem{Hayen2020} L. Hayen, {\it Standard Model $(\alpha)$
  renormalization of $g_A$ and its impact on new physics searches},
  Phys. Rev. D {\bf 103}, 113001 (2021); \\ DOI:
  https://doi.org/10.1103/PhysRevD.103.113001; arXiv: 2010.07262
  [hep-ph].
  
\bibitem{Hayen2021} L. Hayen, {\it Radiative corrections to nucleon
  weak charges and beyond Standard Model impact}, arXiv: 2102.03458
  [hep-ph].

\bibitem{Gorchtein2021} M. Gorchtein and Ch.-Y. Seng, {\it Dispersion
  relation analysis of the radiative corrections to $g_A$ in the
  neutron beta decay}, arXiv: 2106.09185 [hep-ph].

\bibitem{Berman1958} S. M. Berman, {\it Radiative corrections to muon
  and neutron decay}, Phys. Rev. {\bf 112}, 267 (1958); \\ DOI:
  https://doi.org/10.1103/PhysRev.112.267.

\bibitem{Kinoshita1959} T. Kinoshita and A. Sirlin, {\it Radiative
  corrections to Fermi interactions}, Phys. Rev. {\bf 113}, 1652
  (1959); \\ DOI: https://doi.org/10.1103/PhysRev.113.1652.
  
\bibitem{Berman1962} S. M. Berman and A. Sirlin,{\it Some
  considerations on the radiative corrections to muon and neutron
  decay}, Ann. Phys.  (N.Y.)  {\bf 20}, 20 (1962); \\ DOI:
  https://doi.org/10.1016/0003-4916(62)90114-8.


\bibitem{Kaellen1967} G. K\"all${\acute{\rm e}}$n, {\it Radiative
  corrections to beta decay and nucleon form factors}, Nucl. Phys. B
  {\bf 1}, 225 (1967); \\ DOI:
https://doi.org/10.1016/0550-3213(67)90125-3.


\bibitem{Abers1968} E. S. Abers, D. A. Dicus, R. E. Norton, and
  H. R. Queen, {\it Radiative corrections to the Fermi part of
    strangeness-conserving beta decay}, Phys. Rev. {\bf 167}, 1461
  (1968); \\ DOI: https://doi.org/10.1103/PhysRev.167.1461.

\bibitem{Garcia1978} A. Garc\'ia and M. Maya, {\it First-order radiative
  corrections to asymmetry coefficients in neutron decay},
  Phys. Rev. D {\bf 17}, 1376 (1978); \\ DOI:
  https://doi.org/10.1103/PhysRevD.17.1376.

\bibitem{Gaponov1996} Yu. V. Gaponov and R. U. Khafisov, {\it
  Radiative neutron beta decay and its possible experimental
  realization}, Phys. Lett. B {\bf 379}, 7 (1996); \\ DOI:
  https://doi.org/10.1016/0370-2693(96)00398-X.
\bibitem{Bernard2004}
  V. Bernard, S. Gardner, Ulf-G. Mei\ss ner, and Chi Zang, {\it
  Radiative neutron beta decay in effective field theory},
Phys. Lett. B {\bf 593}, 105 (2004); \\ DOI:
https://doi.org/10.1016/j.physletb.2004.04.064

   
\bibitem{Gluck1993} F. Gl\"uck, {\it Measurable distributions of
  unpolarized neutron decay }, Phys. Rev. D {\bf 47}, 2840 (1993);
  \\ DOI: https://doi.org/10.1103/PhysRevD.47.2840. 

\bibitem{Gluck1995} F. Gl\"uck, I. Jo\'o, and J. Last, {\it Measurable
  parameters of neutron decay}, Nucl. Phys. A {\bf 593}, 125 (1995);
  \\ DOI: https://doi.org/10.1016/0375-9474(95)00354-4.
  
\bibitem{Gluck1996} F. Gl\"uck, {\it Monte Carlo type radiative
  corrections for neutron, muon and hyperon semileptonic decays},
  Comput. Phys. Commun. {\bf 95}, 111 (1996); \\ DOI:
  https://doi.org/10.1016/0010-4655(96)00015-X.
  
\bibitem{Gluck1997} F. Gl\"uck, {\it Order-alpha radiative correction
  calculations for unoriented allowed nuclear, neutron and pion beta
  decays}, Comput. Phys. Commun. {\bf 101}, 223 (1997); \\ DOI:
  https://doi.org/10.1016/S0010-4655(96)00168-3.
 

\bibitem{Gluck1998} F. Gl\"uck, {\it Electron spectra and
  electron-proton asymmetries in polarized neutron decay},
  Phys. Lett. B {\bf 436}, 25 (1998); \\ DOI:
  https://doi.org/10.1016/S0370-2693(98)00881-8.


\bibitem{Mund2013} D. Mund, B. M\"arkisch, M. Deissenroth, J. Krempel,
  M. Schumann, and H. Abele, A. Petoukhov, and T. Soldner, {\it
    Determination of the weak axial vector coupling from a measurement
    of the beta-asymmetry parameter A in neutron beta decay},
  Phys. Rev. Lett. {\bf 110}, 172502 (2013).

\bibitem{Mendenhall2013} M. P. Mendenhall, R. W. Pattie, Jr.,
  Y. Bagdasarova, D. B. Berguno, L. J. Broussard, R. Carr, S. Currie
  {\it et al.}, (UCNA Collaboration), {\it Precision measurement of
    the neutron beta decay asymmetry}, Phys. Rev. C {\bf 87}, 032501
  (2013).
  
\bibitem{Brown2018} M. A.-P. Brown {\it et al.} (the UCNA
  Collaboration), {\it New result for the neutron beta-asymmetry
    parameter $A_0$ from UCNA}, Phys. Rev. C {\bf 97}, 035505 (2018).

\bibitem{Serebrov2019} A. P. Serebrov, O. M. Zherebtsov,
  A. N. Murashkin, G. N. Klyushnikov, and A. K. Fomin, {\it On the
    possibility of measuring the ratio $G_A/G_V$ by means of polarized
    ultracold neutrons}, Phys. Atom. Nucl. {\bf 82}, 98 (2019);
  Yad. Fiz. {\bf 82}, 110 (2019); \\ DOI:
  https://doi.org/10.1134/S1063778819020121.

 \bibitem{Serebrov1998} I. A. Kuznetsov, A. P. Serebrov,
  I. V. Stepanenko, A. V. Alduschenkov, M. S. Lasakov, A. A. Kokin,
  Yu. A. Mostovoi, B. G. Yerozolimsky, and M. S. Dewey, {\it
    Measurements of the antineutrino spin asymmetry in beta decay of
    the neutron and restrictions on the mass of a right-handed gauge
    boson}, Phys. Rev. Lett. {\bf 75}, 794 (1998); \\ DOI:
  https://doi.org/10.1103/PhysRevLett.75.794.

  \bibitem{Abele2005} M. Kreuz, T. Soldner, S. B\"a\ss ler, B. Brand,
  F. Gl\"uck, U. Mayer, D. Mund, V. Nesvizhevsky, A. Petoukhov,
  C. Plonka, J. Reich, C. Vogel, and H. Abele, {\it A measurement of
    the antineutrino asymmetry B in free neutron decay}, Phys. Lett. B
  {\bf 619}, 263 (2005); \\ DOI:
  https://doi.org/10.1016/j.physletb.2005.05.074.
  
 \bibitem{Schumann2007} M. Schumann, T. Soldner, M. Deissenroth,
  F. Gl\"uck, J. Krempel, M. Kreuz, B. Märkisch, D. Mund,
  A. Petoukhov, and H. Abele, {\it Measurement of the neutrino
    asymmetry parameter B in neutron decay}, Phys. Rev. Lett. {\bf
    99}, 191803 (2007); \\ DOI:
  https://doi.org/10.1103/PhysRevLett.99.191803.
  
 \bibitem{Nico2017} G. Darius, W. A. Byron, C. R. DeAngelis {\it et
  al.}, {\it Measurement of the electron-antineutrino angular
  correlation in neutron beta decay}, Phys. Rev. Lett. {\bf 119},
  042502 (2017); \\ DOI:
  https://doi.org/10.1103/PhysRevLett.119.042502. 

\bibitem{Beck2020} M. Beck, F. Ayala Guardia, M. Borg {\it et al.},
  {\it Improved determination of the $\beta-\bar{\nu}_e$ angular
    correlation coefficient a in free neutron decay with the aSPECT
    spectrometer}, Phys. Rev. C {\bf 101}, 055506 (2020); \\ DOI:
  https://doi.org/10.1103/PhysRevC.101.055506.

\bibitem{Nico2021} M. T. Hassan, W. A. Byron, G. Darius {\it et al.},
  {\it Measurement of the neutron decay electron-antineutrino angular
    correlation by the aCORN experiment}, Phys. Rev. C {\bf 103},
  045502 (2021); \\ DOI: https://doi.org/10.1103/PhysRevC.103.045502.
 
\bibitem{Schumann2008} M. Schumann, M. Kreuz, M. Deissenroth,
  F. Gl\"uck, J. Krempel, B. M\"arkisch, D. Mund, A. Petoukhov,
  T. Soldner, and H. Abele, {\it Measurement of the proton asymmetry
    parameter in neutron beta decay}, Phys. Rev. Lett. {\bf 100},
  151801 (2008); \\ DOI:
  https://doi.org/10.1103/PhysRevLett.100.151801.
  
\bibitem{Ivanov2013a} A. N. Ivanov, R. H\"ollwieser,
   N. I. Troitskaya, and M. Wellenzohn, {\it Proton recoil energy and
     angular distribution of neutron radiative beta decay },
   Phys. Rev. D {\bf 88}, 065026 (2013); \\ DOI:
   https://doi.org/10.1103/PhysRevD.88.065026; arXiv: 1306.4448
   [hep-ph].
  
\bibitem{Ivanov2017a} A. N. Ivanov, R. H\"ollwieser, N. I. Troitskaya,
  M. Wellenzohn, and Ya. A. Berdnikov, {\it Precision theoretical
    analysis of neutron radiative beta decay}, Phys. Rev. C 95, 033007
  (2017); \\ DOI: https://doi.org/10.1103/PhysRevD.95.033007; arXiv:
  1701.04613 [hep-ph].

\bibitem{Mitchell1949} K. Mitchell, {\it XXXII. Tables of the
  functions $- \int^z_0\frac{{\rm log}(1 - y)}{y}\,dy$, with an
  account for some properties of this and related functions}, The
  London, Edinburgh, and Dublin Philosophical Magazine and Journal of
  Science, {\bf 40}, 351 (1949); \\ DOI:
  https://doi.org/10.1080/14786444908561256.

\bibitem{Bernard1995} V. Bernard, N. Kaiser, and Ulf-G. Mei\ss ner,
  {\it Chiral dynamics in nucleons and nuclei}, Int. J. Mod. Phys. E
  {\bf 4}, 193 (1995); \\ DOI:
  https://doi.org/10.1142/S0218301395000092.
  
\bibitem{Liesenfeld1999} A. Liesenfeld {\it et al.}, {\it A
  measurement of the axial form factor of the nucleon by the
  $p(e,e'\pi^+)n$ reaction at $W = 1125$\,MeV}, Phys. Lett. B {\bf
  468},20 (1999); \\ DOI:
  https://doi.org/10.1016/S0370-2693(99)01204-6.


\bibitem{Ivanov2019c} A. N. Ivanov, R. H\"ollwieser, N. I. Troitskaya,
  M. Wellenzohn, and Ya. A. Berdnikov, {\it Precision analysis of
    pseudoscalar interactions in neutron beta decays}, Nucl. Phys. B
  {\bf 951}, 114891 (2020); \\ DOI:
  https://doi.org/10.1016/j.nuclphysb.2019.114891; arXiv:1905.04147
  [hep-ph].


\bibitem{Shekhter1959} V. M. Shekhter, {\it On the weak-interaction
  types possible in the scheme of Feynman and Gell-Mann}, Soviet
  Physics JETP {\bf 35}, 316 (1959)

  



 
  
  
\end{thebibliography}
\end{document}